\documentclass[english,11pt]{article}

\usepackage{amsmath}
\usepackage{amsthm}
\usepackage{amssymb}
\usepackage{stmaryrd}
\usepackage{graphicx}
\usepackage{setspace}
\usepackage{changepage}
\usepackage{pdflscape}
\usepackage{afterpage}
\usepackage{minitoc}
\usepackage{dcolumn}
\usepackage{multirow}
\usepackage{booktabs}
\usepackage{threeparttable}
\usepackage{subfig}
\usepackage[authoryear]{natbib}

\usepackage[
  verbose
  ,tmargin=1in
  ,bmargin=1in
  ,lmargin=1in
  ,rmargin=1in
]{geometry}

\usepackage[
  pdfusetitle
  ,bookmarks=true
  ,bookmarksnumbered=true
  ,bookmarksopen=false
  ,breaklinks=false
  ,pdfborder={0 0 0}
  ,pdfborderstyle={}
  ,backref=false
  ,colorlinks=false
]{hyperref}

\makeatletter
\theoremstyle{plain}
\newtheorem{assumption}{\protect\assumptionname}
\newtheorem{lem}{\protect\lemmaname}
\newtheorem{prop}{\protect\propositionname}

\newcolumntype{d}[1]{D{.}{.}{#1}}

\renewcommand{\thetable}{\Roman{table}}
\newlength{\bibitemsep}\newlength{\bibparskip}
\let\oldthebibliography\thebibliography
\renewcommand\thebibliography[1]{%
  \oldthebibliography{#1}%
  \setlength{\parskip}{\bibitemsep}%
  \setlength{\itemsep}{\bibparskip}%
}

\ifdefined\showcaptionsetup
 \PassOptionsToPackage{caption=false}{subfig}
\fi
\makeatother

\providecommand{\assumptionname}{Assumption}
\providecommand{\lemmaname}{Lemma}
\providecommand{\propositionname}{Proposition}
\global\long\def\expec#1{\mathbb{E}\left[#1\right]}%
\global\long\def\expecnu#1{\mathbb{E}_{\nu}\left[#1\right]}%
\global\long\def\var#1{\mathrm{Var}\left[#1\right]}%
\global\long\def\cov#1{\mathrm{Cov}\left[#1\right]}%
\global\long\def\one{\mathbf{1}}%
\begin{document}

\title{Estimating Demand with Recentered Instruments}
\author{\vspace{1.25cm}}
\author{
  Kirill Borusyak \\ UC Berkeley and NBER
  \and
  Mauricio Caceres Bravo \\ Brown
  \and
  Peter Hull\\ Brown and NBER\thanks{
    Contact: k.borusyak@berkeley.edu, mauricio\_caceres\_bravo@brown.edu,
    peter\_hull@brown.edu. We are grateful to Dan Ackerberg, Dmitry
    Arkhangelsky, Matthew Backus, Jean-Francois Houde, Jesse Shapiro, and
    numerous seminar participants for helpful comments. Adamson Bryant and
    Jacob Lefler provided excellent research assistance.
  }
}
\date{\vspace{0.5cm}April 2025}
\maketitle

\begin{abstract}
  \noindent
  \begin{singlespace}
    \vspace{-0.5cm}
    \begin{adjustwidth*}{0.8cm}{0.8cm}
      \normalsize We develop a new approach to estimating flexible demand
      models with exogenous supply-side shocks. Our approach avoids
      conventional assumptions of exogenous product characteristics, putting
      no restrictions on product entry, despite using instrumental variables
      that incorporate characteristic variation. The proposed instruments
      are model-predicted responses of endogenous variables to the exogenous
      shocks, recentered to avoid bias from endogenous characteristics. We
      illustrate the approach in a series of Monte Carlo simulations.
    \end{adjustwidth*}
  \end{singlespace}
  \noindent\thispagestyle{empty}\newpage{}
\end{abstract}

\noindent
\setcounter{page}{1}
\setlength{\abovedisplayskip}{6pt plus 3pt minus 2pt}
\setlength{\belowdisplayskip}{6pt plus 3pt minus 2pt}
\vspace{-1.6cm}

\section{Introduction}
\label{sec:Introduction}

Many economic analyses depend on accurate estimates of the demand
for differentiated products. Prominent examples from the field of
industrial organization (IO) include measuring welfare effects of
mergers or new products and testing models of firm conduct; trade
economists similarly use demand estimates to measure welfare effects
of new tariffs and gains from trade or internal migration. Often,
these analyses leverage structural models of demand that allow for
rich and realistic substitution patterns—such as the mixed multinomial
logit model popularized by \citet{Berry1995}. To estimate these models
with market-level data, researchers need multiple instrumental variables
(IVs) that address the endogeneity of prices and other terms capturing
the substitution patterns.

This paper develops a new approach to constructing powerful instruments
for popular demand models by leveraging a set of exogenous supply-side
shocks (for brevity, “cost shocks”) like input price changes,
new taxes or subsidies, markup regulations, or certain productivity
and ownership shocks. Such shocks are increasingly found in empirical
applications, where they serve as a natural instrument for price.\footnote{
  Examples include \citet{Berry1999} and \citet{Goldberg2001} (exchange rate
  shocks), \citet{Market2016} (subsidies), \citet{miller_understanding_2017}
  (merger shocks), and \citet{nakamura_accounting_2010} (productivity shocks).
} However, even when plausibly exogenous cost shocks are available,
researchers typically identify the parameters governing substitution
patterns (so-called “nonlinear” parameters) using other IVs
constructed from observed characteristics of competing products. Prominent
examples include nest size instruments in nested logit models, “BLP
instruments” which average or sum the characteristics of a product's
competitors, and more sophisticated versions like the efficient IVs
of \citet{Berry1999} and the differentiation IVs of \citet{Gandhi2015}.
Such instruments rely on the econometric exogeneity of product characteristics:
a strong assumption that is often inconsistent with natural models
of product entry \citep{ackerberg_estimating_2009,petrin_identification_2022}.
Characteristic-based IVs can also be weak, and lack across-market
variation when all markets have the same products \citep{Reynaert2014,Nevo2001}.\footnote{
  BLP instruments are sometimes also used to estimate the price sensitivity
  parameter, when cost shocks are not available. \citet{Armstrong2016} studies
  the weak instrument problem that arises in that context.
}

We propose instruments that combine product characteristics and cost
shocks in a particular way: to predict the response of the demand
model's endogenous variables to the exogenous shocks. These IVs are
motivated by thinking of the model as structuring “spillover effects,”
of exogenous changes to product prices on the market shares of other
products. For example, the nested logit model structures spillovers
with a parameter $\sigma$ that governs whether, when prices exogenously
rise, consumers substitute primarily to local competitors in a product's
nest or more evenly to all unaffected products in the market. To distinguish
between those cases and thus identify $\sigma$, we propose instruments
which predict how a product's within-nest market share changes in
response to a set of cost shocks. A simple IV in this spirit is the
the cost shock of a product less the average shock in its nest. We
show how such IVs can be generally constructed from first-order approximations
to model-implied responses, yielding shift-share instruments with
cost shocks as the exogenous “shifts.” We also propose a novel
instrument construction from exact model-based predictions. We build
intuition for these constructions in mixed logit models by considering
instruments constructed to predict the impact of shocks via small
“nonlinear” parameters (i.e., in a “local to logit” approximation,
similar to \citet{salanie_fast_2022}).

Instruments constructed this way are generally complex formulas of
the exogenous cost shocks and the likely endogenous product characteristics.
To avoid bias from the latter, we follow \citet{BH1} in recentering
the IVs: i.e., subtracting their expectation over the data-generating
process of exogenous shocks. For example, a researcher may simulate
this process by permuting observed shocks across comparable products;
she could then recenter any formula IV by subtracting from each product's
instrument value the average value across these counterfactual shocks,
holding the characteristics of all products fixed. When the instruments
are constructed from first-order approximations (i.e. as shift-share
IVs) recentering is simpler as it only requires specifying and adjusting
for the conditional mean of the shocks. In general, recentering ensures
our IVs derive their validity only from the exogeneity of cost shocks—even
though they also derive power from product characteristics.

We formalize this approach in a broad class of demand models, which
includes both conventional mixed and nested logit from IO as well
as analogous constant elasticity of substitution (CES) models from
trade. We focus on estimating the parameters that govern own- and
cross-price elasticities, which are central to many important policy
counterfactuals. Consistency of recentered IV estimates follows when
there are either many uncorrelated markets or many uncorrelated shocks
that can affect multiple markets jointly while inducing sufficient
across-product variation in the instruments. Unlike conventional characteristic-based
IVs, our instruments can yield consistent estimates when all markets
have the same products and product fixed effects are included. Asymptotic
normality of our estimators follows with many market clusters; we
further extend results in \citet{adao2019shift} and \citet{BJH2018}
to show how many-shock asymptotic inference can be conducted with
shift-share instruments. We characterize the asymptotically efficient
recentered IVs, building on \citet{chamberlain1987asymptotic,chamberlain1992efficiency},
\citet{newey1994large}, and \citet{borusyak_efficient_2021}. While
our baseline analysis shows how powerful recentered IVs can be derived
for a given demand model, we also consider non-parametric identification
by building on \citet{Erry2014}. Notably, our approach applies even
when the same products are sold in all markets and product fixed effects
are included, as in \citet{Nevo2001}; characteristic-based IVs have
no variation in those settings.

We compare this approach to conventional ones in a series of Monte
Carlo simulations based on the data-generating process in \citet{Gandhi2015}.
When characteristics are exogenous, the power of recentered instruments—whether
derived from first-order approximations or exact model-based predictions—is
comparable to that of Differentiation IVs and much better than that
of BLP instruments. Expectedly, recentered IVs have less power with
a lower variance of cost shocks while the power of characteristics-based
IVs is lower with less variation in choice sets across markets. A
simple model of strategic product entry introduces significant bias
in characteristic-based IV estimates, while recentered IV estimates
remain accurate.

This paper contributes to two main literatures. First, we contribute
to an IO literature studying demand estimation without exogenous characteristics.
The potential bias from endogenous characteristics has been noted
as far back as \citet{Berry1995}. Existing solutions to this concern
broadly fall into two categories: some papers put additional structure
on the model unobservables (i.e., the unobserved taste shifters) by
assuming characteristic endogeneity is captured by controls (e.g.,
product fixed effects in \citet{Nevo2001}) or imposing a particular
statistical process for the unobservables (e.g., \citet{sweeting2013dynamic}
and \citet{moon2018estimation}). Other papers explicitly models characteristic
choice or product entry (e.g., \citet{crawford2019quality} and \citet{petrin_identification_2022}).\footnote{
  See also \citet{fan2013ownership}, who builds BLP-type instruments for
  a firm's endogenous characteristics from particular characteristics of
  the firm's competitors: namely, consumer characteristics in markets where
  competitors operate.
} In contrast, our approach places no restrictions on how the model
unobservables relate to observed characteristics and does not require
a model of entry, relying only on cost shock exogeneity. This solution
relates to an idea in \citet{ackerberg_estimating_2009} of searching
for “orthogonal instruments” to identify own- and cross-price
elasticities while leaving the relationship between characteristics
and taste shifters unidentified. We propose a concrete way to achieve
this goal, via recentered functions of exogenous cost shocks and endogenous
characteristics.\footnote{
  A larger literature improves mixed logit demand estimation in other
  ways, maintaining the assumption of exogenous characteristics. See,
  e.g., \citet{Berry1999}, \citet{Reynaert2014}, and \citet{Gandhi2015}
  on IV power, \citet{salanie_fast_2022} and \citet{lu_semi-nonparametric_2023}
  on alternative estimation methods, and \citet{wang_sieve_2023} on
  allowing for non-parametric distributions of random coefficients.
}

Second, we contribute to a recent econometrics literature on identification
and estimation with shift-share IVs and other “formula” instruments
combining exogenous shocks with other potentially endogenous data
(\citealp{BJH2018,borusyak_design-based_2023}; \citealt{adao2019shift};
\citealp{BH1,borusyak_efficient_2021}). While this literature studies
linear causal or structural models, we focus on nonlinear demand estimation.
In this sense our work is also related to \citet{Adao2018a} who identify
parameters of a quantitative spatial model by the responses of endogenous
variables to exogenous shocks; \citet{borusyak2022understanding}
follow a similar approach with a migration model. Notably, both of
these papers work with linear approximations of their models—introducing
inaccuracies when shocks are large—while we work directly with nonlinear
demand models.\footnote{
  \citet{adao2024putting} develop a similar approach to specification
  testing that applies to nonlinear models, but they do not propose
  an estimation procedure.
}

Recent complementary work by \citet{Andrews2022} shows that recentered
instruments are more robust to demand model misspecification than
characteristic-based IVs, in the sense of recovering more interpretable
causal estimands in a non-parametric potential outcomes framework.
We demonstrate a different advantage of recentered IVs: that they
can be used to relax the assumption of exogenous characteristics in
a given demand specification. Moreover, we propose specific constructions
of powerful recentered IVs tailored to a class of demand models.

Finally, our analysis relates to empirical studies using weighted
sums or other transformations of shocks as instruments to estimate
demand models. Recent examples include \citet{costinot2016evolving},
\citet{Adao2017}, \citet{couture2020urban}, \citet{fajgelbaum2020return},
\citet{adao2022imports}, \citet{barahona2023equilibrium}, and \citet{adao2024putting};
see also \citet{fujiy_production_2024} who estimate input demand
by firms. Typically these instrument constructions arise from intuitive
arguments instead of being derived as model-implied responses to the
shocks, limiting their power. Moreover, while the constructions are
sometimes simple enough to not need recentering, this consideration
is also not typically part of the formal analysis. We show in a general
setting how powerful instruments can be constructed by leveraging
the structure of the model, and how instrument validity may be made
more credible and transparent via explicit recentering.

The rest of this paper is structured as follows. Section \ref{sec:Motivating-Example:}
builds intuition for our approach in a simple nested logit demand
model with randomized cost shocks. Section \ref{sec:General-Approach}
develops our general approach and discusses asymptotic properties.
Section \ref{sec:Monte-Carlo-Simulations} illustrates the approach
with simulations. Section \ref{sec:Conclusion} concludes. All proofs
are collected in the appendix.

\section{Motivating Example: Nested Logit with Random Cost Shocks}
\label{sec:Motivating-Example:}

We start with a simple example that illustrates the main logic of
our approach as well as its advantages over conventional methods.
Here we keep the presentation informal and intuitive, leaving formal
assumptions and results for the more general analysis in Section \ref{sec:General-Approach}.

\subsection{Setting and Conventional Instruments}
\label{sub:setting-and-conventional-instruments}

Consider a set of markets $m$, each with a set of differentiated
products $j\in\mathcal{J}_{m}$ and an outside good $j=0$. The products
are grouped into “nests” $n(j)$; let $d_{jn}=\one\left[n(j)=n\right]$
denote mutually exclusive nest indicators. A researcher observes the
nest allocation, along with the price $p_{jm}$ and quantity share
$s_{jm}$ of each product. Finally, the researcher observes a set
of shocks $g_{jm}$ which increase products' marginal costs (e.g.,
via input prices) but do not directly affect demand.

The researcher correctly assumes that market shares arise from a nested
logit demand model: a mass of consumers $i$ in each market choose
a single product or the outside good to maximize their utility
$u_{ijm}=\alpha p_{jm}+\xi_{jm}+\varepsilon_{ijm}$,
where $\xi_{jm}$ is a common taste shifter and $\varepsilon_{ijm}$ is
an idiosyncratic taste shock that can be correlated across products
in a nest. The outside good has a zero price and taste shifter, such
that utility from it is $u_{i0m}=\varepsilon_{i0m}$. Conditional
on the prices and taste shifters, the idiosyncratic shocks
$\left(\varepsilon_{ijm}\right)_{j\in\mathcal{J}_{m}\cup\left\{ 0\right\}}$
are distributed across consumers in such a way that the market shares satisfy:
\begin{equation}
  \log\left(s_{jm}/s_{0m}\right)
  =
  \alpha p_{jm}
  +
  \sigma\log\left(s_{jm}/s_{n(j)m}\right)
  +
  \xi_{jm},
  \label{eq:nested_logit}
\end{equation}
where $s_{0m}$ is the market share of the outside good in market
$m$ and $s_{nm}$ is the total market share of products in nest $n$
and market $m$.\footnote{
  Appendix \ref{sec:nested_logit_derivations} gives the formula for
  nested logit market shares and derives this expression from it. The
  nested CES model, commonly used in international trade and spatial
  economics, implies similar expressions; see Section \ref{subsec:Extensions}.
}
From this equation the researcher is interested in estimating $\alpha<0$,
which determines the own-price sensitivity of demand, and $\sigma\in[0,1)$
which captures the extent of within-nest correlation in the taste
shocks that governs substitution patterns. These two parameters determine
the matrix of cross-price elasticities, which is a key input to a
variety of policy counterfactuals (e.g., merger analyses) and consumer
welfare calculations.

Estimating $\alpha$ and $\sigma$ generally requires finding two
instruments which are uncorrelated with the unobserved taste shifters
$\xi_{jm}$ but correlated with the two endogenous variables in equation
(\ref{eq:nested_logit}): price $p_{jm}$ and the log within-nest
market share $\log(s_{jm}/s_{n(j)m})$. Here price endogeneity likely
arises because more popular products (with higher $\xi_{jt}$) are
likely of higher quality and therefore more expensive to produce;
firms may moreover optimally charge higher markups for them. Endogeneity
of $\log(s_{jm}/s_{n(j)m})$ further arises from the market shares'
direct dependence on the taste shifters: i.e., more popular products
will have larger within-nest market shares.

A natural instrument for the own-price sensitivity parameter $\alpha$
is the excluded cost shock $g_{jm}$, since higher costs are predicted
to at least partially pass through to higher prices.\footnote{
  We attribute an instrument to a particular endogenous variable informally;
  technically both instruments jointly identify both parameters when they are
  valid.
} To justify this choice simply, suppose the $g_{jm}$ are drawn in
a randomized trial after product entry. Randomization and the natural
exclusion restriction that the cost shocks do not directly affect
demand ensure that $g_{jm}$ is a valid instrument for equation (\ref{eq:nested_logit}),
i.e. that $\cov{g_{jm},\xi_{jm}}=0$.\footnote{
  Throughout, we call instruments “valid” when they are uncorrelated with
  the model error (i.e., the unobserved taste shifter). We consider an
  instrument's relevance, i.e. its correlation with endogenous variables,
  separately.
}

It is more difficult to find an instrument for the substitution parameter
$\sigma$. One popular strategy is to construct instruments from the
observed characteristics of other products, such as the nest indicators
$d_{jn}$. In particular, in nested logit models, it is common to
use the number of products in $j$'s nest, $N_{n(j)m}=\sum_{k\in\mathcal{J}_{m}}d_{kn(j)}$
(or, equivalently, the number of other products excluding $j$; e.g.,
\citet{Goldberg2001}, \citet{town2003welfare}, \citet{miller_understanding_2017}).
This instrument is expected to predict $\log\left(s_{jm}/s_{n(j)m}\right)$
because a product with a larger number of “local” competitors
in its nest should have, on average, a smaller within-nest share.

Such characteristic-based instruments have at least two drawbacks.
First, their validity hinges on the econometric exogeneity of the
characteristics: a strong assumption that can be at odds with natural
models of product entry. For example, the nest size instrument will
be invalid (with $\cov{N_{n(j)m},\xi_{jm}}>0$) when firms introduce
more products in nests for which consumers have a higher preference
in a particular market—a natural tendency of profit-optimizing firms
with at least partial information on consumer tastes (see, e.g., \citet[p.506]{aguiar2018quality}).
Even if the instrument is based on the product entry by firms other
than the producer of $j$, it is not econometrically exogenous. Second,
such instruments have no useful variation in some common empirical
contexts. Specifically, $N_{n(j)m}$ and similar instruments cannot
be used if all products are sold in all markets and product fixed
effects are included (as is commonly done since \citet{Nevo2001}).
Given these issues, we look for different instruments.

\subsection{Proposed Instruments}
\label{sub:proposed-instruments}

Consider an IV that measures the relative cost shock of product $j$
vs. the average shock in its nest:
\begin{align}
  z_{jm}
  &
  =
  g_{jm}
  -
  \frac{1}{N_{n(j)m}}
  \sum_{k\in\mathcal{J}_{m}}d_{kn(j)}g_{km}.
  \label{eq:z_simple}
\end{align}
We next explain why $z_{jm}$ is both relevant and valid, and how
it exemplifies our general approach to constructing instruments as
certain combinations of exogenous cost shocks and potentially endogenous
characteristics (here, nest indicators) of all the products in the
market.

First, this instrument is likely relevant (i.e., correlated with $\log\left(s_{jm}/s_{n(j)m}\right)$):
if product $j$'s local competitors in its nest have relatively low
cost shocks, its within-nest market share is expected to be lower.
This intuitive argument has a formal backing: $z_{jm}$ approximates
the model-predicted response of the endogenous variable $\log\left(s_{jm}/s_{n(j)m}\right)$
to the exogenous shocks. Specifically, consider a hypothetical scenario
in which all products have equal prices $\check{p}_{km}=\check{\pi}_{0}$
and unobserved taste shifters $\check{\xi}_{km}=0$, and thus all
products within the nest have the same market shares. To this scenario
we introduce an exogenous component of price variation, $\hat{p}_{km}=\check{p}_{km}+\check{\pi}g_{km}$
for some auxiliary constant $\check{\pi}\ne0$ that aims to capture
the pass-through of cost shocks into prices in a simple way. Nested
logit demand characterizes the share response to this set of price
changes; Appendix \ref{sec:nested_logit_derivations} shows that,
in a first-order approximation that is precise for small shocks, the
resulting change in $\log\left(s_{jm}/s_{n(j)m}\right)$ is equal
to $z_{jm}$, up to a constant scaling factor. While cost shocks is
only one of the reasons why $\log\left(s_{jm}/s_{n(j)m}\right)$ deviates
from equal shares, $z_{jm}$ captures the component of variation in
the within-nest shares due to the cost shocks and is thus likely relevant.

Second, $z_{jm}$ is a valid instrument (i.e., $\expec{z_{jm}\xi_{jm}}=0$)
when the cost shocks are exogenous in the sense we previously assumed
to justify their use as an instrument for $\alpha$. This claim is
nontrivial because the formula for $z_{jm}$ incorporates not only
the cost shocks but also the nest dummies, which are likely econometrically
endogenous. Nevertheless, $z_{jm}$ is constructed in such a way that
its validity stems only from the exogeneity of the shocks only. Intuitively,
when cost shocks are random, it is also random whether the shock for
a particular product is higher or lower than the average in its nest.
Formally, $z_{jm}$ is a \emph{recentered} instrument, meaning its
expectation over draws of the shocks is zero conditional on the other
variation \citep{BH1}: $
  \expec{z_{jm}\mid\left(d_{kn}\right)_{k\in\mathcal{J}_{m},n}}
  =
  \expec{g_{jm}}
  -
  \frac{1}{N_{n(j)m}}\sum_{k\in\mathcal{J}_{m}}d_{kn(j)}\expec{g_{km}}
  =
  0
$
when cost shocks are drawn randomly. Thus, $z_{jm}$ is guaranteed
to be uncorrelated with any function of the nest allocation which
could create endogeneity problems. Other formula instruments can be
adjusted via a recentering procedure that removes the component correlated
with the characteristics—as we soon illustrate.

To recap, our proposal is to construct instruments as recentered model-based
predictions of how relevant endogenous variables respond to the cost
shocks. This general approach suggests three ways of improving the
simple $z_{jm}$ instrument, in the sense of likely power gains, by
forming better predictions of the endogenous variable. First, consider
an exact model-based prediction of $\log\left(s_{jm}/s_{n(j)m}\right)$
from the price predictions $\left(\hat{p}_{km}\right)_{k\in\mathcal{J}_{m}}$
in place of the first-order approximation in $z_{jm}$. Appendix \ref{sec:nested_logit_derivations}
shows this prediction can be written:
\begin{align}
\widehat{\log}\left(s_{jm}/s_{n(j)m}\right)
  &
  =
  \frac{\check{\alpha}}{1-\check{\sigma}}\check{\pi}g_{jm}
  -
  \log\left(
    \sum_{k\in\mathcal{J}_{m}}d_{kn(j)}
    \exp\left(
      \frac{\check{\alpha}}{1-\check{\sigma}}\check{\pi}g_{km}
    \right)
  \right),
  \label{eq:z_exact}
\end{align}
where $\check{\alpha}$ and $\check{\sigma}$ are some preliminary
estimates of $\alpha$ and $\sigma$ and $\check{\pi}$, as before,
is an estimate of the pass-through of cost shocks into prices.\footnote{
  Preliminary estimates $\check{\alpha},\check{\sigma}$ can be obtained
  using simpler instruments, such as (\ref{eq:z_simple}), in a first
  step, while $\check{\pi}$ can be obtained from a regression of prices
  and the own-product shocks. In Section \ref{sec:General-Approach}
  we show how a continuously updating estimator can bypass the need
  for $\check{\alpha},\check{\sigma}$.
} Unlike $z_{jm}$, exogeneity of the cost shocks does not make this
prediction a valid instrument; instead, like the conventional nest
size instrument $N_{n(j)m}$, its validity also hinges on the econometric
exogeneity of the nest allocation. To see this simply, note that $
  \widehat{\log}\left(s_{jm}/s_{n(j)m}\right)
$
varies over products even in the absence of cost shocks ($g_{km}=\mu_{g}$
for all $k$). In fact, this variation is driven exactly by nest size:
plugging in $g_{km}=\mu_{g}$ for all $k$ yields $
  \widehat{\log}\left(s_{jm}/s_{n(j)m}\right)=-\log N_{n(j)m}
$,
showing that this prediction suffers from exactly the same endogeneity
concerns as the conventional nest size instrument.

Following \citet{BH1}, we propose obtaining valid instruments by
recentering model-based predictions like (\ref{eq:z_exact}), using
knowledge of how the exogenous shocks are drawn. Recentering is straightforward
when the shocks are drawn randomly in an experiment: the researcher
can re-draw many sets of counterfactual shocks from the experimental
protocol, recompute the prediction under each set, average across
shock counterfactuals to measure the expected prediction ($
  \mu_{jm}
  \equiv
  \expec{
    \widehat{\log}\left(s_{jm}/s_{n(j)m}\right)
    \mid
    \left(d_{kn}\right)_{k\in\mathcal{J}_{m},n}
  }
$
in the case of (\ref{eq:z_exact})), and subtract this expectation
from the actual prediction.\footnote{
  We note that $\mu_{jm}$ is related but not equal to the value of
  $z_{jm}$ with no shocks, $-\log N_{n(j)m}$. The difference arises
  because $z_{jm}$ is a nonlinear function of the shocks, and thus
  taking the expectation of $z_{jm}$ across the shock distribution
  is not the same as plugging in expectation of the shocks.
} Like $z_{jm}$, the recentered prediction
\begin{align}
  z_{jm}^{\text{exact}}
  &
  =
  \widehat{\log}\left(s_{jm}/s_{n(j)m}\right)
  -
  \mu_{jm}\label{eq:exact-rc}
\end{align}
is a valid instrument regardless of any econometric endogeneity of
the nest allocation. Below we discuss other ways to recenter predictions
in observational data, where the shock data-generating process is
unknown. Although recentering removes some variation in the prediction,
which was not necessary with $z_{jm}$, starting from a better prediction
still improves the first-stage \citep{borusyak_efficient_2021}. Note
that inaccuracy of the initial $(\check{\alpha},\check{\sigma})$
estimates in equation (\ref{eq:z_exact}) is not an issue for the
validity of $z_{jm}^{\text{exact}}$, given recentering, though it
will likely affect power.

A second type of improvement comes from using additional data to better
predict the endogenous variables' responses to exogenous shocks. One
particularly useful input is the market shares $s_{jm}^{\text{pre}}$
and prices $p_{jm}^{\text{pre}}$ for the same products and the same
market in an earlier period, before the shocks $g_{km}$ were drawn.
To the extent that prices and taste shifters are serially correlated,
this yields better predictions of prices and market shares in the
period of interest.\footnote{
  A second complementary use of such data is that equation (\ref{eq:nested_logit})
  can be estimated in time-differences, yielding more precise estimates
  from the same instruments when $\xi_{jm}$ is serially correlated.
  We discuss this approach in Section \ref{subsec:Consistencyetc}.
} The instrument that results from incorporating this information is
also very intuitive: Appendix \ref{sec:nested_logit_derivations}
shows that the first-order approximation of the model-predicted response
of $\log\left(s_{jm}/s_{n(j)m}\right)$ to the exogenous shocks around
the pre-period shares (rather than equal shares) yields
\begin{align}
  z_{jm}^{\text{weighted}}
  &
  =
  g_{jm}
  -
  \frac{
    \sum_{k\in\mathcal{J}_{m}}d_{kn(j)}s_{km}^{\text{pre}}g_{km}
  }{
    \sum_{k\in\mathcal{J}_{m}}d_{kn(j)}s_{km}^{\text{pre}}
  }.
  \label{eq:simple_z-1}
\end{align}
Like $z_{jm}$, this prediction is mean-zero over draws of random
cost shocks, $
  \expec{z_{jm}^{\text{weighted}}\mid\left(d_{kn}\right)_{k\in\mathcal{J}_{m},n}}=0
$,
making it a recentered instrument without further adjustment.\footnote{
  $z_{jm}$ and $z_{jm}^{\text{weighted}}$ are examples of shift-share
  instruments, which average the exogenous shocks with a set of weights
  capturing differential shock exposure \citep{BJH2018}. As discussed
  below, such instruments are often easier to recenter or require no
  recentering at all because they are linear in the shocks.
} The appendix further shows that the two improvements can be combined:
a researcher can incorporate the lagged share information to improve
the recentered exact prediction (\ref{eq:exact-rc}), too. In either
case, the recentered instrument will again be valid just by virtue
of the exogenous shocks—even though it now draws power from variation
in lagged market shares (as well as the nest allocation) and lagged
market shares by themselves need not be exogenous.\footnote{
  \label{fn:variation-in-Nevo}We note that, unlike the nest size IV,
  $z_{jm}$ and $z_{jm}^{\text{weighted}}$ have cross-market variation
  even if all products are sold in all markets—so long as cost shocks
  vary across markets. Moreover, $z_{jm}^{\text{weighted}}$ can vary
  across markets even if cost shocks do not, if the pre-period shares
  vary due to any unobserved cost or taste differences.
}

Finally, consider an instrument which uses a more realistic prediction
of how prices respond to the full set of exogenous shocks. A researcher
might, for example, specify an auxiliary pricing model which captures
not only the pass-through of product $j$'s cost shock to its own
price but also how $p_{jm}$ responds to competitor cost shocks $g_{km}$
for $k\neq j$ (depending, for instance, on whether $j$ and $k$
are offered by the same firm). Substituting this model's price predictions
into any of the above instrument constructions, in place of the simple
$\hat{p}_{jm}=\check{\pi}g_{jm}$ prediction, yields an instrument
which is likely more powerful when such cost shock spillovers are
important and can be estimated. The better price prediction can also
be recentered and used in place of $g_{jm}$ as an instrument for
identifying $\alpha$. Note that as with the initial $(\check{\alpha},\check{\sigma})$
estimates in $z_{jm}^{\text{exact}}$, the validity of these instruments
does not hinge on the accuracy of the pricing model.

\section{General Approach}
\label{sec:General-Approach}

We now consider a broader class of demand models and formalize our
general approach. Section \ref{subsec:general_setting} develops the
baseline mixed logit model and the introduces key shock exogeneity
assumption. Section \ref{subsec:Recentered-Instruments} defines recentered
IVs and establishes identification with them. Section \ref{subsec:Constructing-IVs}
develops our proposal for constructing powerful recentered IVs from
the structure of the model, while Section \ref{subsec:Consistencyetc}
establishes consistency and asymptotic inference. Section \ref{subsec:Extensions}
discusses several extensions to the baseline model.

\subsection{Setting}
\label{subsec:general_setting}

We consider a class of random utility models—canonical mixed logit
demand—with market-level data as in \citet{Berry1995}; see \citet{Berry2021}
and \citet{Gandhi2021} for more recent treatments. A researcher observes
a set of markets $m$ (which might correspond to regions, periods,
or both) with differentiated products $j\in\mathcal{J}_{m}$, prices
$p_{jm}$, and quantity shares $s_{jm}$.\footnote{
  Here we do not restrict whether the data consist of many markets or
  just a single one, whether the markets are randomly sampled, or whether
  the number of products per market is large. We return to these issues
  in Section \ref{subsec:Consistencyetc}.
} Each product also has a vector of observed characteristics $x_{jm}\in\mathbb{R}^{L}$
(which includes an intercept), as well as an unobserved scalar taste
shifter $\xi_{jm}$. All variables are normalized such that the outside
good in each market, $j=0$, has $p_{0m}=\xi_{0m}=0$ and $x_{0m}=0$.

Consumers $i$ choose among all products and the outside good by maximizing their utility:
\begin{equation}
  \max_{j\in\mathcal{J}_{m}\cup\left\{ 0\right\} }
  \delta_{jm}
  +
  \eta_{0i}p_{jm}
  +
  \sum_{\ell=1}^{L_{1}}\eta_{i\ell}x_{jm\ell}
  +
  \varepsilon_{ijm}.
  \label{eq:UMP}
\end{equation}
Here product $j$'s mean utility $\delta_{jm}$ is determined by its
price, characteristics, and the taste shifter:
\begin{equation}
  \delta_{jm}
  =
  \alpha p_{jm}+\beta'x_{jm}+\xi_{jm}
  \label{eq:delta-meanu}
\end{equation}
with $\alpha<0$ and $\beta\in\mathbb{R}^{L}$. A subvector of characteristics
$x_{jm}^{(1)}=\left(x_{jm1},\dots,x_{jmL_{1}}\right)$, as well as
(potentially) price, also enter utility with mean-zero “random
coefficients” $\eta_{i}=\left(\eta_{i\ell}\right)_{\ell=0}^{L_{1}}$
that capture heterogeneous consumer preferences. This $\eta_{i}$
is \emph{iid} across consumers and follows distribution $\mathcal{P}\left(\cdot;\sigma\right)$
that is known up to a vector of “nonlinear” parameters $\sigma$
(typically Gaussian with independent components and standard deviations
$\sigma_{0},\dots,\sigma_{L_{1}}$). Finally, $\varepsilon_{ijm}$
is an extreme-value shock, \emph{iid} across consumers and products
including the outside good. Integrating out these shocks implies market
shares satisfy
\begin{equation}
  s_{jm}
  =
  \mathcal{S}_{j}(
    \boldsymbol{\delta}_{m};\sigma,\boldsymbol{x}_{m}^{(1)},\boldsymbol{p}_{m}
  )
  \equiv
  \int \frac{
    \exp\left(
      \delta_{jm}+\eta_{0i}p_{jm}+\sum_{\ell=1}^{L_{1}}\eta_{i\ell}x_{jm\ell}
    \right)
  }{
    1+\sum_{k\in\mathcal{J}_{m}}\exp\left(
      \delta_{km}+\eta_{0i}p_{km}+\sum_{\ell=1}^{L_{1}}\eta_{i\ell}x_{km\ell}
    \right)
  } d\mathcal{\mathcal{P}}(\eta_{i};\sigma),
  \label{eq:shares}
\end{equation}
where bold symbols denote a collection of variables for all products
in the market: $\boldsymbol{v}_{m}=\left(v_{jm}\right)_{j\in\mathcal{J}_{m}}$
for any variable $v_{jm}$. The distribution $\mathcal{P}$ determines
the patterns of product substitutability: for instance, the pure multinomial
logit model corresponds to no variation in random coefficients (typically
captured by $\sigma=0$). The nested logit model considered in Section
\ref{sec:Motivating-Example:} is also a special case, with nest dummies
as characteristics and with a particular choice of $\mathcal{\mathcal{P}}(\cdot;\sigma)$
\citep{mcfadden_modelling_1978}. \citet{berry_estimating_1994} and
\citet{Berry1995} famously show that the share function $
  \mathcal{\mathcal{S}}_{j}(\boldsymbol{\delta}_{m};\sigma,\boldsymbol{x}_{m}^{(1)},\boldsymbol{p}_{m})
$
is invertible, such that mean utilities can be derived from $\sigma$
and observed data:
\begin{equation}
  \delta_{jm}
  =
  \mathcal{D}_{j}\left(
    \boldsymbol{s}_{m};\sigma,\boldsymbol{x}_{m}^{(1)},\boldsymbol{p}_{m}
  \right),
  \label{eq:share-inv}
\end{equation}
for functions $\mathcal{D}_{j}(\cdot)$ that generally do not have
a closed-form but can be computed numerically.

We focus on estimating parameters $\theta=(\alpha,\sigma')'$, which
are central to a number of important policy counterfactuals that do
not involve changing product characteristics. Indeed, the model implies
that own- and cross-price elasticities can be characterized in terms
of $\theta$ and observed data:
\begin{equation}
  \frac{ds_{jm}}{dp_{km}}
  =
  \int
    \left(\alpha+\eta_{0i}\right)
    s_{ji}
    \left(\one\left[j=k\right]-s_{ki}\right)
    d\mathcal{\mathcal{P}}\left(\eta_{i};\sigma\right)
  \label{eq:elasticities}
\end{equation}
where
\[
  s_{ji}
  =
  \frac{
    \exp\left(
      \mathcal{D}_{j}\left(
        \boldsymbol{s}_{m};\sigma,\boldsymbol{x}_{m}^{(1)},\boldsymbol{p}_{m}
      \right)
      +
      \eta_{0i}p_{jm}
      +
      \sum_{\ell=1}^{L_{1}}\eta_{i\ell}x_{jm\ell}
    \right)
  }{
    1
    +
    \sum_{k\in\mathcal{J}_{m}}
      \exp\left(
        \mathcal{D}_{k}\left(\boldsymbol{s}_{m};\sigma,\boldsymbol{x}_{m}^{(1)},\boldsymbol{p}_{m}\right)
        +
        \eta_{0i}p_{km}
        +
        \sum_{\ell=1}^{L_{1}}\eta_{i\ell}x_{km\ell}
      \right)
  }.
\]
In contrast, a consistent estimate of the causal effect of characteristics
on mean utility, $\beta$, is not needed when analyzing a merger (\citet{ackerberg_estimating_2009}).
Similarly, the welfare gains from a new product can be computed without
$\beta$.\footnote{
  \label{fn:beta-non-causal}While such gains require a prediction of
  the new product's mean utility, naturally based on its characteristics,
  this constitutes a prediction problem and not a causal problem which
  would require the structural parameter $\beta$. We are not aware
  of previous work making this point.
}

To identify $\theta$ we leverage share inversion, which implies a
structural equation that is additive in the unobserved taste shifter:
\[
  \mathcal{D}_{j}\left(\boldsymbol{s}_{m};\sigma,\boldsymbol{x}_{m}^{(1)},\boldsymbol{p}_{m}\right)
  =
  \alpha p_{jm}
  +
  \beta'x_{jm}
  +
  \xi_{jm}.
\]
This representation permits estimation of $\theta$ from moment conditions of the form
\begin{align}
  \expec{Z_{jm}\cdot\left(
    \mathcal{D}_{j}\left(
      \boldsymbol{s}_{m};\sigma,\boldsymbol{x}_{m}^{(1)},\boldsymbol{p}_{m}
    \right)
    -
    \alpha p_{jm}
    -
    \beta'x_{jm}
  \right)}
  &
  =
  0
  \label{eq:moments}
\end{align}
for some vector of instruments $Z_{jm}$ that are uncorrelated with
$\xi_{jm}$. If the instruments are also uncorrelated with $x_{jm}$
(as ours will be), equation (\ref{eq:moments}) will hold for any
value of $\beta$.

To build instruments, we assume the researcher observes a set of supply-side
shocks $g_{jm}$ that vary by product and market. Shocks to input
prices is a natural source of supply-side shocks commonly used in
the IO literature (e.g., \citet{Villas-Boas2007}, \citet{backus_common_2021},
\citet{barahona2023equilibrium}); other studies have used exchange
rate shocks (e.g., \citet{Berry1999,Goldberg2001}), shocks from weather
(as a productivity shifter; e.g., \citet{nakamura_accounting_2010}),
product-specific subsidies \citep[e.g.,][]{Market2016}, taxes \citep[e.g.,][]{Dearing2022},
and markup shocks due to mergers \citep[e.g.,][]{miller_understanding_2017}.\footnote{
  We do not require the shocks to be independent across products and
  markets. For instance, all cars produced in the same country are assigned
  the same exchange rate shock, and input price shocks affect all products
  using this input, to different extents. For now we assume that the
  researcher can assign each product to the corresponding country of
  production or shares of different inputs but we relax this assumption
  in Section \ref{subsec:Extensions}.
} We generically call the $g_{jm}$ “cost shocks” and assume
that they are exogenous in the following sense:
\begin{equation}
  \expec{\xi_{jm}\mid\boldsymbol{g}_{m},\boldsymbol{x}_{m}
  }=
  \expec{\xi_{jm}\mid\boldsymbol{x}_{m}}
  \qquad
  \text{\ensuremath{\forall m,j\in\mathcal{J}_{m}}}.
  \label{eq:exog-shocks0}
\end{equation}
That is, we assume the taste shifter of each product $j$ is mean-independent
of cost shocks (both for $j$ and its competitors), conditionally
on the observed characteristics of all products.

To understand the economic content of the assumption, note that the
unobserved taste shifter captures both objective characteristics of
the product chosen by firms and the subjective preferences of consumers.
Thus, several conditions have to be met for equation (\ref{eq:exog-shocks0})
to hold. First, cost shocks should not affect product entry decisions
or firm choices of unobserved characteristics. A sufficient condition
is that cost shocks are realized after those decisions have been made
(but before prices are set—a condition required for relevance of cost
shocks as instruments). Second, cost shocks should not influence consumer
preferences. Such influence could be possible when cost shocks influence
firms' advertising decisions or stem from the prices of inputs that
affect consumer earnings.

In addition to these exclusion restrictions, cost shocks may not be
\emph{correlated with} variables that affect product entry or consumer
preferences. This can be viewed as an independence condition, in the
sense of \citet{imbens1994identification}, that is automatically
satisfied in randomized experiments. Examples from the literature
show how it can also hold in observational data. Several IO papers
use the exchange rate in the country of production as a cost shock
for automobiles \citep[e.g.,][]{Berry1999,Goldberg2001,Grieco2021}
and other industries \citep[e.g.,][]{nakamura_accounting_2010}. While
these studies use the \emph{level} of the exchange rate, their \emph{changes}
over time\emph{ }may be particularly attractive as the $g_{jm}$ because
exchange rates are known to roughly follow a random walk (e.g., \citet{kilian2003so}).
The same argument applies to many commodity price changes, for goods
using these commodities as inputs (e.g., coffee in \citet{nakamura_accounting_2010}).
For inputs traded on futures markets, \citet{ackerberg_estimating_2009}
point out that the difference between the realized input price at
the time when downstream firms set prices and the price of a futures
contract at the earlier moment when product entry has been decided
is guaranteed to be unrelated to the characteristics.

While the assumption in (\ref{eq:exog-shocks0}) is restrictive, it
is important to highlight what it does \emph{not} entail: it allows
unobserved taste shifters $\xi_{jm}$ to be arbitrarily correlated
with the observed characteristics of both product $j$ and its competitors.\footnote{
  An additional feature of (\ref{eq:exog-shocks0}) is that it allows
  the cost shocks to be mutually correlated. Mutual correlations of
  taste shifters are also allowed, as when unobserved market-level demand
  and cost conditions affect the choice of unobserved quality of all
  firms in the market.
} This is in contrast to the prevalent approach in the literature
\citep[e.g.,][]{Berry1995,Berry1999,Gandhi2015} which imposes a stronger assumption:
\begin{equation}
  \expec{\xi_{jm}\mid\boldsymbol{x}_{m},\boldsymbol{g}_{m}}
  =
  0,
  \label{eq:assu-conventional}
\end{equation}
equivalent to imposing $\expec{\xi_{jm}\mid\boldsymbol{x}_{m}}=0$
in addition to equation (\ref{eq:exog-shocks0}). Under this stronger
condition, instruments constructed as functions of own and competing
products' characteristics are valid. This includes BLP instruments,
computed as sums or averages of competitor characteristics, efficient
instruments from \citet{Berry1999}, as well as the differentiation
IVs of \citet{Gandhi2015} which capture the average distance between
$x_{jm}$ and characteristics of competitors or the number of products
in the market with characteristics sufficiently close to $x_{jm}$.

As recognized as far back as \citet{Berry1995}, however, the econometric
exogeneity of observed characteristics is an unappealing restriction
on product entry. It is often arbitrary which objective characteristics
are observed or unobserved by the econometrician, such that there
is no reason why the two groups should be uncorrelated with each other.\footnote{
  Note, however, that it is generally necessary for characteristics
  with random coefficients to be observed. An interesting exception
  is provided by \citet{Adao2017} who show identification in a model
  with a random coefficient on unobserved mean utility; our approach
  applies to that model as well.
} Moreover, in natural models of product entry, $\xi_{jm}$ can be
related to characteristics of competing products. This can happen,
for instance, when all firms observe some information about the cost
or demand conditions common to the market when making product entry
decisions. Similar to the discussion in Section \ref{sec:Motivating-Example:},
in a market where consumers like small and fuel-efficient cars (i.e.
their $\xi_{jm}$ is predicted to be higher), we expect all firms
to pivot towards these characteristics, whether in ways observed or
unobserved to the econometrician. In this situation, BLP and differentiation
IVs need not be valid.

In the rest of our analysis we maintain a modified version of (\ref{eq:exog-shocks0})
that allows shock exogeneity to be conditional on some other observed
data $\boldsymbol{q}_{m}$ (as well as product characteristics):
\begin{assumption}[Exogenous cost shocks]
\label{assu:exogenous-shocks}
  $
    \expec{\xi_{jm}\mid\boldsymbol{g}_{m},\boldsymbol{x}_{m},\boldsymbol{q}_{m}}
    =
    \expec{\xi_{jm}\mid\boldsymbol{x}_{m},\boldsymbol{q}_{m}},
  $
  $
    \forall m,j\in\mathcal{J}_{m}
  $.
\end{assumption}
\noindent Both condition (\ref{eq:exog-shocks0}) and Assumption
\ref{assu:exogenous-shocks} hold when the shocks are unconditionally
as-if randomly assigned; the modified assumption will then be helpful
to construct more powerful instruments that use information in $\boldsymbol{q}_{m}$
(e.g., lagged market shares and prices as in Section \ref{sec:Motivating-Example:}).
In other settings, conditioning on potential confounders in $\boldsymbol{q}_{m}$
may help make shock exogeneity more plausible, such as when the $g_{jm}$
are systematically correlated with some observed $q_{jm}$.

\subsection{Recentered Instruments}
\label{subsec:Recentered-Instruments}

We say $Z_{jm}$ is a vector of \emph{formula} instruments when it
can be written as $Z_{jm}=f_{jm}(\boldsymbol{g}_{m},\boldsymbol{x}_{m},\boldsymbol{q}_{m})$
for some non-stochastic and known vector-valued functions $\left(f_{jm}\right)_{m,j\in\mathcal{J}_{m}}$.
We further say that $Z_{jm}$ consists of \emph{recentered} formula
instruments (or just recentered IVs) when the formulas satisfy:
\begin{align}
  \expec{f_{jm}(\boldsymbol{g}_{m},\boldsymbol{x}_{m},\boldsymbol{q}_{m})\mid\boldsymbol{x}_{m},\boldsymbol{q}_{m}}
  &
  =
  0,
  \quad
  \forall m,j\in\mathcal{J}_{m}.
  \label{eq:recentered}
\end{align}
Our first result shows this property characterizes the complete set
of valid instruments in our setup:
\begin{lem}
\label{prop:only_recentering}
Assumption \ref{assu:exogenous-shocks} implies $\expec{Z_{jm}\xi_{jm}}=0$ if
and only if $Z_{jm}$ consists of recentered IVs.
\end{lem}
\noindent This result follows immediately from Proposition 1 in \citet{borusyak_efficient_2021}.\footnote{
  Like there, the “only if” part of Lemma 1 should be understood
  as follows: unless $Z_{jm}$ consists of recentered instruments, it
  is possible to find a conditional distribution of $\xi_{jm}$ such
  that Assumption \ref{assu:exogenous-shocks} holds but $\expec{Z_{jm}\xi_{jm}}\ne0$.
} It shows that when researchers are only willing to assume that cost
shocks are exogenous, in the sense of Assumption~\ref{assu:exogenous-shocks},
the only justified instruments are recentered IVs. More positively,
it shows that any candidate formula instrument $
  h_{jm}(\boldsymbol{g}_{m},\boldsymbol{x}_{m},\boldsymbol{q}_{m})
$
for some functions $\left(h_{jm}\right)_{m,j\in\mathcal{J}_{m}}$
can be made a valid IV by recentering: i.e., by subtracting off its
conditional mean $
  \expec{h_{jm}(\boldsymbol{g}_{m},\boldsymbol{x}_{m},\boldsymbol{q}_{m})\mid\boldsymbol{x}_{m},\boldsymbol{q}_{m}}
$.

\citet{BH1} and \citet{borusyak_design-based_2023} discuss several
strategies for recentering formula instruments. One general approach
follows when it is possible to generate a set of counterfactual shock
vectors $\boldsymbol{g}_{m}^{(c)}$, for $c=1,\dots,C$, which are
either drawn from the same distribution as $\boldsymbol{g}_{m}$ (conditional
on $\boldsymbol{x}_{m}$ and $\boldsymbol{q}_{m}$) or otherwise as
likely to have been realized. For example, the $\boldsymbol{g}_{m}^{(c)}$
could be generated by redrawing from the same randomization protocol
that generated $\boldsymbol{g}_{m}$ in a randomized trial, as in
the Section \ref{sec:Motivating-Example:} motivating example. In
observational data, the counterfactual shocks can instead be generated
by reshuffling the set of observed $g_{jm}$ across comparable products,
markets, or both. Given such $\boldsymbol{g}_{m}^{(c)}$, any formula
instrument $h_{jm}(\boldsymbol{g}_{m},\boldsymbol{x}_{m},\boldsymbol{q}_{m})$
can be recentered by recomputing instrument values under each counterfactual
(holding fixed $\boldsymbol{x}_{m}$ and $\boldsymbol{q}_{m}$) and
subtracting the average across these values, $
  \frac{1}{C}\sum_{c}h_{jm}(\boldsymbol{g}_{m}^{(c)},\boldsymbol{x}_{m},\boldsymbol{q}_{m})
$,
for each $j$ and $m$.\footnote{
  The number of counterfactuals $C$ does not matter for recentered
  IV consistency, though it generally affects the asymptotic variance.
  See footnotes 19 and 21 in \citet{BH1}.
} A second general approach follows when the formula instrument is
linear in the shocks: i.e., when $
  h_{jm}(\boldsymbol{g}_{m},\boldsymbol{x}_{m},\boldsymbol{q}_{m})
  =
  \sum_{k\in\mathcal{J}_{m}}w_{jkm}g_{km}
$
for some exposure weights (or “shares”) $w_{jkm}$ that are
functions of $(\boldsymbol{x}_{m},\boldsymbol{q}_{m})$. Recentering
such shift-share IVs only requires de-meaning the cost shocks (or
“shifts”) by their conditional expectations,
$\expec{g_{km}\mid\boldsymbol{x}_{m},\boldsymbol{q_{m}}}$.

Lemma \ref{prop:only_recentering} implies that a recentered IV vector
$Z_{jm}$ locally identifies model parameters $\theta$ under a rank
condition \citep{rothenberg1971identification}: that the matrix $\expec{Z_{jm}\nabla_{jm}^{\prime}}$
is full column rank, where
\begin{equation}
  \nabla_{jm}
  =
  \frac{\partial}{\partial\theta}
  \left(
    \mathcal{D}_{j}\left(\boldsymbol{s}_{m};\sigma,\boldsymbol{x}_{m}^{(1)},\boldsymbol{p}_{m}\right)
    -
    \alpha p_{jm}
    -
    \beta'x_{jm}
  \right)
  =
  \left(\begin{array}{c} - p_{jm} \\ \nabla_{jm}^{\sigma} \end{array}\right),
  \label{eq:nabla}
\end{equation}
for $
  \nabla_{jm}^{\sigma}
  =
  \partial\mathcal{D}_{j}\left(\boldsymbol{s}_{m};\sigma,\boldsymbol{x}_{m}^{(1)},\boldsymbol{p}_{m}\right)
  /
  \partial\sigma
$
with the derivative evaluated at true parameter values.\footnote{
  As \citet{newey1994large} note, establishing global identification
  with nonlinear moment conditions is generally challenging; the objective
  function for mixed logit estimation is known to not be globally convex
  \citep{Conlon2020}.
} Note again that $\theta$ is identified while $\beta$ is not, since
recentering makes $Z_{jm}$ uncorrelated with $x_{jm}$. But this
is unimportant since $\beta$ does not directly enter price elasticities
or important policy counterfactuals. We next discuss how likely powerful
IVs can be constructed.

\subsection{Constructing Model-Based IVs}
\label{subsec:Constructing-IVs}

We propose constructing recentered IVs that predict how the vector
of model's residual derivatives $\nabla_{jm}$ responds to the exogenous
cost shocks. Specifically, we consider an instrument vector of length
$\dim(\theta)$ approximating:
\begin{align}
  \tilde{Z}_{jm}
  &
  =
  \expec{\nabla_{jm}\mid\boldsymbol{g}_{m},\boldsymbol{x}_{m},\boldsymbol{q}_{m}}
  -
  \expec{\nabla_{jm}\mid\boldsymbol{x}_{m},\boldsymbol{q}_{m}}.
  \label{eq:Z*}
\end{align}
The first term of $\tilde{Z}_{jm}$ is the best predictor of the residual
derivatives given the cost shocks, product characteristics, and other
data in $\boldsymbol{q}_{m}$, where the expectation is taken over
the conditional distribution of $(\boldsymbol{p}_{m},\boldsymbol{s}_{m})$
that corresponds to different realizations of unobserved demand and
cost shocks. The second term recenters this best predictor by its
expectation over the shocks, $
  \expec{\expec{\nabla_{jm}\mid\boldsymbol{g}_{m},\boldsymbol{x}_{m},\boldsymbol{q}_{m}}\mid\boldsymbol{x}_{m},\boldsymbol{q}_{m}}
  =
  \expec{\nabla_{jm}\mid\boldsymbol{x}_{m},\boldsymbol{q}_{m}}
$.
Hence $\tilde{Z}_{jm}$ captures how the model residual's derivative
is affected by the specific draw of shocks. Note that this $\tilde{Z}_{jm}$
is guaranteed to satisfy the rank condition when cost shocks are relevant
(i.e. when $\tilde{Z}_{jm}\neq0$) since then $
  \expec{\tilde{Z}_{jm}\nabla_{jm}^{\prime}}
  =
  \expec{\tilde{Z}_{jm}\tilde{Z}_{jm}^{\prime}}
$
which is generally fully rank.\footnote{
  Formally, $
    \expec{\tilde{Z}_{jm}\nabla_{jm}^{\prime}}
    =
    \expec{\tilde{Z}_{jm}\expec{\nabla_{jm}\mid\boldsymbol{g}_{m},\boldsymbol{x}_{m},\boldsymbol{q}_{m}}^{\prime}}
    =
    \expec{\tilde{Z}_{jm}\tilde{Z}_{jm}^{\prime}}
  $
  by the law of iterated expectations and the fact that $
    \expec{\tilde{Z}_{jm}\expec{\nabla_{jm}\mid\boldsymbol{x}_{m},\boldsymbol{q}_{m}}^{\prime}}
    =
    0
  $
  by virtue of recentering.
} Below we show $\tilde{Z}_{jm}$ is closely related to the asymptotically
efficient recentered IV vector.

The overall logic of our approximations to $\tilde{Z}_{jm}$ is as
follows. Instead of integrating over the unobserved shocks, we predict
$\nabla_{jm}$ in a single “no-shock” scenario that would prevail
in the absence of unexpected $\boldsymbol{g}_{m}$ shocks, i.e. when
$
  \boldsymbol{g}_{m}
  =
  \expec{\boldsymbol{g}_{m}\mid\boldsymbol{x}_{m},\boldsymbol{q}_{m}}
$.
This scenario is constructed using the data in $(\boldsymbol{x}_{m},\boldsymbol{q}_{m})$
only. We then predict how prices and shares would deviate because
of the realized shocks. Specifically: cost shocks affect prices and
the researcher can construct an unexpected component of prices using
an auxiliary model of cost shock pass-through, as in Section \ref{sec:Motivating-Example:}.
In turn, prices affect market shares; using preliminary values of
demand parameters, the researcher can then measure the impact of unexpected
price changes on shares. Finally, $\partial\mathcal{D}_{j}/\partial\sigma$
is a function of the shares so it can be predicted for the scenario
with shocks, as a function of $(\boldsymbol{g}_{m},\boldsymbol{x}_{m},\boldsymbol{q}_{m})$.
To approximate $\tilde{Z}_{jm}$, it then remains to recenter this
prediction as in Section \ref{subsec:Recentered-Instruments}.

The IV construction proceeds in four steps. First, the researcher
picks some preliminary values of the parameters $\check{\alpha}<0$
and $\check{\sigma}$. For now, we view these as non-stochastic though
it is without loss to allow them to be functions of $\left(\boldsymbol{x}_{m},\boldsymbol{q}_{m}\right)$.
We discuss in-sample estimation in Section \ref{subsec:Consistencyetc}.

Second, the researcher constructs the no-shock scenario which comprises
of a prediction of prices and shares $(\check{\boldsymbol{p}}_{m},\check{\boldsymbol{s}}_{m})$
based on the information in $\left(\boldsymbol{x}_{m},\boldsymbol{q}_{m}\right)$
only. With panel data and persistent cost and demand shocks, a natural
choice is the prices and shares in a period prior to the realization
of the $\boldsymbol{g}_{m}$ shocks (collected in $\boldsymbol{q}_{m}$).
In a single cross-section, predicted prices and mean utilities may
be the fitted values from regressing prices and $
  \mathcal{D}_{j}\left(
    \boldsymbol{s}_{m};\check{\sigma},\boldsymbol{x}_{m}^{(1)},\boldsymbol{p}_{m}
  \right)
$
on characteristics, respectively, while predicted shares may follow
from the model (i.e., equation (\ref{eq:shares})) at the parameters
$\check{\sigma}$ and implied mean utilities.\footnote{
  One could also include recentered shocks in these regressions to improve
  predictive power but take fitted values corresponding to the characteristics
  only. Nonlinear predictions, such as with machine learning algorithms,
  may improve precision, too.
}

Third, the researcher forms price predictions as deviations from $\check{\boldsymbol{p}}_{m}$
due to the exogenous cost-shocks. For clarity here we work with the
simplest predictions:
\begin{equation}
  \hat{p}_{jm}
  =
  \check{p}_{jm}+\check{\pi}\tilde{g}_{jm},
\end{equation}
where $\tilde{g}_{jm}=g_{jm}-\expec{g_{jm}\mid\boldsymbol{x}_{m},\boldsymbol{q}_{m}}$
is product $j$'s recentered cost shock and $\check{\pi}\ne0$ is
a pass-through coefficient that is again assumed non-stochastic or
measurable with respect to $(\boldsymbol{x}_{m},\boldsymbol{q}_{m})$,
for now. These predictions need not be correct for the resulting instruments
to be valid, as in Section \ref{sec:Motivating-Example:}. We consider
extensions with more elaborate shock pass-through models below.

The price prediction immediately suggests the first instrument (recall
equation (\ref{eq:nabla})): $-\check{\pi}\tilde{g}_{jm}$, or equivalently
the recentered shock $\tilde{g}_{jm}$. The price prediction also
implies a prediction for mean utilities that will shortly prove helpful:
\[
  \hat{\delta}_{jm}
  =
  \check{\delta}_{jm}+\check{\alpha}\check{\pi}\tilde{g}_{jm}
  \quad\text{for }
  \check{\delta}_{jm}
  =
  \mathcal{D}_{j}\left(
    \check{\boldsymbol{s}}_{m};\check{\sigma},\boldsymbol{x}_{m}^{(1)},\check{\boldsymbol{p}}_{m}
  \right).
\]

Finally, the researcher generates predictions for the market shares
and ultimately $\nabla_{jm}^{\sigma}$. We propose two versions: a
first-order approximation which yields a recentered shift-share IV,
and an exact prediction that may yield a more powerful instrument
but generally requires further recentering. Using the first-order
approximation, one predicts market shares as
\begin{align*}
  \hat{s}_{jm}
  &
  =
    \check{s}_{jm}
    +
    \sum_{k\in\mathcal{J}_{m}}
      \frac{\partial}{\partial\delta_{km}}
      \mathcal{S}_{j}(\check{\boldsymbol{\delta}}_{m};\check{\sigma},\boldsymbol{x}_{m}^{(1)},\check{\boldsymbol{p}}_{m})
      \left(\hat{\delta}_{km}-\check{\delta}_{km}\right)
  \\
  &
  =
  \check{s}_{jm}
  +
  \sum_{k\in\mathcal{J}_{m}}
    \frac{\partial}{\partial\delta_{km}}
    \mathcal{S}_{j}(\check{\boldsymbol{\delta}}_{m};\check{\sigma},\boldsymbol{x}_{m}^{(1)},\check{\boldsymbol{p}}_{m})
    \check{\alpha}\check{\pi}\tilde{g}_{km}
\end{align*}
and the model residual's derivative with respect to the vector $\sigma$ as
\begin{align}
  \hat{\nabla}_{jm}^{\sigma}
  &
  =
    \check{\nabla}_{jm}^{\sigma}
    +
    \sum_{k\in\mathcal{J}_{m}}
      \frac{\partial^{2}}{\partial p_{km}
      \partial\sigma}\mathcal{D}_{j}\left(\check{\boldsymbol{s}}_{m};\check{\sigma},\boldsymbol{x}_{m}^{(1)},\check{\boldsymbol{p}}_{m}\right)
      (\hat{p}_{km}-\check{p}_{km})\nonumber
  \\
  &
  \phantom{==}
  +
  \sum_{k\in\mathcal{J}_{m}}
    \frac{\partial^{2}}{\partial s_{km}\partial\sigma}
    \mathcal{D}_{j}\left(\check{\boldsymbol{s}}_{m};\check{\sigma},\boldsymbol{x}_{m}^{(1)},\check{\boldsymbol{p}}_{m}\right)
    (\hat{s}_{km}-\check{s}_{km})\nonumber
  \\
  &
  =
  \check{\nabla}_{jm}^{\sigma}
  +
  \sum_{k\in\mathcal{J}_{m}}w_{jkm}\tilde{g}_{km}.
  \label{eq:nabla_firstorder}
\end{align}
Here $
  \check{\nabla}_{jm}^{\sigma}
  =
  \frac{\partial}{\partial\sigma}
  \mathcal{D}_{j}\left(\check{\boldsymbol{s}}_{m};\check{\sigma},\boldsymbol{x}_{m}^{(1)},\check{\boldsymbol{p}}_{m}\right)
$
predicts the residual's derivative in the absence of shocks and
\begin{align}
  w_{jkm}
  &
  =
  \check{\pi}
  \frac{\partial^{2}}{\partial p_{km}\partial\sigma}
  \mathcal{D}_{j}\left(\check{\boldsymbol{s}}_{m};\check{\sigma},\boldsymbol{x}_{m}^{(1)},\check{\boldsymbol{p}}_{m}\right)
  \nonumber
  \\
  &
  \phantom{==}
  +
  \check{\alpha}\check{\pi}
  \sum_{k^{\prime}\in\mathcal{J}_{m}}
    \frac{\partial^{2}}{\partial s_{k^{\prime}m}\partial\sigma}
    \mathcal{D}_{j}\left(\check{\boldsymbol{s}}_{m};\check{\sigma},\boldsymbol{x}_{m}^{(1)},\check{\boldsymbol{p}}_{m}\right)
    \cdot
    \frac{\partial}{\partial\delta_{km}}
    \mathcal{S}_{k^{\prime}}(\check{\boldsymbol{\delta}}_{m};\check{\sigma},\boldsymbol{x}_{m}^{(1)},\check{\boldsymbol{p}}_{m})
  \label{eq:w_jmk}
\end{align}
is the predicted first-order effect of $\tilde{g}_{km}$ on that derivative.
Recentering the prediction in (\ref{eq:nabla_firstorder}) eliminates
the first term, resulting in the $\dim(\theta)\times1$ vector of
shift-share instruments
\[
  Z_{jm}^{SSIV}
  =
  \left(
    \begin{array}{c}
    -
    \check{\pi}\tilde{g}_{jm}
    \\
    \sum_{k\in\mathcal{J}_{m}}w_{jkm}\tilde{g}_{km}
    \end{array}
  \right).
\]
Again, an advantage of $Z_{jm}^{SSIV}$ is that it only requires specification
of the conditional shock means $\expec{\boldsymbol{g}_{m}\mid\boldsymbol{x}_{m},\boldsymbol{q}_{m}}$
for recentering. Also conveniently, the choice of $\check{\pi}$
is immaterial with this approach, as it only rescales the instruments.
Without a random coefficient in price the first term in equation (\ref{eq:w_jmk})
drops out, making $\check{\alpha}$ immaterial too.

Alternatively, the researcher can obtain the exact prediction of how
changes in mean utilities due to the cost shocks affect shares:
\[
  \hat{s}_{jm}
  =
  \mathcal{S}_{j}(\hat{\boldsymbol{\delta}}_{m};\check{\sigma},\boldsymbol{x}_{m}^{(1)},\hat{\boldsymbol{p}}_{m}),
\]
and how those changes affect $\nabla_{jm}^{\sigma}$:
\[
  \hat{\nabla}_{jm}^{\sigma}
  =
  \frac{\partial}{\partial\sigma}
  \mathcal{D}_{j}\left(\hat{\boldsymbol{s}}_{m};\check{\sigma},\boldsymbol{x}_{m}^{(1)},\hat{\boldsymbol{p}}_{m}\right).
\]
This $\hat{\nabla}_{jm}^{\sigma}$ is a nonlinear function of the
shocks which needs to be recentered, generally by specifying shock
counterfactuals as described in Section \ref{subsec:Recentered-Instruments}.\footnote{
  This $\hat{\nabla}_{jm}^{\sigma}$ relates to the efficient IV construction
  of \citet{Berry1999}: the two would be equivalent if the price predictions
  $\hat{p}_{jm}$ were taken from an equilibrium pricing model and mean
  utilities were set to $
    \hat{\delta}_{jm}
    =
    \check{\alpha}\hat{p}_{jm}+\check{\beta}^{\prime}x_{jm}
  $
  for an initial value $\check{\beta}$ of $\beta$. Our instrument
  differs in three ways: it is based on a shock pass-through model that
  need not be correct, it uses additional information (in particular,
  when lagged prices and shares are available), and it is recentered
  to avoid bias from endogenous characteristics. \citet{Conlon2020}
  propose an improvement on the \citet{Berry1999} instrument that integrates
  over an empirical distribution of $\xi_{jm}$ rather than setting
  unobserved demand shocks to zero. Our approach could be similarly
  extended.
} The recentered formula IV vector is then:
\[
  Z_{jm}^{FIV}
  =
  \left(
    \begin{array}{c}
      -\check{\pi}\tilde{g}_{jm}
    \\
      \hat{\nabla}_{jm}^{\sigma}
      -
      \expec{\hat{\nabla}_{jm}^{\sigma}\mid\boldsymbol{x}_{m},\boldsymbol{q}_{m}}
    \end{array}
  \right).
\]

Appendix Proposition \ref{prop:sw_approx} builds intuition for these
instruments by considering the case where the nonlinear parameters
$\sigma$ are the standard deviations of random coefficients and $\check{\sigma}$
is close to zero, which would correspond to a pure multinomial logit
model.\footnote{
  \citet{salanie_fast_2022} use an approximation of mixed logit around
  $\sigma=0$ to simplify share inversion and estimation. Our proof
  of Appendix Proposition \ref{prop:sw_approx} offers a new derivation
  that yields additional intuition, discussed in Appendix \ref{subsec:appx-Local-to-Logit}.
} In this “local-to-logit” approximation, the shift-share IV
corresponding to the standard deviation $\sigma_{\ell}$ of a non-price
characteristic $x_{jm\ell}$ can be written, up to a scaling factor, as:
\[
  z_{jm\ell}^{SSIV}
  \approx
  x_{jm\ell}\cdot\sum_{k\in\mathcal{J}_{m}}\check{s}_{km}\left(x_{km\ell}-\bar{x}_{m\ell}\right)\tilde{g}_{km}
  \quad\text{for }
  \bar{x}_{m\ell}
  =
  \sum_{k\in\mathcal{J}_{m}}\check{s}_{km}x_{km\ell}.
\]
This IV an interaction between product $j$'s own characteristic $x_{jm\ell}$
and a market-specific aggregate of the shocks: the share-weighted
covariance across the products in the market between $x_{km\ell}$
and the recentered cost shock (including the outside good with a shock
set to zero). The covariance is positive when the cost shocks unexpectedly
make products with high $x_{km\ell}$ more expensive relative to other
products in the market. Thus, $z_{jm\ell}^{SSIV}$ is performing an
analysis similar to difference-in-differences: it compares changes
in market shares for high-$x_{jm\ell}$ vs. low-$x_{jm\ell}$ products
in markets where high-$x_{km\ell}$ vs. low-$x_{km\ell}$ products
became less competitive because of the exogenous cost shocks. Identification
with this instrument is therefore based on the core property of mixed
logit models: that, after a cost shock, market shares are reallocated
towards products with similar characteristics when $\sigma_{\ell}$
is large, but to all products evenly (in proportion of their market
shares) when $\sigma=0$. The instrument for the random coefficient
in price has additional terms related to how cost shocks affect prices
directly; see Appendix \ref{subsec:appx-Local-to-Logit}.

\subsection{Estimation and Asymptotics}
\label{subsec:Consistencyetc}

We use a generalized method of moments (GMM) procedure to estimate $\theta$.
Specifically, we generalize the moment condition (\ref{eq:moments}) to write:
\begin{equation}
  \expec{
    Z_{jm}\left(\check{\theta},\check{\pi}\right)
    \cdot
    \left(
      \mathcal{D}_{j}\left(\boldsymbol{s}_{m};\sigma,\boldsymbol{x}_{m}^{(1)},\boldsymbol{p}_{m}\right)
      -
      \alpha p_{jm}
      -
      \mathcal{B}_{j}(\boldsymbol{x}_{m},\boldsymbol{q}_{m};\gamma,\check{\theta})
    \right)
  }
  =
  0,
  \label{eq:moments_with_B}
\end{equation}
where $Z_{jm}\left(\check{\theta},\check{\pi}\right)$ is a vector
of recentered instruments of the same dimensionality as $\theta$,
constructed as above, now with the dependence on preliminary parameter
values $(\check{\theta},\check{\pi})$ made explicit. The
$\mathcal{B}_{j}(\boldsymbol{x}_{m},\boldsymbol{q}_{m};\gamma,\check{\theta})$
term is included to reduce residual variation in the error
$\Xi_{jm}=\beta'x_{jm}+\xi_{jm}$. Since recentered IVs are uncorrelated
with any function of $(\boldsymbol{x}_{m},\boldsymbol{q}_{m})$,
the moment conditions (\ref{eq:moments_with_B}) hold at the true
parameter values regardless of $\mathcal{B}_{j}(\cdot)$ and for any $\gamma$.

Two examples of $\mathcal{B}_{j}(\cdot)$ are illuminating. First,
it can be set to $\gamma'x_{jm}$ with $\gamma$ estimated as a projection
coefficient (i.e., from regressing the estimate of $\Xi_{jm}$ on
$x_{jm}$). While $\gamma$ need not coincide with the causal effect
of characteristics on demand, $\beta$, this is not a problem for
many important policy counterfactuals (as discussed in Section \ref{subsec:general_setting}).
Second, in panel data, one can set $\mathcal{B}_{j}$ to be the lagged
value of $\Xi_{jm}$ obtained from the lagged prices and shares using
the initial parameter values, $\check{\theta}$ (and with no additional
parameters, $\gamma=\emptyset$). This corresponds to estimating the
model “in differences,” as commonly done in linear models and
sometimes also for nonlinear demand models (e.g., \citet{Adao2017}).
Such differencing helps reduce residual variation when the unobserved
demand shifters are strongly serially correlated.

Our baseline estimator is the $(\hat{\theta},\hat{\gamma})$ that
solves the sample analog of condition (\ref{eq:moments_with_B}),
averaging across product-market pairs, along with the sample analog
of a moment condition for $\gamma$:
\begin{equation}
  \expec{
    \frac{\partial\mathcal{B}_{j}(\boldsymbol{x}_{m},\boldsymbol{q}_{m};\gamma,\check{\theta})}{\partial\gamma}
    \cdot
    \left(
      \mathcal{D}_{j}\left(\boldsymbol{s}_{m};\sigma,\boldsymbol{x}_{m}^{(1)},\boldsymbol{p}_{m}\right)
      -
      \alpha p_{jm}
      -
      \mathcal{B}_{j}(\boldsymbol{x}_{m},\boldsymbol{q}_{m};\gamma,\check{\theta})
    \right)
  }
  =
  0.
  \label{eq:moments_for_B}
\end{equation}
This condition defines $\gamma$ as the coefficient giving the least-squares
fit of $\Xi_{jm}$ on $\mathcal{B}_{j}(\boldsymbol{x}_{m},\boldsymbol{q}_{m};\gamma,\check{\theta})$
(e.g. a projection on $x_{jm}$ in the above linear case). It can
be understood in the same way as how, in regression analyses of randomized
control trials, predetermined controls are included to soak up residual
outcome variation with the coefficients on them not interpreted causally.
Here we assume the researcher has obtained initial values of $\theta$
by, say, an initial GMM procedure with $\tilde{g}_{jm}$ and conventional
characteristic-based IVs as instruments. The researcher has also obtained
an initial pass-through constant $\check{\pi}$, for example from
least-squares estimation of $p_{jm}=\pi_{0}+\pi\tilde{g}_{jm}+\epsilon_{jm}$.

We also consider a “continuously updating” estimator which,
unlike the baseline estimator, does not require initial values of
$\theta$. This estimator replaces $\check{\theta}$ in the moment
conditions (\ref{eq:moments_with_B})–(\ref{eq:moments_for_B}) with
$\theta$; that is, the instruments $Z_{jm}$ (and, if applicable,
the $\mathcal{B}_{j}$ term) are updated when searching for the parameter
estimate.\footnote{
  We use the term “continuously updating” differently to the standard
  continuously updating estimator, where it refers to the choice of the
  GMM weighting matrix. That choice is not relevant to our just-identified
  setting.
} We use this estimator in our simulations, below. Another alternative
is to use a two-step or iterative GMM procedure.

Consistency and asymptotic normality of these estimators follow from
standard GMM theory (e.g. \citet[Theorems 2.6 and 3.1]{newey1994large})
when there are many \emph{iid} markets $m$ (or, more generally, many
\emph{iid }market clusters: e.g., in a panel with many regions and a
small number of time periods). In other cases, such as when there
are only a few markets or when across-market linkages create dependences
in the instruments and GMM residuals, asymptotic properties can be
established from many \emph{iid} shocks $g_{jm}$ (or, more generally,
many \emph{iid} shock clusters\emph{}) following \citet{adao2019shift},
\citet{BJH2018}, and \citet{BH1}. This strategy is helpful, for
instance, when the markets represent regions and the shocks arise
from exchange rate fluctuations, which affect all regions at once.
Other shocks can affect the demand for similar products across multiple
regions, too. We develop this approach in a setting where $\gamma=\emptyset$
(e.g. the differencing case discussed above), $Z_{jm}$ consists of
shift-share instruments, and the preliminary values $(\check{\theta},\check{\pi})$
are non-stochastic.\footnote{
  For recentered instruments that do not have a shift-share structure,
  \citet{BH1} provide sufficient conditions for consistency in linear
  IV settings. Adapting them to our current setting is left to future
  work. No general asymptotic inference results are currently known
  for such instruments, even for linear IV settings.
} The estimator $\hat{\theta}$ then solves:
\begin{align}
  0
  &
  =
  \sum_{m}\sum_{j\in\mathcal{J}_{m}}
    \left(\sum_{k\in\mathcal{J}_{m}}w_{jkm}\tilde{g}_{km}\right)
    \left(
      \mathcal{D}_{j}\left(\boldsymbol{s}_{m};\hat{\sigma},\boldsymbol{x}_{m}^{(1)},\boldsymbol{p}_{m}\right)
      -
      \hat{\alpha}p_{jm}
      -
      \mathcal{B}_{j}(\boldsymbol{x}_{m},\boldsymbol{q}_{m},\check{\theta})
    \right)
  \nonumber
  \\
  &
  =
  \sum_{m}\sum_{k\in\mathcal{J}_{m}}
  \tilde{g}_{km}\mathcal{R}_{km}(\hat{\theta})
  \label{eq:blpaggregate}
\end{align}
where $
  \mathcal{R}_{km}(\hat{\theta})
  =
  \sum_{j\in\mathcal{J}_{m}}w_{jkm}
  \left(
    \mathcal{D}_{j}\left(\boldsymbol{s}_{m};\hat{\sigma},\boldsymbol{x}_{m}^{(1)},\boldsymbol{p}_{m}\right)
    -
    \hat{\alpha}p_{jm}
    -
    \mathcal{B}_{j}(\boldsymbol{x}_{m},\boldsymbol{q}_{m},\check{\theta})
  \right)
$
is an “aggregated” shock-level residual in the sense of \citet{adao2019shift}
and \citet{BJH2018}.\footnote{
  Note that we include the “price instrument” $-\check{\pi}\tilde{g}_{jm}$
  as a shift-share IV here, with $w_{jkm}=-\check{\pi}\mathbf{1}[j=k]$.
} Equation (\ref{eq:blpaggregate}) represents $\hat{\theta}$ as the
solution of a “shock-level” GMM procedure, with an estimable
variance of $\sum_{m}\sum_{k\in\mathcal{J}_{m}}\tilde{g}_{km}\mathcal{R}_{km}(\theta)$
given many \emph{iid} shocks or many shock clusters that allow for
shock correlations across markets. Standard GMM expressions can then
be applied, as before, regardless of the correlation structure in
the residual $
  \mathcal{D}_{j}\left(\boldsymbol{s}_{m};\sigma,\boldsymbol{x}_{m}^{(1)},\boldsymbol{p}_{m}\right)
  -
  \alpha p_{jm}
  -
  \mathcal{B}_{j}(\boldsymbol{x}_{m},\boldsymbol{q}_{m},\check{\theta})
$
across both products and markets. Convergence of $\hat{\theta}$ only
requires the shocks to induce sufficient variation across product-market
pairs in the IVs.

Results on the asymptotic efficiency of recentered IVs for linear
structural equations, developed by \citet{borusyak_efficient_2021},
can also be extended to characterize the optimal IV matrix,
$Z^{*}=\left(Z_{jm}^{*\prime}\right)_{m,j\in\mathcal{J}_{m}}$
without assuming many independent markets. Under appropriate regularity
conditions, it takes the form:
\begin{align*}
  Z^{*}
  &
  =
  \expec{\xi\xi^{\prime}\mid\boldsymbol{x},\boldsymbol{q}}^{-1}
  \left(\expec{\nabla\mid\boldsymbol{g},\boldsymbol{x},\boldsymbol{q}}-\expec{\nabla\mid\boldsymbol{x},\boldsymbol{q}}\right),
\end{align*}
where $\xi=\left(\xi_{jm}\right)_{m,j\in\mathcal{J}_{m}}$,
$\nabla=\left(\nabla_{jm}^{\prime}\right)_{m,j\in\mathcal{J}_{m}}$,
and $\boldsymbol{v}=\left(\boldsymbol{v}_{m}\right)_{m}$ for any
variable $\boldsymbol{v}$. The inner term in parentheses stacks the
recentered best predictors of the model's residual derivatives, $\tilde{Z}_{jm}$.
The recentered predictor is then adjusted by $\expec{\xi\xi^{\prime}\mid\boldsymbol{x},\boldsymbol{q}}^{-1}$,
which can be understood as combining a partial residualization of
$\tilde{Z}_{jm}$ on $\expec{\xi\mid\boldsymbol{x},\boldsymbol{q}}$
and a reweighting by $\var{\xi\mid\boldsymbol{x},\boldsymbol{q}}^{-1}$
(see Proposition 3 in \citet{borusyak_efficient_2021}). This characterization
provides further motivation for our focus on approximating $\tilde{Z}_{jm}$
as well as for the adjustment for $\mathcal{B}_{j}(\boldsymbol{x}_{m},\boldsymbol{q}_{m},\check{\theta})$
in (\ref{eq:moments_with_B}), as a proxy for $\expec{\xi_{jm}\mid\boldsymbol{x},\boldsymbol{q}}$,
in estimation. Weighting by an estimate of the residual's inverse
variance, as in feasible generalized least squares, is less popular
in practice and not pursued here.

\subsection{Extensions}
\label{subsec:Extensions}

We now develop several extensions of the baseline model. We consider
them one by one to avoid notational clutter, but in practice they
can be combined.

\paragraph{Observed Consumer Characteristics.}

In some applications, the researcher observes the distribution of
consumer characteristics in each market and allows these consumer
characteristics correlate with tastes for product characteristics
in $\boldsymbol{x}_{m}^{(1)}$. Our results extend immediately to
that case. Specifically, the consumer with characteristics
$c_{i}=\left(c_{ir}\right)_{r=1}^{R}$ solves:
\[
  \max_{j\in\mathcal{J}_{m}\cup\left\{ 0\right\} }
  \delta_{jm}
  +
  \left(\sum_{r=1}^{R}\gamma_{r0}c_{ir}
  +
  \eta_{i0}\right)p_{jm}
  +
  \sum_{\ell=1}^{L_{1}}
    \left(\sum_{r=1}^{R}\gamma_{r\ell}c_{ir}+\eta_{i\ell}\right)
    x_{jm\ell}+\varepsilon_{ijm},
\]
where $\gamma=\left(\gamma_{r\ell}\right)$ serve as additional nonlinear
parameters and extreme-value shocks $\varepsilon_{ijm}$ are independent
from $(c_{i},\eta_{i})$. The distribution $\mathcal{\mathcal{P}}_{m}\left(\cdot;\sigma\right)$
of $(c_{i},\eta_{i})$ is known: typically $c_{i}$ is assumed independent
of $\eta_{i}$ with the market-specific distribution taken from the
data. Again, the model is invertible (see, e.g., \citet{Gandhi2021}),
and the rest of the analysis goes through without change.

\paragraph{Using Lagged Prices and Shares with Product Entry and Exit.}

When lagged prices and shares are available, our baseline recommendation
is to use them as $\left(\check{\boldsymbol{p}}_{m},\check{\boldsymbol{s}}_{m}\right)$
when constructing the instruments. This approach requires a modification
when some products have recently entered the market and their lagged
information is not available. Moreover, if many products have exited,
lagged shares may be a poor prediction of the current period's share
in the absence of cost shocks.

In such cases, our proposal is to predict prices $\check{p}_{jm}$
and mean utilities $\check{\delta}_{jm}$ for all products in the
current period and construct shares from them, as $
  \check{s}_{jm}
  =
  \mathcal{S}_{j}(\check{\boldsymbol{\delta}}_{m};\check{\sigma},\boldsymbol{x}_{m}^{(1)},\check{\boldsymbol{p}}_{m})
$.
For continuing products, lagged price can serve as $\check{p}_{jm}$,
while mean utility can be obtained from the inversion of lagged shares,
given $\check{\sigma}$. For new products, one may proceed as in a
single cross-section, taking fitted values from regressions of realized
price and implied mean utility $
  \mathcal{D}_{j}\left(\boldsymbol{s}_{m};\check{\sigma},\boldsymbol{x}_{m}^{(1)},\boldsymbol{p}_{m}\right)
$
on characteristics.

\paragraph{Identification of $\beta$ via Instruments for Product Entry.}

Our baseline analysis assumes that the $\boldsymbol{g}_{m}$ shocks
do not affect the characteristics of available products, which leaves
the causal effect of characteristics on mean utility, $\beta$, unidentified.
This is in contrast to the price coefficient $\alpha$ which is identified
because price is affected by the shocks and the researcher is able
to construct a relevant recentered instrument, $\check{\pi}\tilde{g}_{jm}$.
If the researcher has access to shocks that affect some characteristics
in a predictable way, those characteristics can be treated in the
same way as the baseline model treats price, and the corresponding
components of $\beta$ become identified.

\paragraph{Incorporating a Pricing Model.}

Our baseline analysis focuses solely on demand estimation, leaving
the supply side flexible. Although we require an auxiliary model of
cost shock pass-through, it need not be correct and is indeed very
simple in our baseline proposal. This modular approach has the advantage
that demand can be estimated with fewer assumptions. However, some
counterfactuals, such as a merger between two firms, require predicting
how prices would change, and thus taking a stand on how firms set
prices. In that case, the researcher may consider leveraging the pricing
model to obtain more powerful—albeit less robust—demand estimates,
too, as in \citet{Berry1995}. While we leave the technical presentation
to future drafts, this extension should be straightforward.

\paragraph{Estimated Mapping from Inputs to Products.}

Our baseline analysis assumes that the cost shocks are product-specific.
When the shocks originate from input prices or exchange rate fluctuations,
this requires (at least partial) knowledge of the mapping from inputs
to products: e.g., the percentages of wheat and corn among the ingredients
of ready-to-eat cereals \citep{barahona2023equilibrium} or the country
of assembly for each car model \citep{Grieco2021}. However, in some
settings where input prices are observed, the mapping to products
is not available. \citet{Villas-Boas2007} addresses this problem
by using interactions between market-specific input prices and product
dummies as instruments. The intuition is that, when the same product
is observed in sufficiently many markets (e.g., time periods), the
sensitivity of each product's price to all inputs prices is revealed.

This insight can be adapted to our setting, in a two-step approach.
First, the exposure of each product to the set of inputs is estimated
by a product-specific (e.g., time-series) regression of price on recentered
input price shocks. This regression should have sufficiently many
observations per product, such that the parameters converge to some
pseudo-true values, which need not reflect the true production function.
Second, a product-specific cost shock is constructed as a shift-share
aggregate of input price shocks as shifts with estimated exposures
as shares and used in the rest of the analysis. We leave the precise
asymptotic analysis of this approach to future drafts.

\paragraph{Alternative Demand Models.}

While we have focused on nested and mixed logit, our approach extends
to other popular parametric demand systems. Most directly, nested
and mixed constant elasticity of substitution (CES) models are closely
related, with prices replaced with log prices and quantity shares
replaced with expenditure shares. Our instrument construction then
goes through. Similarly, our analysis extends directly to variations
on the mixed logit model used in IO, such as the Hotelling model of
spatial product differentiation \citep[e.g.][]{houde_spatial_2012}
and the “principles of differentiation” model of \citet{bresnahan1997market}
which combines multiple nest groupings.

\paragraph{Non-Parametric Demand.}

We follow \citet{Erry2014} in considering non-parametric identification
of demand under Assumption \ref{assu:exogenous-shocks} instead of
the conventional stronger assumption (our equation (\ref{eq:assu-conventional}))
that they impose. As they show, identification requires an index restriction:
that at least one component of $(p_{jm},x_{jm}')'$ enters demand
without a random coefficient. Moreover, exogenous cost shocks are
only sufficient for non-parametric identification if price satisfies
this property. We show that this result extends to our weaker assumption, too:
\begin{prop}\label{prop:nonparametric}
  Consider the non-parametric inverse demand model with an index restriction on price:
  \begin{equation}
    p_{jm}
    +
    \xi_{jm}
    =
    \mathcal{D}_{j}(\boldsymbol{s}_{m},\boldsymbol{x}_{m})
    \label{eq:np-model}
  \end{equation}
  for an unknown set of functions $\mathcal{D}_{j}$. Suppose Assumption
  \ref{assu:exogenous-shocks} holds with $\boldsymbol{q}_{m}=\emptyset$
  and the cost shocks satisfy a completeness property: for any function
  $h(\boldsymbol{s}_{m},\boldsymbol{x}_{m})$ with finite expectation,
  $
    \expec{h(\boldsymbol{s}_{m},\boldsymbol{x}_{m})\mid\boldsymbol{g}_{m},\boldsymbol{x}_{m}}
    =
    0
  $
  a.s. implies $h(\boldsymbol{s}_{m},\boldsymbol{x}_{m})=0$ a.s. Then
  $\mathcal{D}_{j}(\cdot)$ and the unobserved demand shifter $\xi_{jm}$
  are identified up to an additive term $\beta_{j}(\boldsymbol{x}_{m})$.
  Moreover, cross-price elasticities are point-identified.
\end{prop}

\section{Monte Carlo Simulations}
\label{sec:Monte-Carlo-Simulations}

We now analyze the bias and variance properties of the recentered
IV approach, relative to conventional alternatives, in a Monte Carlo
simulation that largely follows \citet{Gandhi2015}. Section \ref{subsec:baseline_sim}
describes the baseline data-generating process, where both conventional
and recentered IVs are valid, and shows what data features drive the
variance of estimates in the two approaches. Section \ref{subsec:endog_sim}
then introduces product characteristic endogeneity, demonstrating
that our proposed IVs remain accurate while conventional characteristic-based
IVs are significantly biased. Details of the algorithms used for estimation
are reported in Appendix \ref{sec:monte_carlo_dgp_estimation}.

\subsection{Mixed Logit with Exogenous Characteristics}
\label{subsec:baseline_sim}

\paragraph{Simulation Design.}

We simulate a set of regions $r=1,\dots,100$ in two periods $t\in\left\{ 1,2\right\} $;
hence $m=(r,t).$ In each period, consumers choose between products
$j\in\mathcal{J}_{m}=\{1,\ldots,15\}$ and the outside good to maximize
their utility, according to equation (\ref{eq:UMP}). We consider
$L_{1}=2$ observed time-invariant characteristics $x_{jr\ell}\stackrel{iid}{\sim}N(0,1)$,
in addition to the intercept $x_{jr0}=1$. Random coefficients are
placed on both characteristics, $x_{jm}^{(1)}=\left(x_{jm1},x_{jm2}\right)$
but not on price. The random coefficients $\eta_{i\ell}\stackrel{iid}{\sim}N(0,\sigma_{\ell}^{2})$
have true standard deviations of $\sigma_{\ell}=4$ for $\ell=1,2$.
Product $j$'s mean utility $\delta_{jm}$ is determined each period
according to equation (\ref{eq:delta-meanu}) with persistent unobserved
demand shifters: $\xi_{jr1}\stackrel{iid}{\sim}N(0,1)$ and $\xi_{jr2}=0.9\xi_{jm1}+\sqrt{1-0.9^{2}}\cdot e_{jm}$
for $e_{jm}\stackrel{iid}{\sim}N(0,1)$. We set $\beta_{0}=35$, $\beta_{1}=\beta_{2}=2$,
and $\alpha=-0.2-4\exp(0.5)$. Market shares are simulated with 1,000 independent draws:
\begin{equation}
  s_{jm}
  =
  \dfrac{1}{1000}\sum_{i=1}^{1000}
  \frac{
    \exp\left(\delta_{jm}+\eta_{i}^{\prime}x_{jm}^{(1)}\right)
  }{
    1+\sum_{k\in\mathcal{J}_{m}}\exp\left(\delta_{km}+\eta_{i}'x_{km}^{(1)}\right)
  }.
  \label{eq:share-1000draws}
\end{equation}

Prices are set by a simultaneous Bertrand-Nash game where each product
is produced by a single firm. In each period the price vector for
each region $\boldsymbol{p}_{mt}$ is the solution to the following
system of equations derived from the firms' first-order conditions:
\begin{equation}
  \boldsymbol{p}_{m}
  =
  \boldsymbol{c}_{m}
  -
  \left[
    \dfrac{d\boldsymbol{S}(\boldsymbol{\delta}_{m};\sigma,\boldsymbol{x}_{m}^{(1)},\boldsymbol{p}_{m})}{d\boldsymbol{p}_{m}'}
  \right]^{-1}
  \cdot
  \boldsymbol{S}(\boldsymbol{\delta}_{m};\sigma,\boldsymbol{x}_{m}^{(1)},\boldsymbol{p}{}_{m}).
  \label{eq:price-foc}
\end{equation}
where $c_{jm}=\gamma^{\prime}x_{jm}+\omega_{jm}+g_{jm}$ is firm $j$'s
marginal cost (and the derivative with respect to price includes the
effect through $\boldsymbol{\delta}_{m}$). We set $\gamma_{0}=5$
and $\gamma_{1}=\gamma_{2}=1$ and generate persistent unobserved
cost shocks: $\omega_{jr1}\stackrel{iid}{\sim}N(0,1)$ and
$\omega_{jr2}=0.9\omega_{jr1}+\sqrt{1-0.9^{2}}\cdot w_{jr}$
for $w_{jr}\stackrel{iid}{\sim}N(0,1)$. The observed cost shocks
only happen in the second period, such that $g_{jr1}=0$ and
$g_{jr2}\stackrel{iid}{\sim}N(0,0.2^{2})$.\footnote{
  The parameters of our simulation were picked to follow the simulation
  in \citet{Gandhi2015} as much as possible. The deviations arise for
  three reasons: we have two periods, we distinguish between observed
  and unobserved costs shocks, and we do not have a random coefficient
  on price. Our value for $\alpha$ is picked as the average price coefficient:
  they set the linear coefficient on price to $-0.2$ and have (the
  negative of) log-normal random coefficients with the mean $-4e^{0.5}$.
  We also set $L_{1}=2$ instead of $4$ and $\beta_{0}=35$ instead of $50$.
} Note that the variance of $g_{jr2}$ is only 4\% of the variance
of unobserved cost shocks, consistent with the limited exogenous shock
variation expected in typical applications.

We estimate this model for two alternative sets of moment conditions.
For the conventional characteristic-based IVs, let $Z_{jr2}^{C}$
be a vector collecting $g_{jr2}$, $x_{jr}$, and a set of two instruments
for $\sigma=(\sigma_{1},\sigma_{2})$: either BLP (sum of competitor
characteristics) instruments or the local or quadratic differentiation
IVs for proposed by \citet{Gandhi2015}. These instruments are given by:
\begin{align*}
  \text{BLP Sum of Characteristics IV}:   & \quad z_{jr2\ell} = \sum_{k\in\mathcal{J}_{r},k\ne j} x_{kr\ell}\\
  \text{GH Quadratic Differentiation IV}: & \quad z_{jr2\ell} = \sum_{k\in\mathcal{J}_{r},k\ne j} \left(x_{jr\ell}-x_{kr\ell}\right)^{2} \\
  \text{GH Local Differentiation IV}:     & \quad z_{jr2\ell} = \sum_{k\in\mathcal{J}_{r},k\ne j} \one\left[|x_{jr\ell}-x_{jr\ell}|<\kappa_{\ell}\right]
\end{align*}
with a proximity threshold $\kappa_{\ell}$; we follow \citet{Gandhi2015}
and use the standard deviation of $x_{jm\ell}$. We then estimate
$(\alpha,\beta,\sigma)$ via GMM using data from the second period
only (when the cost shock is available) and the moment condition:
\begin{align*}
  \expec{Z_{jr2}^{C}\xi_{jr2}} & = 0.
\end{align*}

For the recentered instruments, let $Z_{jr2}^{R}$ be a vector collecting
$g_{jr2}$ and either the shift-share or recentered exact prediction
IVs proposed in Section \ref{sec:General-Approach}. We recenter by
permuting the cost-shocks 20 times across both products and markets.
We use the continuously updating procedure proposed in Section \ref{subsec:Consistencyetc}
to bypass the need for initial estimates (iterative GMM yields very
similar results). The moment condition is:
\begin{align*}
  \expec{Z_{jr2}^{R}\Delta\xi_{jr}} & = 0
\end{align*}
for $\Delta\xi_{jr}=\xi_{jr2}-\xi_{jr1}$. Differencing corresponds
to setting $\mathcal{B}_{j}$ defined in Section \ref{subsec:Consistencyetc}
to the lagged values $\beta'x_{jr}+\xi_{jr1}$ (since characteristics
are time-invariant they drop from the estimation) which helps reduce
residual variation since the demand shifters are serially correlated.
Since we simulate a panel with two periods, it is natural to set the
prediction for prices and shares that we will use to construct $Z_{jr2}^{R}$
to their pre-cost shock (i.e. period one) values, $
  (\check{p}_{jr2},\check{s}_{jr2})
  =
  (p_{jr1},s_{jr1})
$.
See Appendix \ref{sec:monte_carlo_dgp_estimation} for additional
details on our estimation.

\paragraph{Baseline Results.}

\begin{figure}[!tp]
  \centering
  \caption{Baseline Simulation Results}
  \label{fig:monte-carlo-baseline}
  \subfloat[Price Coefficient $\alpha$]{%
    \includegraphics[width=0.485\columnwidth]{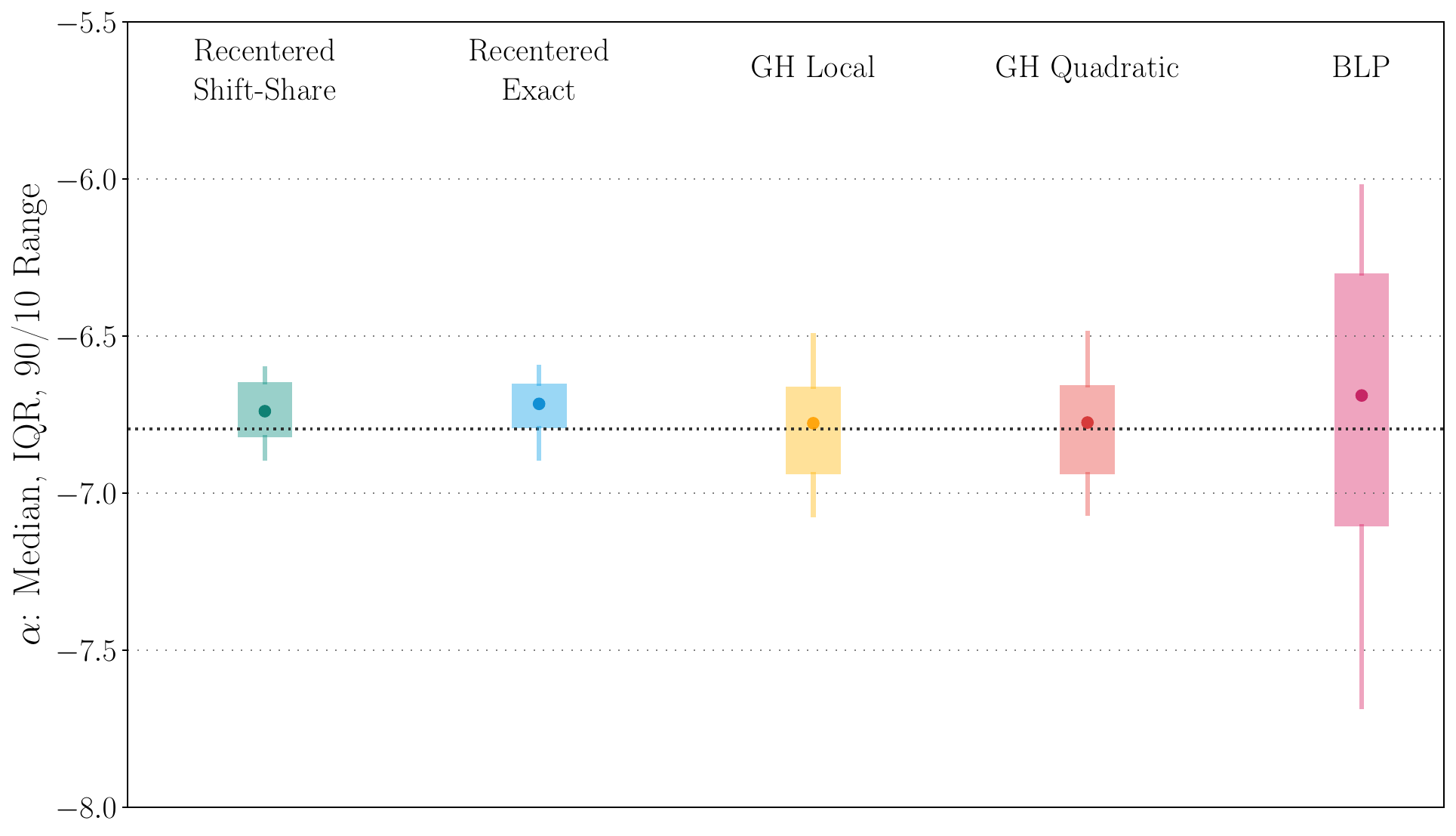}}
  \hfill{}
  \subfloat[Non-Linear Parameters $\sigma_{1},\sigma_{2}$]{%
    \includegraphics[width=0.485\columnwidth]{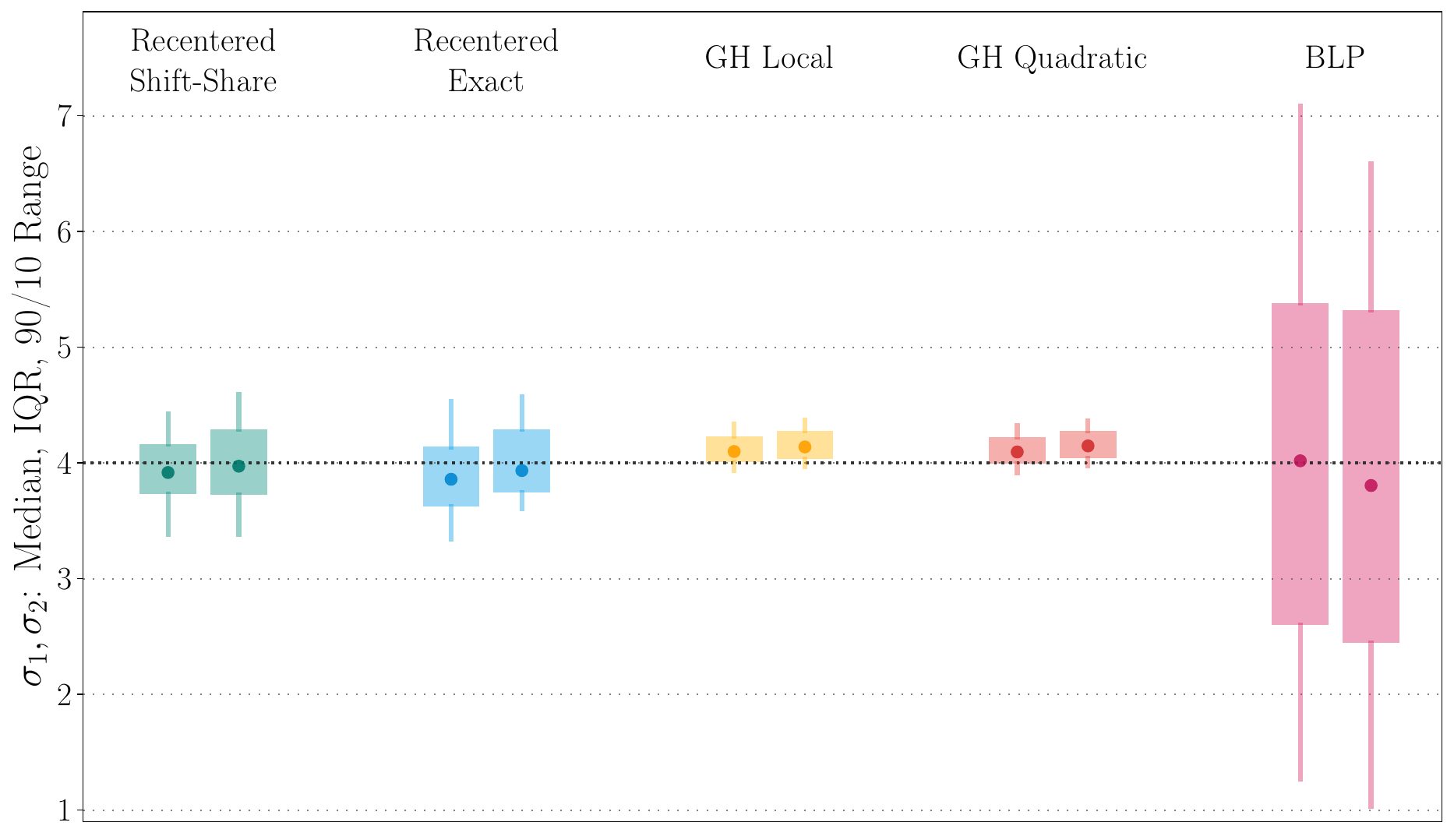}} \\
  \begin{minipage}{0.95\linewidth}
    \footnotesize {\itshape Notes.}
    The two panels show the simulated distributions for the GMM estimates
    of $\alpha$ and $( \sigma_1,\sigma_2)$ across 100 simulations of the
    data-generating process described in Section \ref{subsec:baseline_sim}. The
    ``Recentered Shift-Share'' estimates use the shift-share IV described in
    Section \ref{subsec:Constructing-IVs}; ``Recentered Exact'' estimates use the
    exact prediction IV (recentered around the average of 20 permutations of the
    cost shock); ``GH Local,'' ``GH Quadratic,'' and ``BLP'' correspond to the
    characteristic IVs described in Section \ref{subsec:baseline_sim}. For each
    set of estimates, we plot the median, a box delineating the 25th and 75th
    percentiles, lines denoting the 10th and 90th percentiles, and a horizontal
    dashed line denoting the true value of the parameters.
  \end{minipage}
\end{figure}

Figure \ref{fig:monte-carlo-baseline} shows the results of our estimation
for 100 Monte Carlo simulations. As expected, each set of instruments
yields approximately unbiased estimates for each of the parameters.
The recentered IVs tend to estimate the price coefficient somewhat
more precisely than the differentiation IVs, with a tighter distribution
of estimates, while the reverse is true for the nonlinear parameters.
BLP instrument estimates are considerably noisier for all parameters;
we drop them going forward to focus on the leading characteristic-based IVs.

\paragraph{Sensitivity.}

\begin{figure}[!tp]
  \centering
  \caption{Role of Cost Shock Variation}
  \label{fig:monte-carlo-varyshocks}
  \subfloat[Price Coefficient $\alpha$]{%
    \includegraphics[width=0.485\columnwidth]{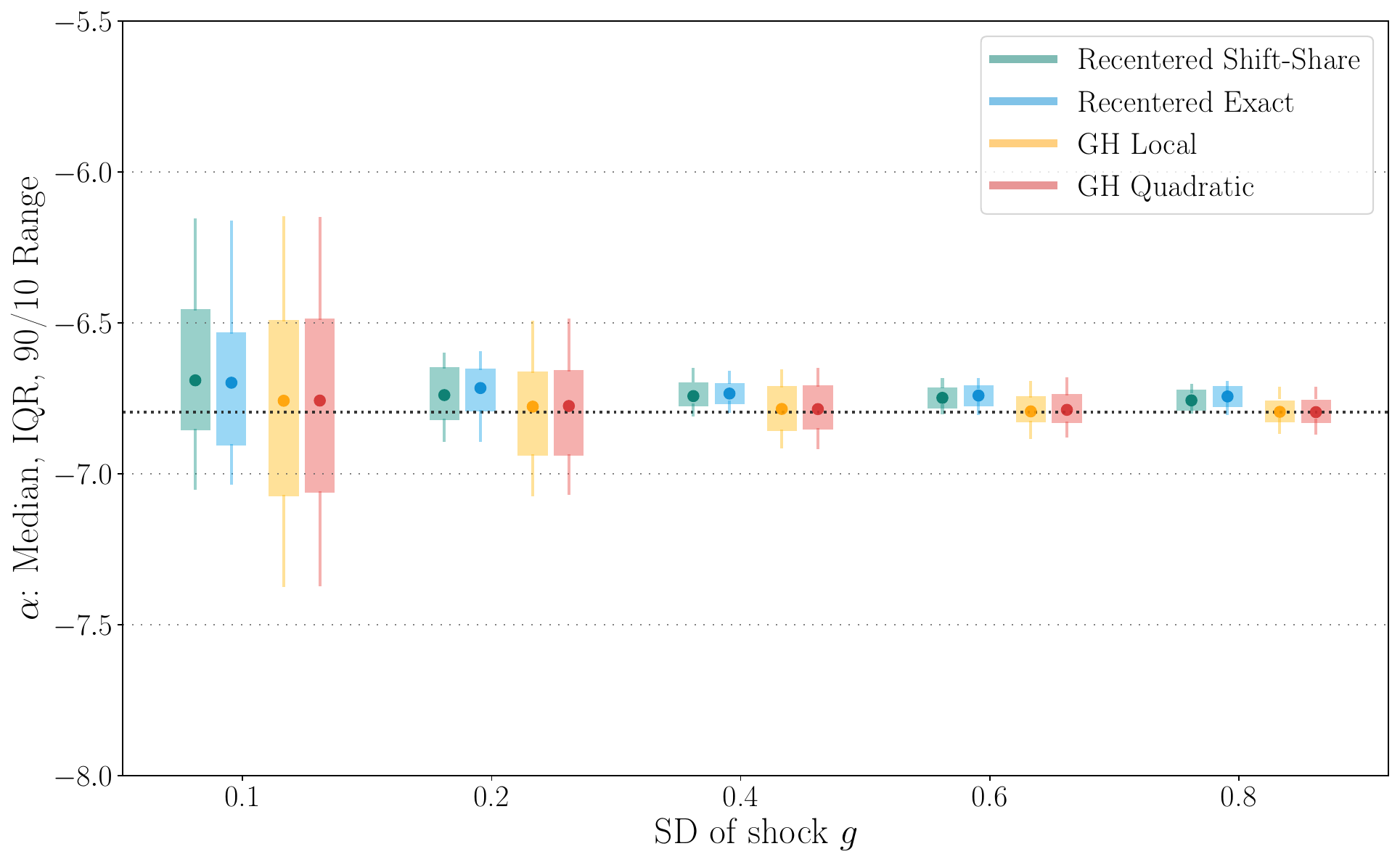}}
  \hfill{}
  \subfloat[Non-Linear Parameter $\sigma_{1}$]{%
    \includegraphics[width=0.485\columnwidth]{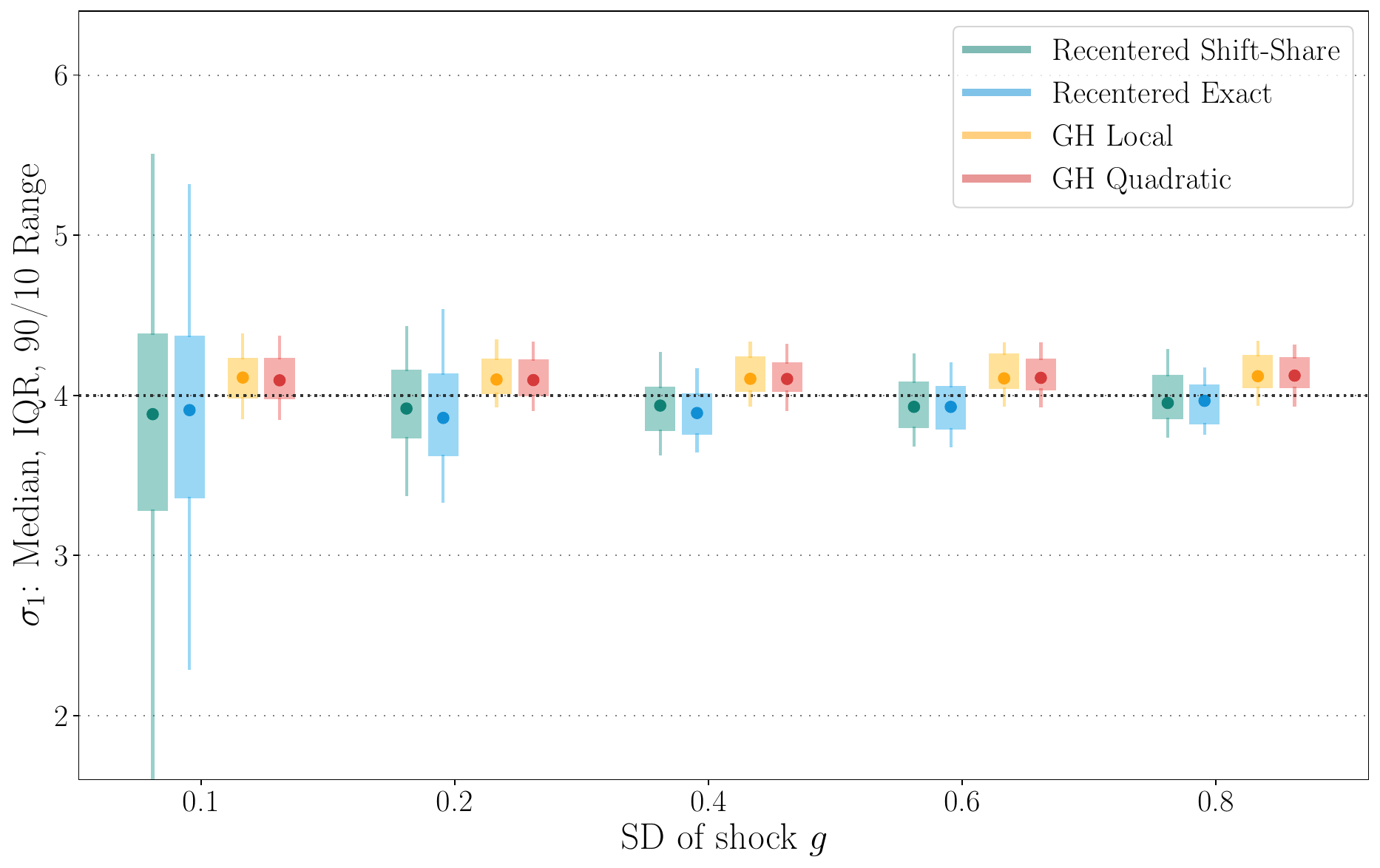}} \\
  \begin{minipage}{0.95\linewidth}
    \footnotesize {\itshape Notes.}
    The two panels show the distributions of the indicated parameter estimates for
    different values of the standard deviation of the cost shock $g_{jr2}$. The
    data-generarating process is otherwise unchanged; see notes to Figure
    \ref{fig:monte-carlo-baseline}.
  \end{minipage}
\end{figure}

\begin{figure}[!tp]
  \centering
  \caption{Role of Cross-Market Characteristic Variation}
  \label{fig:monte-carlo-varyxcommon}
  \subfloat[Price Coefficient $\alpha$]{
    \includegraphics[width=0.485\columnwidth]{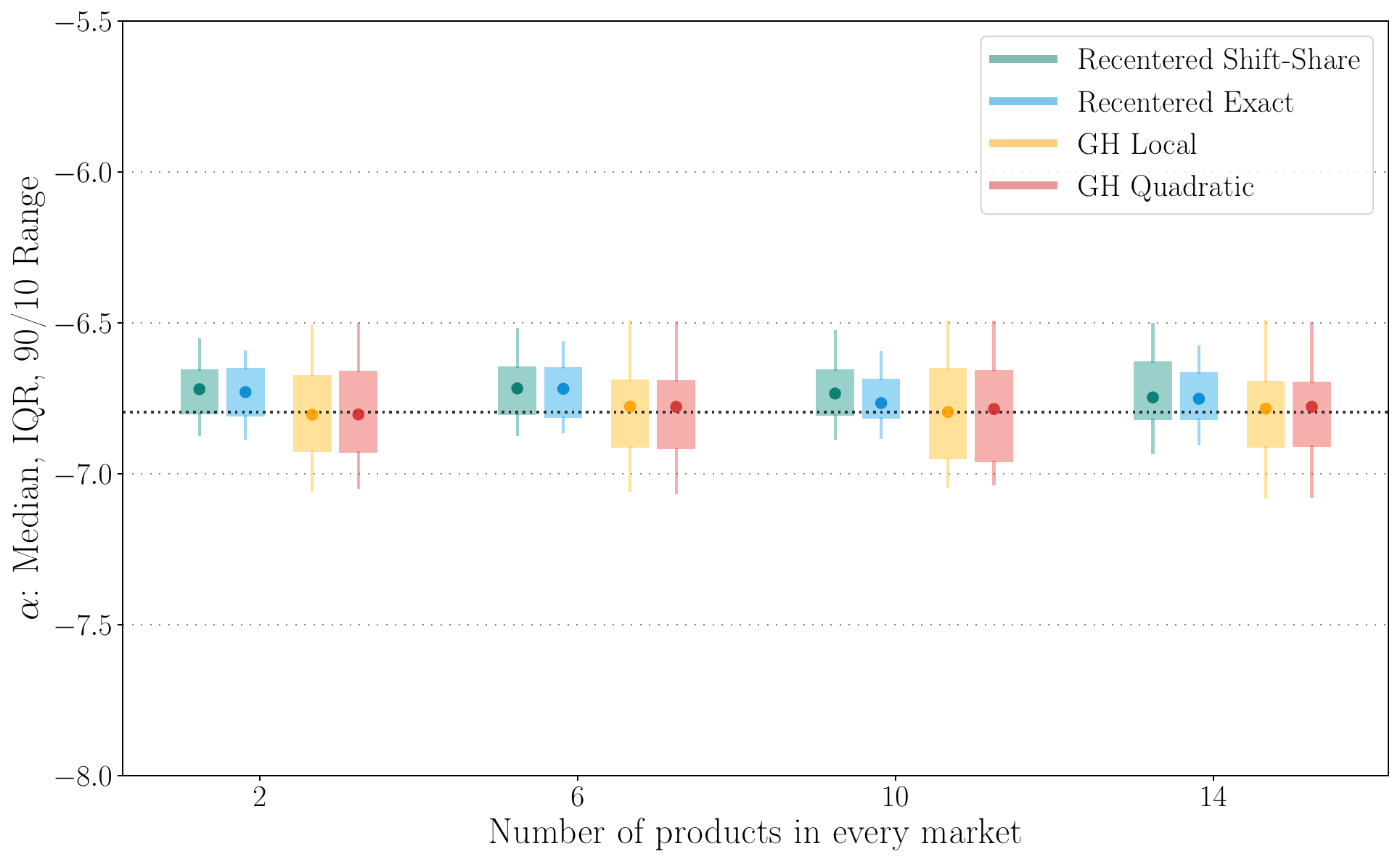}}
  \hfill{}
  \subfloat[Non-Linear Parameter $\sigma_{1}$]{
    \includegraphics[width=0.485\columnwidth]{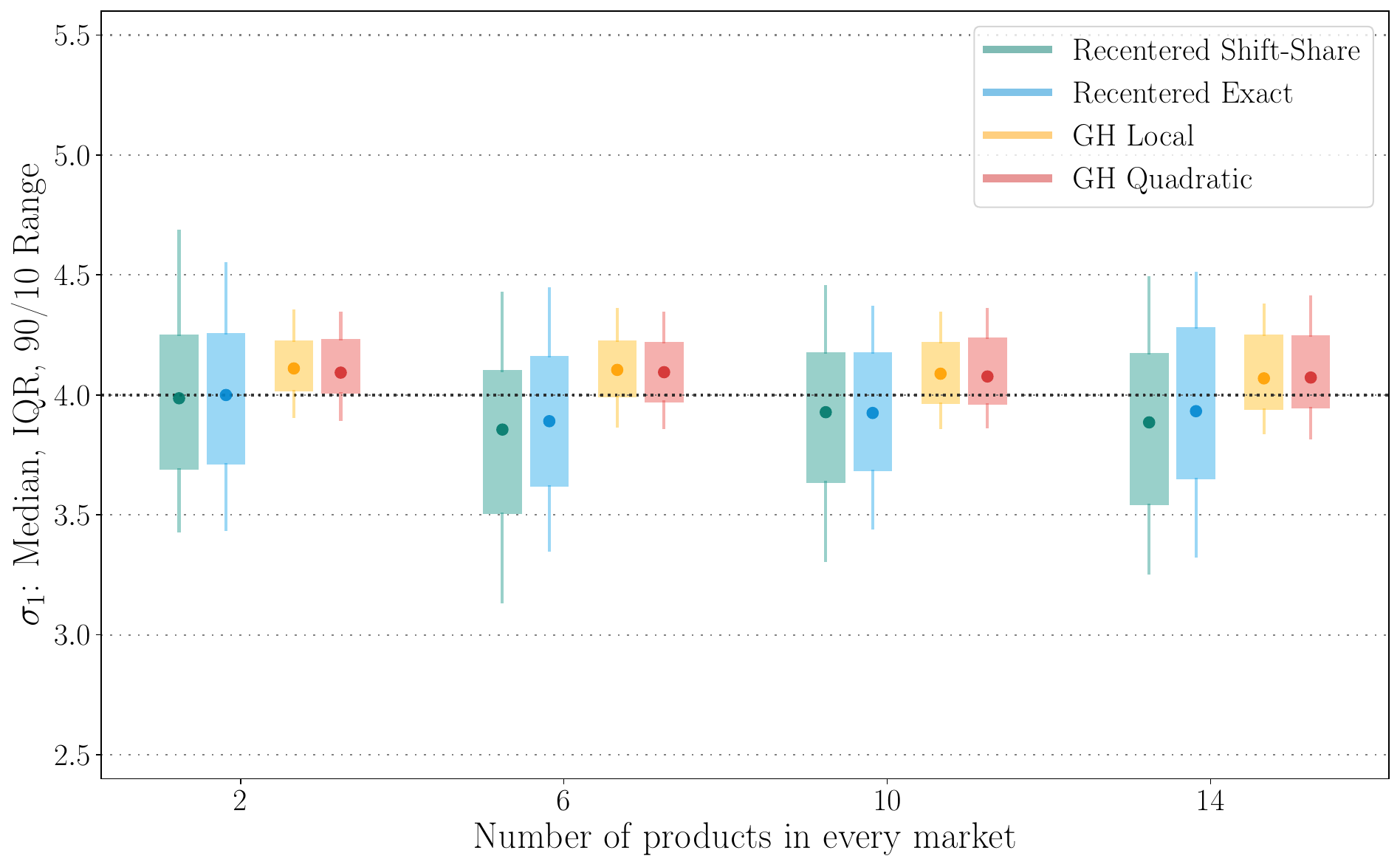}} \\
  \begin{minipage}{0.95\linewidth}
    \footnotesize {\itshape Notes.}
    The two panels show the distributions of the indicated parameter estimates as
    we vary the number of common products across markets. In each simulation we set
    $x_{jr\ell} = x_{j1\ell}$ for indicated number of common products $j = 1,
    \ldots, C$. The data-generarating process is otherwise unchanged; see  notes to
    Figure \ref{fig:monte-carlo-baseline}.
  \end{minipage}
\end{figure}

We next study how the precision of the recentered IV and differentiation IV
estimates varies with two key features of the data-generating process.  Figure
\ref{fig:monte-carlo-varyshocks} shows that recentered IVs have less power
to estimate the nonlinear parameters $\sigma$ with a lower variance of cost
shocks (both IV approaches benefit from more variable shocks for estimating
the price coefficient). In turn, Figure~\ref{fig:monte-carlo-varyxcommon}
shows that differentiation IVs have lower power for $\sigma$ with less
variation in choice sets across markets. While in our baseline simulation
each market has an independent draw of product characteristics, here we make
a subset of products the same across all markets, as when sold nationally. In
the extreme case where all products are common across markets, differentiation
IVs only have variation because product fixed effects are not included in our
estimation procedure.

\subsection{Endogenous Product Characteristics}
\label{subsec:endog_sim}

We now show that our proposed IVs continue to accurately estimate
the price elasticity parameters even when product characteristics
are endogenous, while the differentiation IVs do not. We consider
a simple model of characteristic endogeneity that assumes each region
has a time-invariant “bliss point” $B_{r}$ for the first product
characteristic.\footnote{This simulation is in spirit to \citet{Gandhi2015}, Section 4.4.}
Consumers dislike products far from the bliss point, which we model
by subtracting $3(x_{jr1}-B_{r})^{2}$ from $\xi_{jrt}$. Realizing
this, firms introduce more products near the bliss point, which we
model by centering the distribution of $x_{jr1}$ around $B_{r}$:
$x_{jm1}\stackrel{iid}{\sim}N(B_{m},1)$. Here the differentiation
IVs of \citet{Gandhi2015} are invalid because popular products are
in the dense part of the distribution of product characteristics.
By contrast, our proposed IVs only require exogeneity of the cost
shocks.

\begin{figure}[!tp]
  \centering
  \caption{Endogenous Characteristics}
  \label{fig:monte-carlo-bliss}
  \subfloat[Price Coefficient]{
    \includegraphics[width=0.485\columnwidth]{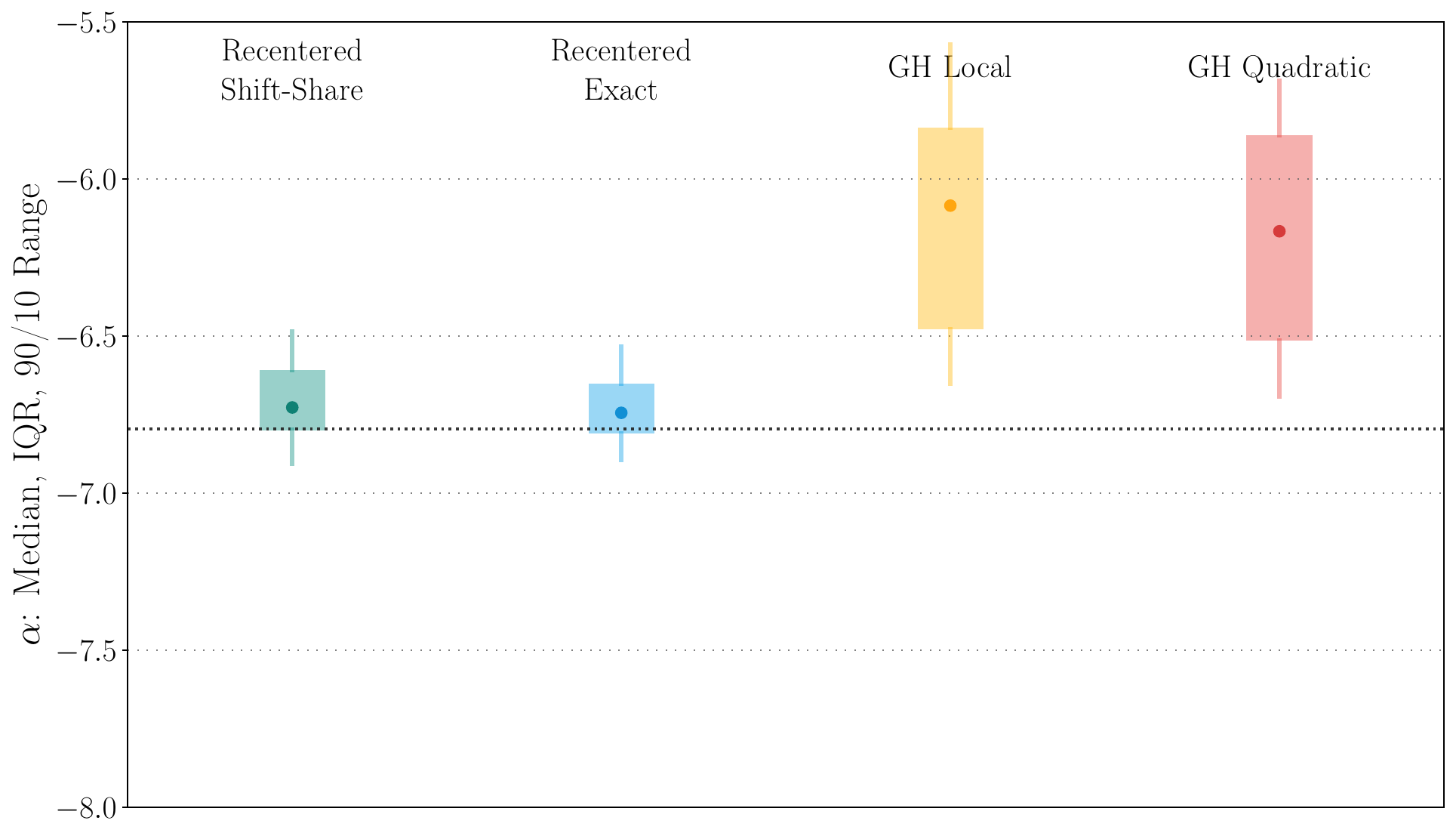}}
  \hfill{}
  \subfloat[Non-linear Parameters $\sigma_{1},\sigma_{2}$]{
    \includegraphics[width=0.485\columnwidth]{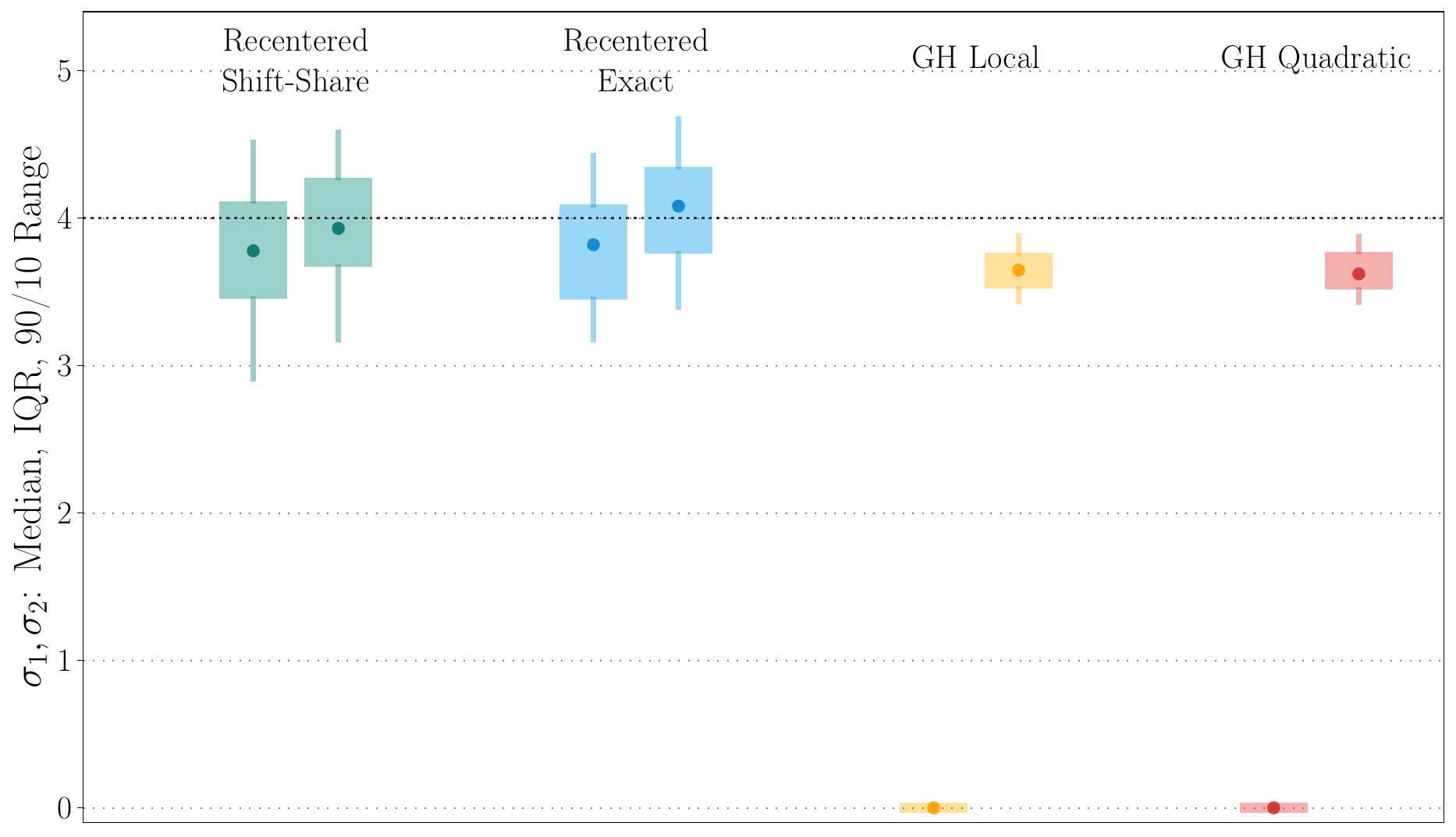}} \\
  \begin{minipage}{0.95\linewidth}
    \footnotesize {\itshape Notes.}
    The two panels show the distributions of the indicated parameter estimates
    as we introduce a bliss point $B_r$ for the first characteristic in each
    region. The data-generarating process is otherwise unchanged; see notes to
    Figure \ref{fig:monte-carlo-baseline}.
  \end{minipage}
\end{figure}

Figure \ref{fig:monte-carlo-bliss} summarizes the results: there
is little bias when using our proposed IV and a moderate decrease
in power relative to Figure \ref{fig:monte-carlo-baseline}. However,
using either of the two differentiation IV strategies yields substantially
biased estimates for $\sigma_{1}$: the estimates are equal to zero
in all simulation draws. The reason is that, under exogenous entry,
mixed logit predicts a negative correlation between market shares
and the degree of local competition as proxied by the differentiated
IVs. This negative correlation is especially strong when the variance
of random coefficients is high. However, with endogenous entry, as
in our simulations, market shares tend to correlate positively with
being in the dense part of the characteristic space (because that
means being close to the bliss point). This generates a strong negative
bias in $\sigma_{1}$ when using the differentiation IVs.\footnote{
  We thank Jean-Francois Houde for pointing this out to us.
}
Notably, this bias seems to also affect differentiation IV estimation
of the price coefficient $\alpha$ and the other nonlinear parameter
$\sigma_{2}$; recentered IV estimates remain unbiased for these parameters
as well.

\section{Conclusion}
\label{sec:Conclusion}

Modern demand models give a flexible yet tractable structure for substitution
across goods. We develop new tools for bringing this structure to
data by leveraging its predictions of how key endogenous variables
respond to a set of exogenous supply-side shocks. Our recentered IV
approach avoids the widespread but often implausible assumption of
exogenous product characteristics, letting us “reuse” the exogenous
shocks to construct multiple powerful instruments targeted at each
of the nonlinear parameters of the model. Simulations suggest recentered
IVs can have comparable power to leading characteristic-based IVs
while avoiding severe bias from characteristic endogeneity. Future
drafts will illustrate this approach in a real setting.

Several open paths remain in this agenda. First, we have only considered
here demand estimation with market-level data; the role of recentered
IV with individual choice-level data is an interesting question for
future research. Second, while we have characterized non-parametric
identification of demand with recentered IVs, flexible estimation
(using, e.g., modern machine learning tools) is worth further study.
Finally, we expect recentered IVs to be useful for identifying structural
models beyond demand—such as for dynamic choice or strategic entry
in IO, or other phenomena in macroeconomics, international trade,
and spatial economics. Developing these extensions may yield practical
new ways to improve the credibility and transparency of structural
estimation.

\begin{singlespace}
\noindent\bibliographystyle{econ-econometrica}
\bibliography{sbartik}

@article{kilian2003so,
	author = {Kilian, Lutz and Taylor, Mark P},
	journal = {Journal of International Economics},
	number = {1},
	pages = {85--107},
	publisher = {Elsevier},
	title = {Why is it so difficult to beat the random walk forecast of exchange rates?},
	volume = {60},
	year = {2003}}

@article{town2003welfare,
	author = {Town, Robert and Liu, Su},
	journal = {RAND Journal of Economics},
	pages = {719--736},
	publisher = {JSTOR},
	title = {The welfare impact of Medicare HMOs},
	year = {2003}}

@article{borusyak2022understanding,
	author = {Borusyak, Kirill and Dix-Carneiro, Rafael and Kovak, Brian},
	journal = {Working Paper},
	title = {Understanding migration responses to local shocks},
	year = {2022}}

@article{rothenberg1971identification,
	author = {Rothenberg, Thomas J},
	journal = {Econometrica: Journal of the Econometric Society},
	pages = {577--591},
	publisher = {JSTOR},
	title = {Identification in parametric models},
	year = {1971}}

@incollection{mcfadden_modelling_1978,
	author = {McFadden, Daniel},
	booktitle = {Spatial {Interaction} {Theory} and {Panning} {Models}},
	pages = {75--96},
	publisher = {North Holland: Amsterdam},
	title = {Modelling the {Choice} of {Residential} {Location}},
	year = {1978}}

@article{Berry1995,
	author = {Berry, Steven and Levinsohn, James and Pakes, Ariel},
	journal = {Econometrica},
	number = {4},
	pages = {841--890},
	title = {Automobile {Prices} in {Market} {Equilibrium}},
	volume = {63},
	year = {1995}}

@article{Adao2017,
	author = {Ad{\~a}o, Rodrigo and Costinot, Arnaud and Donaldson, Dave},
	issn = {00028282},
	journal = {American Economic Review},
	number = {3},
	pages = {633--689},
	title = {Nonparametric counterfactual predictions in neoclassical models of international trade},
	volume = {107},
	year = {2017}}

@article{Adao2018a,
	author = {Ad{\~a}o, Rodrigo and Arkolakis, Costas and Esposito, Federico},
	journal = {Working Paper},
	title = {General {Equilibrium} {Effects} in {Space}: {Theory} and {Measurement}},
	year = {2023}}

@article{BJH2018,
	author = {Borusyak, Kirill and Hull, Peter and Jaravel, Xavier},
	journal = {Review of Economic Studies},
	number = {1},
	pages = {181--213},
	title = {Quasi-{Experimental} {Shift}-{Share} {Research} {Designs}},
	volume = {89},
	year = {2022}}

@article{Gandhi2015,
	author = {Gandhi, A and Houde, J F},
	journal = {Working Paper},
	title = {Measuring {Substitution} {Patterns} in {Differentiated} {Products} {Industries}: {The} {Missing} {Instruments}},
	year = {2020}}

@article{Berry1999,
	author = {Berry, Steven and Levinsohn, James and Pakes, Ariel},
	journal = {American Economic Review},
	number = {3},
	pages = {400--430},
	title = {Voluntary {Export} {Restraints} on {Automobiles}: {Evaluating} a {Trade} {Policy}},
	volume = {89},
	year = {1999}}

@article{Erry2014,
	author = {Berry, Steven and Haile, Philip},
	issn = {0012-9682},
	journal = {Econometrica},
	number = {5},
	pages = {1749--1797},
	title = {Identification in {Differentiated} {Products} {Markets} {Using} {Market} {Level} {Data}},
	volume = {82},
	year = {2014}}

@article{BH1,
	author = {Borusyak, Kirill and Hull, Peter},
	journal = {Econometrica},
	number = {6},
	pages = {2155--2185},
	title = {Non-{Random} {Exposure} to {Exogenous} {Shocks}},
	volume = {91},
	year = {2023}}

@incollection{Berry2021,
	author = {Berry, Steven T and Haile, Philip A},
	booktitle = {Handbook of industrial organization},
	number = {1},
	pages = {1--62},
	publisher = {Elsevier},
	title = {Foundations of demand estimation},
	volume = {4},
	year = {2021}}

@article{Nevo2001,
	author = {Nevo, Aviv},
	issn = {00129682},
	journal = {Econometrica},
	number = {2},
	pages = {307--342},
	title = {Measuring market power in the ready-to-eat cereal industry},
	volume = {69},
	year = {2001}}

@article{Villas-Boas2007,
	author = {Villas-Boas, Sofia Berto},
	issn = {1467937X},
	journal = {Review of Economic Studies},
	number = {2},
	pages = {625--652},
	title = {Vertical relationships between manufacturers and retailers: {Inference} with limited data},
	volume = {74},
	year = {2007}}

@incollection{Gandhi2021,
	author = {Gandhi, Amit and Nevo, Aviv},
	booktitle = {Handbook of industrial organization},
	number = {1},
	pages = {63--139},
	publisher = {Elsevier},
	title = {Empirical models of demand and supply in differentiated products industries},
	volume = {4},
	year = {2021}}

@article{Andrews2022,
	author = {Andrews, Isaiah and Barahona, Nano and Gentzkow, Matthew and Rambachan, Ashesh and Shapiro, Jesse M},
	journal = {Forthcoming Quarterly Journal of Economics},
	pages = {1--41},
	title = {Causal {Interpretation} of {Structural} {IV} {Estimands}},
	year = {2025}}

@article{fan2013ownership,
	author = {Fan, Ying},
	journal = {American Economic Review},
	number = {5},
	pages = {1598--1628},
	publisher = {American Economic Association},
	title = {Ownership consolidation and product characteristics: A study of the US daily newspaper market},
	volume = {103},
	year = {2013}}

@article{Conlon2020,
	author = {Conlon, Christopher and Gortmaker, Jeff},
	issn = {17562171},
	journal = {RAND Journal of Economics},
	number = {4},
	pages = {1108--1161},
	title = {Best practices for differentiated products demand estimation with {PyBLP}},
	volume = {51},
	year = {2020}}

@article{Reynaert2014,
	author = {Reynaert, Mathias and Verboven, Frank},
	issn = {18726895},
	journal = {Journal of Econometrics},
	number = {1},
	pages = {83--98},
	title = {Improving the performance of random coefficients demand models: {The} role of optimal instruments},
	volume = {179},
	year = {2014}}

@article{Armstrong2016,
	author = {Armstrong, Timothy B.},
	issn = {0012-9682},
	journal = {Econometrica},
	number = {5},
	pages = {1961--1980},
	title = {Large {Market} {Asymptotics} for {Differentiated} {Product} {Demand} {Estimators} {With} {Economic} {Models} of {Supply}},
	volume = {84},
	year = {2016}}

@article{Dearing2022,
	author = {Adam Dearing},
	journal = {Journal of Public Economics},
	pages = {104561},
	title = {Estimating structural demand and supply models using tax rates as instruments},
	volume = {205},
	year = {2022}}

@article{Market2016,
	author = {Li, Jing},
	journal = {Working Paper},
	title = {Compatibility and {Investment} in the {U}.{S}. {Electric} {Vehicle} {Market}},
	year = {2016}}

@article{Goldberg2001,
	author = {Goldberg, Pinelopi Koujianou and Verboven, Frank},
	issn = {00346527},
	journal = {Review of Economic Studies},
	number = {4},
	pages = {811--848},
	title = {The evolution of price dispersion in the european car market},
	volume = {68},
	year = {2001}}

@article{Grieco2021,
	author = {Grieco, Paul L.E. and Murry, Charles and Yurukoglu, Ali},
	journal = {Quarterly Journal of Economics},
	number = {2},
	pages = {1201--1253},
	title = {The {Evolution} of {Market} {Power} in the {US} {Auto} {Industry}},
	volume = {139},
	year = {2024}}

@techreport{petrin_identification_2022,
	author = {Petrin, Amil and Ponder, Mark and Seo, Boyoung},
	institution = {Working Paper},
	title = {Identification and estimation of discrete choice demand models when observed and unobserved characteristics are correlated},
	year = {2022}}

@article{miller_understanding_2017,
	author = {Miller, Nathan H. and Weinberg, Matthew C.},
	issn = {0012-9682},
	journal = {Econometrica},
	note = {Publisher: The Econometric Society},
	number = {6},
	pages = {1763--1791},
	title = {Understanding the {Price} {Effects} of the {MillerCoors} {Joint} {Venture}},
	volume = {85},
	year = {2017}}

@article{newey_instrumental_2003,
	author = {Newey, Whitney K. and Powell, James L.},
	issn = {0012-9682, 1468-0262},
	journal = {Econometrica},
	language = {en},
	month = sep,
	number = {5},
	pages = {1565--1578},
	title = {Instrumental {Variable} {Estimation} of {Nonparametric} {Models}},
	urldate = {2023-01-23},
	volume = {71},
	year = {2003}}

@article{berry_estimating_1994,
	author = {Berry, Steven T},
	journal = {The RAND Journal of Economics},
	pages = {242--262},
	publisher = {JSTOR},
	title = {Estimating discrete-choice models of product differentiation},
	year = {1994}}

@article{salanie_fast_2022,
	author = {Salanie, Bernard and Wolak, Frank A},
	journal = {Working Paper},
	language = {en},
	title = {Fast, {Detail}-free, and {Approximately} {Correct}: {Estimating} {Mixed} {Demand} {Systems}},
	year = {2022}}

@article{borusyak_design-based_2023,
	author = {Borusyak, Kirill and Hull, Peter and Jaravel, Xavier},
	doi = {10.1093/ectj/utae003},
	issn = {1368-4221},
	journal = {The Econometrics Journal},
	month = {01},
	number = {1},
	pages = {83-108},
	title = {Design-Based Identification with Formula Instruments: a Review},
	url = {https://doi.org/10.1093/ectj/utae003},
	volume = {28},
	year = {2024}}

@article{lu_semi-nonparametric_2023,
	author = {Lu, Zhentong and Shi, Xiaoxia and Tao, Jing},
	issn = {03044076},
	journal = {Journal of Econometrics},
	language = {en},
	month = aug,
	number = {2},
	pages = {2245--2265},
	title = {Semi-nonparametric estimation of random coefficients logit model for aggregate demand},
	urldate = {2024-03-07},
	volume = {235},
	year = {2023}}

@article{wang_sieve_2023,
	author = {Wang, Ao},
	issn = {03044076},
	journal = {Journal of Econometrics},
	language = {en},
	month = aug,
	number = {2},
	pages = {325--351},
	shorttitle = {Sieve {BLP}},
	title = {Sieve {BLP}: {A} semi-nonparametric model of demand for differentiated products},
	urldate = {2024-03-07},
	volume = {235},
	year = {2023}}

@article{ackerberg_estimating_2009,
	author = {Ackerberg, Daniel A and Crawford, Gregory S},
	journal = {Unpublished},
	language = {en},
	title = {Estimating {Price} {Elasticities} in {Differentiated} {Product} {Demand} {Models} with {Endogenous} {Characteristics}},
	year = {2009}}

@article{houde_spatial_2012,
	author = {Houde, Jean-Fran{\c c}ois},
	issn = {0002-8282},
	journal = {American Economic Review},
	language = {en},
	month = aug,
	number = {5},
	pages = {2147--2182},
	title = {Spatial {Differentiation} and {Vertical} {Mergers} in {Retail} {Markets} for {Gasoline}},
	urldate = {2024-06-21},
	volume = {102},
	year = {2012}}

@article{nakamura_accounting_2010,
	author = {Nakamura, Emi and Zerom, Dawit},
	journal = {The review of economic studies},
	number = {3},
	pages = {1192--1230},
	publisher = {Wiley-Blackwell},
	title = {Accounting for incomplete pass-through},
	volume = {77},
	year = {2010}}

@article{backus_common_2021,
	author = {Backus, Matthew and Conlon, Christopher and Sinkinson, Michael},
	journal = {Working Paper},
	language = {en},
	title = {Common {Ownership} and {Competition} in the {Ready}-{To}-{Eat} {Cereal} {Industry}},
	year = {2021}}

@article{fujiy_production_2024,
	author = {Fujiy, Brian C. and Ghose, Devaki and Khanna, Gaurav},
	journal = {Working Paper},
	series = {Policy {Research} {Working} {Papers}},
	title = {Production {Networks} and {Firm}-{Level} {Elasticities} of {Substitution}},
	year = {2024}}

@article{borusyak_efficient_2021,
	author = {Borusyak, Kirill and Hull, Peter},
	journal = {Working Paper},
	title = {Optimal Formula Instruments},
	year = {2025}}

@article{adao2019shift,
	author = {Ad{\~a}o, Rodrigo and Koles{\'a}r, Michal and Morales, Eduardo},
	journal = {The Quarterly Journal of Economics},
	number = {4},
	pages = {1949--2010},
	publisher = {Oxford University Press},
	title = {Shift-share designs: Theory and inference},
	volume = {134},
	year = {2019}}

@article{chamberlain1987asymptotic,
	author = {Chamberlain, Gary},
	journal = {Journal of Econometrics},
	number = {3},
	pages = {305--334},
	publisher = {Elsevier},
	title = {Asymptotic efficiency in estimation with conditional moment restrictions},
	volume = {34},
	year = {1987}}

@article{chamberlain1992efficiency,
	author = {Chamberlain, Gary},
	journal = {Econometrica: Journal of the Econometric Society},
	pages = {567--596},
	publisher = {JSTOR},
	title = {Efficiency bounds for semiparametric regression},
	year = {1992}}

@article{newey1994large,
	author = {Newey, Whitney K and McFadden, Daniel},
	journal = {Handbook of econometrics},
	pages = {2111--2245},
	publisher = {Elsevier},
	title = {Large sample estimation and hypothesis testing},
	volume = {4},
	year = {1994}}

@article{sweeting2013dynamic,
	author = {Sweeting, Andrew},
	journal = {Econometrica},
	number = {5},
	pages = {1763--1803},
	publisher = {Wiley Online Library},
	title = {Dynamic product positioning in differentiated product markets: The effect of fees for musical performance rights on the commercial radio industry},
	volume = {81},
	year = {2013}}

@article{moon2018estimation,
	author = {Moon, Hyungsik Roger and Shum, Matthew and Weidner, Martin},
	journal = {Journal of Econometrics},
	number = {2},
	pages = {613--644},
	publisher = {Elsevier},
	title = {Estimation of random coefficients logit demand models with interactive fixed effects},
	volume = {206},
	year = {2018}}

@article{crawford2019quality,
	author = {Crawford, Gregory S and Shcherbakov, Oleksandr and Shum, Matthew},
	journal = {American Economic Review},
	number = {3},
	pages = {956--995},
	publisher = {American Economic Association 2014 Broadway, Suite 305, Nashville, TN 37203},
	title = {Quality overprovision in cable television markets},
	volume = {109},
	year = {2019}}

@article{costinot2016evolving,
	author = {Costinot, Arnaud and Donaldson, Dave and Smith, Cory},
	journal = {Journal of Political Economy},
	number = {1},
	pages = {205--248},
	publisher = {University of Chicago Press Chicago, IL},
	title = {Evolving comparative advantage and the impact of climate change in agricultural markets: Evidence from 1.7 million fields around the world},
	volume = {124},
	year = {2016}}

@article{couture2020urban,
	author = {Couture, Victor and Handbury, Jessie},
	journal = {Journal of Urban Economics},
	pages = {103267},
	publisher = {Elsevier},
	title = {Urban revival in America},
	volume = {119},
	year = {2020}}

@article{fajgelbaum2020return,
	author = {Fajgelbaum, Pablo D and Goldberg, Pinelopi K and Kennedy, Patrick J and Khandelwal, Amit K},
	journal = {The Quarterly Journal of Economics},
	number = {1},
	pages = {1--55},
	publisher = {Oxford University Press},
	title = {The return to protectionism},
	volume = {135},
	year = {2020}}

@article{adao2022imports,
	author = {Ad{\~a}o, Rodrigo and Carrillo, Paul and Costinot, Arnaud and Donaldson, Dave and Pomeranz, Dina},
	journal = {The Quarterly Journal of Economics},
	number = {3},
	pages = {1553--1614},
	publisher = {Oxford University Press},
	title = {Imports, exports, and earnings inequality: Measures of exposure and estimates of incidence},
	volume = {137},
	year = {2022}}

@article{aguiar2018quality,
	author = {Aguiar, Luis and Waldfogel, Joel},
	journal = {Journal of Political Economy},
	number = {2},
	pages = {492--524},
	publisher = {University of Chicago Press Chicago, IL},
	title = {Quality predictability and the welfare benefits from new products: Evidence from the digitization of recorded music},
	volume = {126},
	year = {2018}}

@article{barahona2023equilibrium,
	author = {Barahona, Nano and Otero, Crist{\'o}bal and Otero, Sebasti{\'a}n},
	journal = {Econometrica},
	number = {3},
	pages = {839--868},
	publisher = {Wiley Online Library},
	title = {Equilibrium effects of food labeling policies},
	volume = {91},
	year = {2023}}

@article{imbens1994identification,
	author = {Guido W. Imbens and Joshua D. Angrist},
	issn = {00129682, 14680262},
	journal = {Econometrica},
	number = {2},
	pages = {467--475},
	publisher = {[Wiley, Econometric Society]},
	title = {Identification and Estimation of Local Average Treatment Effects},
	urldate = {2025-03-14},
	volume = {62},
	year = {1994}}

@techreport{dekle2008global,
	author = {Dekle, Robert and Eaton, Jonathan and Kortum, Samuel},
	institution = {IMF Staff Papers},
	title = {Global rebalancing with gravity: Measuring the burden of adjustment},
	year = {2008}}

@article{adao2024putting,
	author = {Ad{\~a}o, Rodrigo and Costinot, Arnaud and Donaldson, Dave},
	eprint = {https://academic.oup.com/qje/advance-article-pdf/doi/10.1093/qje/qjae041/60925124/qjae041.pdf},
	issn = {0033-5533},
	journal = {The Quarterly Journal of Economics},
	month = {11},
	pages = {qjae041},
	title = {Putting Quantitative Models to the Test: An Application to the U.S.-China Trade War},
	year = {2024}}

@article{bresnahan1997market,
	author = {Bresnahan, Timothy F and Stern, Scott and Trajtenberg, Manuel},
	journal = {The Rand Journal of Economics},
	pages = {S17},
	publisher = {Rand Corporation},
	title = {Market segmentation and the sources of rents from innovation: Personal computers in the late 1980s},
	volume = {28},
	year = {1997}}

@article{kociswhiten1997halton,
	author = {Kocis, Ladislav and Whiten, William J},
	journal = {ACM Transactions on Mathematical Software (TOMS)},
	number = {2},
	pages = {266--294},
	publisher = {ACM New York, NY, USA},
	title = {Computational investigations of low-discrepancy sequences},
	volume = {23},
	year = {1997}}

@article{Halchenko2020grid,
	author = {Halchenko, Volodymyr and Trembovetska, Ruslana and Tychkov, Volodymyr and Storchak, AV},
	journal = {Applied Computer Systems},
	number = {1},
	pages = {70--76},
	publisher = {Applied Computer Systems},
	title = {The construction of effective multidimensional computer designs of experiments based on a quasi-random additive recursive Rd--sequence},
	volume = {25},
	year = {2020}}

@article{Varadhan2008Simple,
	author = {Varadhan, Ravi and Roland, Christophe},
	journal = {Scandinavian Journal of Statistics},
	number = {2},
	pages = {335--353},
	title = {Simple and {Globally} {Convergent} {Methods} for {Accelerating} the {Convergence} of {Any} {EM} {Algorithm}},
	volume = {35},
	year = {2008}}
\newpage{}
\end{singlespace}

\appendix

\setcounter{table}{0}
\setcounter{figure}{0}
\setcounter{equation}{0}

\renewcommand{\thetable}{A\arabic{table}}
\renewcommand{\thefigure}{A\arabic{figure}}
\renewcommand{\theequation}{A\arabic{equation}}

\counterwithin{prop}{section} \renewcommand{\theprop}{A\arabic{prop}}
\counterwithin{lem}{section}  \renewcommand{\thelem}{A\arabic{lem}}

\begin{center}
{\huge\textbf{Online Appendix}}{\huge\par}
\par\end{center}

\section{Theoretical Appendix}

\subsection{Derivations for Section \ref{sec:Motivating-Example:}}
\label{sec:nested_logit_derivations}

\paragraph{Market Shares with Nested Logit Demand. }

Let $\delta_{jm}=\alpha p_{jm}+\xi_{jm}$ be product $j$'s mean utility
and $D_{nm}=\sum_{j\in\mathcal{J}_{m}}d_{jn}\exp\left(\delta_{jm}/(1-\sigma)\right)$.
Then, as is well known (e.g, \citet{berry_estimating_1994}), nested
logit market shares satisfy
\begin{align}
  \frac{s_{jm}}{s_{n(j)m}}
         & = \frac{\exp\left(\delta_{jm}/(1-\sigma)\right)}{D_{n(j)m}},\label{eq:nested_within_share} \\
  s_{nm} & = \frac{D_{nm}^{1-\sigma}}{1+\sum_{n'}D_{n^{\prime}m}^{1-\sigma}},                         \\
  s_{0m} & = \frac{1}{1+\sum_{n'}D_{n^{\prime}m}^{1-\sigma}}.
\end{align}
Equation (\ref{eq:nested_logit}) follows from simple manipulation of these terms.

\paragraph{Exact Prediction of Within-Nest Share. }

Let $\hat{p}_{jm}=\check{\pi}_{0}+\check{\pi}g_{jm}$ where $\check{\pi}_{0}$
is an intercept suppressed in Section \ref{sec:Motivating-Example:}.
Plugging $\delta_{jm}=\check{\alpha}\hat{p}_{jm}$ into the within-nest
share expression (\ref{eq:nested_within_share}) yields (\ref{eq:z_exact}):
\begin{align*}
  \widehat{\log}\frac{s_{jm}}{s_{n(j)m}}
  & = \frac{\check{\alpha}}{1-\check{\sigma}}
    \left(\check{\pi}_{0}+\check{\pi}g_{jm}\right)
    -
    \log\sum_{k\in\mathcal{J}_{m}}
      d_{kn(j)}
      \exp\left(\frac{\check{\alpha}}{1-\check{\sigma}}\left(\check{\pi}_{0}+\check{\pi}g_{km}\right)\right)
  \\
  & = \frac{\check{\alpha}}{1-\check{\sigma}}
      \check{\pi} g_{jm}
      -
      \log\sum_{k\in\mathcal{J}_{m}}
      d_{kn(j)}
      \exp\left(\frac{\check{\alpha}}{1-\check{\sigma}}\check{\pi}g_{km}\right).
\end{align*}

\paragraph{First-Order Approximation.}

We now take a first-order approximation of (\ref{eq:z_exact}) around
$g_{km}=\mu_{g}$ for all $k$:
\begin{align*}
  \widehat{\log}\frac{s_{jm}}{s_{n(j)m}}
  & \approx\frac{\check{\alpha}}{1-\check{\sigma}}\check{\pi}\mu_{g}-\log\sum_{k\in\mathcal{J}_{m}}d_{kn}\exp\left(\frac{\check{\alpha}}{1-\check{\sigma}}\check{\pi}\mu_{g}\right)\\
  & \quad+\frac{\check{\alpha}}{1-\check{\sigma}}\check{\pi}\left(g_{jm}-\mu_{g}\right)-\frac{\sum_{k\in\mathcal{J}_{m}}d_{kn}\exp\left(\frac{\check{\alpha}}{1-\check{\sigma}}\check{\pi}\mu_{g}\right)\frac{\check{\alpha}}{1-\check{\sigma}}\check{\pi}\left(g_{km}-\mu_{g}\right)}{\sum_{k\in\mathcal{J}_{m}}d_{kn}\exp\left(\frac{\check{\alpha}}{1-\check{\sigma}}\check{\pi}\mu_{g}\right)}\\
  & =-\log N_{n(j)m}+\frac{\check{\alpha}}{1-\check{\sigma}}\check{\pi}\left(g_{jm}-\frac{1}{N_{n(j)m}}\sum_{k\in\mathcal{J}_{m}}d_{kn}g_{km}\right).
\end{align*}
We interpret the second term as the response of $\log(s_{jm}/s_{n(j)m})$
to the set of cost shocks (while the first term does not depend on
the shocks and is eliminated by recentering). This response is therefore
equal to $z_{jm}$ in (\ref{eq:z_simple}), up to a non-zero scaling
factor that does not affect IV estimation.

\paragraph{Exact Prediction Using Lagged Shares.}

Suppose the choice set $\mathcal{J}_{m}$ has not changed since the
pre-period and $g_{jm}$ are shocks to price changes, such that
$p_{jm}=p_{jm}^{\text{pre}}+\pi g_{jm}+\omega_{jm}$.
Let $\xi_{jm}^{\text{pre}}$ be the unobserved taste shifter in the
pre-period\footnote{
  Given parameters, $\xi_{jm}^{\text{pre}}$ can be inverted from the
  pre-period prices and shares; however, this is not necessary.
} such that $
  \delta_{jm}^{\text{pre}}
  =
  \alpha p_{jm}^{\text{pre}}+\xi_{jm}^{\text{pre}}
$
and, according to (\ref{eq:nested_within_share}),
\[
  \frac{s_{jm}^{\text{pre}}}{s_{n(j)m}^{\text{pre}}}
  =
  \frac{
    \exp\left(\delta_{jm}^{\text{pre}}/(1-\sigma)\right)
  }{
    \sum_{k\in\mathcal{J}_{m}}d_{kn(j)}\exp\left(\delta_{jm}^{\text{pre}}/(1-\sigma)\right)
  }.
\]

We predict prices in the period of interest as $
  \hat{p}_{jm}
  =
  p_{jm}^{\text{pre}}+\check{\pi}g_{jm}
$
and predict mean utilities by using the predicted price and the pre-period
taste shifter:
\[
  \hat{\delta}_{jm}
  =
  \check{\alpha}\hat{p}_{jm}+\xi_{jm}^{\text{pre}}
  =
  \delta_{jm}^{\text{pre}}
  +
  \check{\alpha}\check{\pi}g_{jm}.
\]
Substituting these into (\ref{eq:nested_within_share}) analogously
to the exact hat algebra technique of \citet{dekle2008global} yields
the exact prediction:
\begin{align}
  \widehat{\log}\frac{s_{jm}}{s_{n(j)m}}
    & = \log\frac{\exp\left(\hat{\delta}_{jm}/(1-\check{\sigma})\right)}{\sum_{k\in\mathcal{J}_{m}}d_{kn(j)}\exp\left(\hat{\delta}_{km}/(1-\check{\sigma})\right)}\nonumber \\
    & = \log\frac{\exp\left(\delta_{jm}^{\text{pre}}/(1-\sigma)\right)\cdot\exp\left(\check{\alpha}\check{\pi}g_{jm}/(1-\check{\sigma})\right)}{\sum_{k\in\mathcal{J}_{m}}d_{kn(j)}\exp\left(\delta_{km}^{\text{pre}}/(1-\check{\sigma})\right)\cdot\exp\left(\check{\alpha}\check{\pi}g_{km}/(1-\check{\sigma})\right)}\nonumber \\
    & = \log\frac{\exp\left(\delta_{jm}^{\text{pre}}/(1-\check{\sigma})\right)}{\sum_{k\in\mathcal{J}_{m}}d_{kn(j)}\exp\left(\delta_{km}^{\text{pre}}/(1-\check{\sigma})\right)}+\log\frac{\exp\left(\check{\alpha}\check{\pi}g_{jm}/(1-\check{\sigma})\right)}{\sum_{k\in\mathcal{J}_{m}}d_{kn(j)}\frac{s_{km}^{\text{pre}}}{s_{n(j)m}^{\text{pre}}}\exp\left(\check{\alpha}\check{\pi}g_{km}/(1-\check{\sigma})\right)}\nonumber \\
    & = \log\frac{s_{jm}^{\text{pre}}}{s_{n(j)m}^{\text{pre}}}+\frac{\check{\alpha}\check{\pi}}{1-\check{\sigma}}g_{jm}-\log\sum_{k\in\mathcal{J}_{m}}d_{kn(j)}\frac{s_{km}^{\text{pre}}}{s_{n(j)m}^{\text{pre}}}\exp\left(\frac{\check{\alpha}\check{\pi}}{1-\check{\sigma}}g_{km}\right).\label{eq:logit-exact-diffs}
\end{align}
As with exact hat algebra, lagged shares serve as sufficient statistics
in this prediction, while lagged prices and taste shifters need not
be observed or computed.

Recentering this prediction over the distribution of shocks yields
a recentered exact prediction
\begin{multline*}
  \frac{\check{\alpha}\check{\pi}}{1-\check{\sigma}}
  \left(g_{jm}-\mu_{g}\right)
  -
  \log\sum_{k\in\mathcal{J}_{m}}
    d_{kn(j)}
    \frac{s_{km}^{\text{pre}}}{s_{n(j)m}^{\text{pre}}}
    \exp\left(\frac{\check{\alpha}\check{\pi}}{1-\check{\sigma}}g_{km}\right)
  \\
  +
  \expec{\log\sum_{k\in\mathcal{J}_{m}}d_{kn(j)}\frac{s_{km}^{\text{pre}}}{s_{n(j)m}^{\text{pre}}}\exp\left(\frac{\check{\alpha}\check{\pi}}{1-\check{\sigma}}g_{km}\right)\mid\left(d_{kn},s_{km}^{\text{pre}}\right)_{k\in\mathcal{J}_{m},n}}.
\end{multline*}

\paragraph{First-Order Approximation Using Lagged Shares.}

Linearizing (\ref{eq:logit-exact-diffs}) around $g_{km}=\mu_{g}$
for all $k$ yields
\begin{align*}
  \widehat{\log}\frac{s_{jm}}{s_{n(j)m}}
  & \approx\log\frac{s_{jm}^{\text{pre}}}{s_{n(j)m}^{\text{pre}}}+\frac{\check{\alpha}\check{\pi}}{1-\check{\sigma}}g_{jm}-\frac{\check{\alpha}\check{\pi}}{1-\check{\sigma}}\mu_{g}-\sum_{k\in\mathcal{J}_{m}}d_{kn(j)}\frac{s_{km}^{\text{pre}}}{s_{n(j)m}^{\text{pre}}}\frac{\check{\alpha}\check{\pi}}{1-\check{\sigma}}\left(g_{km}-\mu_{g}\right)\\
  & = \log\frac{s_{km}^{\text{pre}}}{s_{n(j)m}^{\text{pre}}}+\frac{\check{\alpha}\check{\pi}}{1-\check{\sigma}}\left(g_{jm}-\sum_{k\in\mathcal{J}_{m}}d_{kn(j)}\frac{s_{km}^{\text{pre}}}{s_{n(j)m}^{\text{pre}}}g_{km}\right).
\end{align*}
Recentering this expression eliminates the first term, while the second
term is equal to $z_{jm}^{\ast}$ from (\ref{eq:simple_z-1}) up to
a non-zero scaling factor.

\subsection{Local-to-Logit Approximation}
\label{subsec:appx-Local-to-Logit}

In this section we consider a model in which the random coefficients
on $x_{jm}^{(1)}$, and potentially also a random coefficient on price,
are equal to $\eta_{i\ell}=\sigma_{\ell}\nu_{i\ell}$ for mutually
uncorrelated $\nu_{i\ell}$ with $\expec{\nu_{i\ell}}=0$ and $\expec{\nu_{i\ell}^{2}}=1$
(but any marginal distributions). We suppress the dependence of the
$\mathcal{S}_{j}$ and $\mathcal{D}_{j}$ on $\boldsymbol{x}_{m}^{(1)}$
and $\boldsymbol{p}_{j}$ to simplify notation. We first state a lemma
characterizing market shares and the share inversion function in the
“local-to-logit” approximation to the mixed logit model, which
provides a new intuition for the mechanics of mixed logit and for
the approximations derived by \citet[Theorem 2]{salanie_fast_2022}.
We then characterize $Z_{jm}^{SSIV}$ from Section \ref{subsec:Constructing-IVs}
when the preliminary parameter values are local-to-logit, regardless
of the true parameters.
\begin{lem}
  \label{lem:SW-inversion}
  In the above model, without a random coefficient
  on price, the following Taylor expansions in $
    \sigma
    =
    \left(\sigma_{\ell}\right)_{\ell=1}^{L_{1}}
  $
  hold around $\sigma=0$:
  \begin{multline}
    \mathcal{S}_{j}(\boldsymbol{\delta}_{m};\sigma)
    =
    \mathcal{S}_{j}(\boldsymbol{\delta}_{m};0)
    \cdot
    \left[
      1+\sum_{\ell=1}^{L_{1}}\frac{\sigma_{\ell}^{2}}{2}\left(
        \left(x_{jm\ell}-\bar{x}_{m\ell}\right)^{2}
        -
        \sum_{k\in\mathcal{J}_{m}\cup\left\{ 0\right\} }
          s_{km}\left(x_{km\ell}-\bar{x}_{m\ell}\right)^{2}
      \right)
    \right] \\
    +
    O(\sigma^{3}),
    \label{eq:SW-S}
  \end{multline}
  \begin{equation}
    \log\frac{\mathcal{S}_{j}(\boldsymbol{\delta}_{m};\sigma)}{\mathcal{S}_{0}(\boldsymbol{\delta}_{m};\sigma)}
    =
    \delta_{jm}+\sum_{\ell=1}^{L_{1}}a_{jm\ell}\frac{\sigma_{\ell}^{2}}{2}+O(\sigma^{3}),\label{eq:SW-logS}
  \end{equation}
  \begin{equation}
    \mathcal{D}_{j}\left(\boldsymbol{s}_{m};\sigma\right)
    =
    \log\frac{s_{jm}}{s_{0m}}
    -
    \sum_{\ell=1}^{L_{1}}a_{jm\ell}\frac{\sigma_{\ell}^{2}}{2}+O(\sigma^{3}),
    \label{eq:SW-D}
  \end{equation}
  where $O(\sigma^{q})$ indicates $q$th- and higher-order terms and
  \begin{equation}
    a_{jm\ell}
    =
    \left(x_{jm\ell}-\sum_{k\in\mathcal{J}_{m}}s_{km}x_{km\ell}\right)^{2}
    -
    \left(0-\sum_{k\in\mathcal{J}_{m}}s_{km}x_{km\ell}\right)^{2},
    \label{eq:a_jm}
  \end{equation}
  with $s_{km}$ understood as $\mathcal{S}_{k}\left(\boldsymbol{\delta}_{m};0\right)$
  in equations (\ref{eq:SW-S}) and (\ref{eq:SW-logS}). With a random
  coefficient on price, the same expressions apply with price viewed
  as another characteristic: i.e., with $x_{jm0}\equiv p_{jm}$ and
  with the summations including $\ell=0$.
\end{lem}
Intuitively, equation (\ref{eq:SW-S}) shows that—in the vicinity
of simple multinomial logit—increasing $\sigma_{\ell}$ raises the
share of good $j$ if and only if the $\ell$th characteristic of
this good is relatively unusual in its market, in the sense that its
further away from the market average than the typical product is (where
by “further” we mean averaging square distances with market
share weights, counting the outside good as one of the products).

We now characterize the shift-share construction from Section
\ref{subsec:Constructing-IVs} in this approximation:
\begin{prop}
  \label{prop:sw_approx}
  In the model of this section, the $\ell$th shift-share instrument from
  (\ref{eq:nabla_firstorder}) corresponding to the random coefficient on a
  non-price characteristic satisfies:
  \begin{equation}
    \sum_{k\in\mathcal{J}_{m}}
      w_{jkm\ell}\tilde{g}_{km}
    =
    \left(-2\check{\alpha}\check{\pi}\check{\sigma}_{\ell}\right)
    \cdot
    x_{jm\ell}
    \cdot
    \sum_{k\in\mathcal{J}_{m}}
      \check{s}_{km}
      \left(x_{km\ell}-\bar{x}_{m\ell}\right)
      \tilde{g}_{km}
    +
    O(\check{\sigma}^{2}).
    \label{eq:ssiv_sw_nonprice}
  \end{equation}
  The instrument for the random coefficient on price (if included in
  the model) is
  \begin{align}
    \sum_{k\in\mathcal{J}_{m}}
      w_{jkm0}\tilde{g}_{km}
      =
      &
      2\check{\sigma}_{0}\check{\pi}
      \left(\check{p}_{jm}-\bar{p}_{m}\right)
      \tilde{g}_{jm}
      -
      2\check{\sigma}_{0}\check{\pi}\check{p}_{jm}
      \sum_{k\in\mathcal{J}_{m}}\check{s}_{km}\tilde{g}_{km}
      \label{eq:ssiv_sw_price}
    \\
    &
      -
      2\check{\alpha}\check{\pi}\check{\sigma}_{0}
      \cdot
      p_{jm}
      \cdot
      \sum_{k\in\mathcal{J}_{m}}
        \check{s}_{km}
        \left(\check{p}_{jm}-\bar{p}_{m}\right)
        \tilde{g}_{km}
      +
      O(\check{\sigma}^{2}),
      \nonumber
  \end{align}
  where $\bar{p}_{m}=\sum_{k\in\mathcal{J}_{m}}\check{s}_{km}\check{p}_{km}$.
\end{prop}
Intuition for these results follow from the above discussion of equation
(\ref{eq:SW-S}). Focus first on non-price characteristics. While
cost shocks cannot affect product entry or characteristics under our
assumptions, they can still make certain products more or less unusual
in the market by reallocating market shares and thus shifting the
share-weighted average of characteristics $\bar{x}_{m\ell}$. Under
our simple model of cost shock pass-through, $\bar{x}_{m\ell}$ increases
whenever products $k$ with higher $x_{km\ell}$ receive a lower shock—as
captured by the covariance term in Proposition \ref{prop:sw_approx}.
When $\bar{x}_{m\ell}$ moves up, high-$x_{jm\ell}$ products become
less unusual and lose market shares when $\sigma_{\ell}$ is higher
while low-$x_{jm\ell}$ products become more unusual and gain market
power in that case. Our instrument identifies $\sigma_{\ell}$ by
tracing such differential responses, provided there is enough variation
in the $\ell$th market-level aggregate shock. When there is a random
coefficient on price, there are additional effects captured by the
first two terms in (\ref{eq:ssiv_sw_price}): cost shocks move the
price of good $j$ and prices of its competitors. For instance, the
first term reflects that if $j$ is priced higher than the market
average, a high cost shock that increases its price makes the product
vertically more unusual and raises the market share if $\sigma_{0}$
is larger.

We finally specialize equation (\ref{eq:ssiv_sw_nonprice}) to the
common case where a random coefficient is included for the intercept
$x_{jm\ell}=1$. Given the normalization $x_{0m}=0$, this captures
heterogeneous preferences for all inside goods vs. the outside good.
In that case,
\[
  \sum_{k\in\mathcal{J}_{m}}
    w_{jkm\ell}\tilde{g}_{km}
  =
  \left(-2\check{\alpha}\check{\pi}\check{\sigma}_{\ell}\right)
  \cdot
  \check{s}_{0m}
  \left(1-\check{s}_{0m}\right)
  \cdot
  \frac{\sum_{k\in\mathcal{J}_{m}}\check{s}_{km}\tilde{g}_{km}}{\sum_{k\in\mathcal{J}_{m}}\check{s}_{km}}
  +
  O(\check{\sigma}^{2}).
\]
That is, the instrument is based on the share-weighted average shock
to all inside products in the market, which move prices of all inside
goods relative to the outside good. The average shock is scaled to
place a higher weight on markets with the share of the outside good
closer to 0.5.

\paragraph{Proof of Lemma \ref{lem:SW-inversion}.}

For this lemma, random coefficients on price can be handled simply
by viewing price as another characteristic. We therefore omit random
coefficients on price without loss of generality.

For consumer $i$ in market $m$, let
\[
  s_{ji}
  =
  \frac{
    \exp\left(\delta_{jm}+\sum_{\ell=1}^{L_{1}}\sigma_{\ell}\nu_{i\ell}x_{jm\ell}\right)
  }{
    1+\sum_{k\in\mathcal{J}_{m}}\exp\left(\delta_{km}+\sum_{\ell=1}^{L_{1}}\sigma_{\ell}\nu_{i\ell}x_{km\ell}\right)
  }
\]
denote the probability of choosing product $j\in\mathcal{J}_{m}\cup\left\{ 0\right\}$
given her $\nu_{i}=(\nu_{i\ell})_{\ell=1}^{L_{1}}$. Then $s_{jm}=\expecnu{s_{ji}}$
where $\expecnu{\cdot}$ denotes the expectation with respect to the
distribution of $\nu_{i}$. Also, denote $
  \bar{x}_{i\ell}
  =
  \sum_{k\in\mathcal{J}_{m}}s_{ki}x_{km\ell}
$
and $\bar{x}_{m\ell}=\sum_{k\in\mathcal{J}_{m}}s_{km}x_{km\ell}$. Then we have:
\begin{align*}
  \frac{\partial\mathcal{S}_{j}}{\partial\sigma_{\ell}}
  &
  =
  \expecnu{\frac{\partial s_{ji}}{\partial\sigma_{\ell}}}
  =
  \expecnu{\nu_{i\ell}s_{ji}\left(x_{jm\ell}-\bar{x}_{i\ell}\right)}
\end{align*}
and
\[
  \frac{\partial^{2}\mathcal{S}_{j}}{\partial\sigma_{\ell}\sigma_{\ell'}}
  =
  \expecnu{\frac{\partial^{2}s_{ji}}{\partial\sigma_{\ell}\partial\sigma_{\ell'}}}
  =
  \expecnu{
    \nu_{i\ell}s_{ji}\left(
      \nu_{i\ell'}\left(
        x_{jm\ell}
        -
        \bar{x}_{i\ell}
      \right)\left(
        x_{jm\ell'}
        -
        \bar{x}_{i\ell'}
      \right)
      -
      \sum_{k\in\mathcal{J}_{m}}
        x_{km\ell} s_{ki} \nu_{i\ell'}
        \left(x_{km\ell'}-\bar{x}_{i\ell'}\right)
    \right)
  }.
\]
At $\sigma=0$, all consumers have the same conditional choice probabilities, $s_{ji}=s_{jm}$, and thus
\begin{align*}
  \frac{\partial\mathcal{S}_{j}}{\partial\sigma_{\ell}}\mid_{\sigma=0} & =\expecnu{\nu_{i\ell}}\cdot s_{jm}\left(x_{jm\ell}-\bar{x}_{m\ell}\right)\\
   & = 0,\\
  \frac{\partial^{2}\mathcal{S}_{j}}{\partial\sigma_{\ell}\partial\sigma_{\ell'}}\mid_{\sigma=0} & \propto\expecnu{\nu_{i\ell}\nu_{i\ell'}}=0\qquad\text{for }\ell'\ne\ell\\
  \frac{\partial^{2}\mathcal{S}_{j}}{\left(\partial\sigma_{\ell}\right)^{2}}\mid_{\sigma=0} & =\expecnu{\nu_{i\ell}^{2}}s_{jm}\left[\left(x_{jm\ell}-\bar{x}_{m\ell}\right)^{2}-\sum_{k\in\mathcal{J}_{m}}s_{km}x_{km\ell}\left(x_{km\ell}-\bar{x}_{m\ell}\right)\right]\\
   & = s_{jm}\left[\left(x_{jm\ell}-\bar{x}_{m\ell}\right)^{2}-\sum_{k\in\mathcal{J}_{m}\cup\left\{ 0\right\} }s_{km}\left(x_{km\ell}-\bar{x}_{m\ell}\right)^{2}\right].
\end{align*}
By a second-order Taylor expansion,
\[
  \mathcal{S}_{j}(\boldsymbol{\delta}_{m};\sigma)
  =
  \mathcal{S}_{j}(\boldsymbol{\delta}_{m};0)
  \cdot
  \left[
    1 + \sum_{\ell}
      \frac{\sigma_{\ell}^{2}}{2}
      \left(
        \left(x_{jm\ell}-\bar{x}_{m\ell}\right)^{2}
        -
        \sum_{k\in\mathcal{J}_{m}\cup\left\{ 0\right\} }
          s_{km}\left(x_{km\ell'}-\bar{x}_{m\ell'}\right)^{2}
      \right)
  \right]
  +
  O(\sigma^{3}),
\]
establishing (\ref{eq:SW-S}). The Taylor approximation in logs then implies
\[
  \log\mathcal{S}_{j}(\boldsymbol{\delta}_{m};\sigma)
  =
  \log\mathcal{S}_{j}(\boldsymbol{\delta}_{m};0)
  +
  \sum_{\ell}\frac{\sigma_{\ell}^{2}}{2}
  \left(
    \left(x_{jm\ell}-\bar{x}_{m\ell}\right)^{2}
    -
    \sum_{k\in\mathcal{J}_{m}\cup\left\{ 0\right\} }
      s_{km}\left(x_{km\ell'}-\bar{x}_{m\ell'}\right)^{2}
  \right)
  +
  O(\sigma^{3}).
\]
Subtracting an analogous expression that holds for the outside good,
\begin{align*}
   \log\frac{\mathcal{S}_{j}(\boldsymbol{\delta}_{m};\sigma)}{\mathcal{S}_{0}(\boldsymbol{\delta}_{m};\sigma)}
   &
   =
   \log\frac{\mathcal{S}_{j}(\boldsymbol{\delta}_{m};0)}{\mathcal{S}_{0}(\boldsymbol{\delta}_{m};0)}
   +
   \sum_{\ell}\frac{\sigma_{\ell}^{2}}{2}\left(\left(x_{jm\ell}-\bar{x}_{m\ell}\right)^{2}-\left(0-\bar{x}_{m\ell}\right)^{2}\right)
   +
   O(\sigma^{3})
   \\
   &
   =
   \delta_{jm}
   +
   \sum_{\ell}\frac{\sigma_{\ell}^{2}}{2}
     \left(\left(x_{jm\ell}-\bar{x}_{m\ell}\right)^{2}-\left(0-\bar{x}_{m\ell}\right)^{2}\right)
   +
   O(\sigma^{3}),
\end{align*}
where the second line uses the standard result on share inversion
with simple multinomial logit, yielding equation (\ref{eq:SW-logS}).

Finally, plugging in $
  \mathcal{S}_{j}(\boldsymbol{\delta}_{m};0)
  =
  \mathcal{S}_{j}(\boldsymbol{\delta}_{m};\sigma)
  +
  O(\sigma)
$
into equation (\ref{eq:SW-logS}) yields (\ref{eq:SW-D}), as $\mathcal{D}_{j}$
is the mapping from the shares to $\delta_{jm}$.

\paragraph{Proof of Proposition \ref{prop:sw_approx}.}

We apply the steps of the shift-share IV construction
in Section \ref{subsec:Constructing-IVs} to this
setting. Because the above derivation shows that
$\frac{\partial\mathcal{S}_{j}}{\partial\sigma_{\ell}}\mid_{\sigma=0}=0$,
\[
  \frac{\partial\mathcal{S}_{j}(\check{\boldsymbol{\delta}}_{m};\sigma)}{\partial\delta_{km}}
  =
  \frac{\partial\mathcal{S}_{j}(\check{\boldsymbol{\delta}}_{m};0)}{\partial\delta_{km}}+O(\sigma^{2})
  =
  s_{jm}\left(\one\left[j=k\right]-s_{km}\right)+O(\sigma^{2})
\]
for $s_{km}=\mathcal{S}_{j}\left(\check{\boldsymbol{\delta}}_{m};0\right)$.
Since $
  \mathcal{S}_{j}\left(\check{\boldsymbol{\delta}}_{m};0\right)
  =
  \mathcal{S}_{j}\left(\check{\boldsymbol{\delta}}_{m};\check{\sigma}\right)
  +
  O(\sigma^{2})
$
by equation (\ref{eq:SW-S}), we also have
\[
  \frac{\partial\mathcal{S}_{j}(\check{\boldsymbol{\delta}}_{m};\sigma)}{\partial\delta_{km}}
  =
  \check{s}_{jm}\left(\one\left[j=k\right]-\check{s}_{km}\right)
  +
  O(\sigma^{2}),
\]
and thus
\begin{align*}
  \hat{s}_{jm}-\check{s}_{jm}
  & = \check{\alpha}\check{\pi}\sum_{k\in\mathcal{J}_{m}}\frac{\partial\mathcal{S}_{j}(\check{\boldsymbol{\delta}}_{m};\check{\sigma})}{\partial\delta_{km}}\tilde{g}_{km}\\
  & = \check{\alpha}\check{\pi}s_{jm}\left(\tilde{g}_{jm}-\sum_{k\in\mathcal{J}_{m}}s_{km}\tilde{g}_{km}\right)+O(\sigma^{2}).
\end{align*}
Next, by equation (\ref{eq:SW-D}) and writing equation (\ref{eq:a_jm}) as
\begin{equation}
  a_{jm\ell}
  =
  x_{jm\ell}^{2}-2x_{jm\ell}\cdot\sum_{k\in\mathcal{J}_{m}}s_{km}x_{km\ell},
  \label{eq:SW-a-another-way}
\end{equation}
we have
\[
  \frac{\partial^{2}\mathcal{D}_{j}\left(\check{\boldsymbol{s}}_{m};\sigma\right)}{\partial s_{km}\partial\sigma_{\ell}}
  =
  \sigma_{\ell}\frac{\partial a_{jm\ell}}{\partial s_{km}}+O(\sigma^{2})
  =
  -2\sigma_{\ell}x_{jm\ell}x_{km\ell}+O(\sigma^{2}).
\]
Also, recalling that $p_{jm}\equiv x_{jm0}$, equation (\ref{eq:SW-S})
implies that $
  \frac{\partial^{2}\mathcal{D}_{j}\left(\check{\boldsymbol{s}}_{m};\sigma\right)}{\partial p_{km}\partial\sigma_{\ell}}
  =
  O(\sigma^{2})
$
for $\ell\ne0$ while for $\ell=0$ equations (\ref{eq:SW-D}) and (\ref{eq:SW-a-another-way}) yield
\[
  \frac{\partial^{2}\mathcal{D}_{j}\left(\check{\boldsymbol{s}}_{m};\sigma\right)}{\partial p_{km}\partial\sigma_{0}}
  =
  \sigma_{0}\frac{\partial a_{jm0}}{\partial p_{km}}+O(\sigma^{2})
  =
  2\sigma_{0}\left[\left(\check{p}_{jm}-\bar{p}_{m}\right)\one\left[j=k\right]-\check{p}_{jm}\check{s}_{km}\right]
  +
  O(\sigma^{2}).
\]
By (\ref{eq:nabla_firstorder}), the instrument for $\sigma_{\ell}$,
which is the recentered $\ell$th component of the predicted residual
derivative evaluated at $\check{\boldsymbol{s}}_{m}$, $\check{\boldsymbol{p}}_{m}$,
and $\check{\sigma}$, equals:
\begin{align*}
  \hat{\nabla}_{jm\ell}^{\sigma}-\check{\nabla}_{jm\ell}^{\sigma}
  &
  =
  \sum_{k\in\mathcal{J}_{m}}\frac{\partial^{2}\mathcal{D}_{j}\left(\check{\boldsymbol{s}}_{m};\check{\sigma}\right)}{\partial p_{km}\partial\sigma_{\ell}}(\hat{p}_{km}-\check{p}_{km})
  +
  \sum_{k\in\mathcal{J}_{m}}\frac{\partial^{2}\mathcal{D}_{j}\left(\check{\boldsymbol{s}}_{m};\check{\sigma}\right)}{\partial s_{km}\partial\sigma_{\ell}}(\hat{s}_{km}-\check{s}_{km}).
\end{align*}
For $\ell\ne0$, the first term is equal to $O(\check{\sigma}^{2})$ and thus
\begin{align*}
  \hat{\nabla}_{jm\ell}^{\sigma}-\check{\nabla}_{jm\ell}^{\sigma}
  & = -2\check{\sigma}_{\ell}\check{\alpha}\check{\pi}x_{jm\ell}\sum_{k\in\mathcal{J}_{m}}\check{s}_{km}x_{km\ell}\left(\tilde{g}_{km}-\sum_{k'\in\mathcal{J}_{m}}\check{s}_{k'm}\tilde{g}_{k'm}\right)+O(\check{\sigma}^{2})\\
  & = -2\check{\sigma}_{\ell}\check{\alpha}\check{\pi}x_{jm\ell}\sum_{k\in\mathcal{J}_{m}}\check{s}_{km}\left(x_{km\ell}-\bar{x}_{m\ell}\right)\tilde{g}_{km}+O(\check{\sigma}^{2}),
\end{align*}
where the second line used the properties of covariances. In turn, for $\ell=0$,
\begin{align*}
  \hat{\nabla}_{jm0}^{\sigma}-\check{\nabla}_{jm0}^{\sigma} & =2\check{\sigma}_{0}\check{\pi}\sum_{k\in\mathcal{J}_{m}}\left[\left(\check{p}_{jm}-\bar{p}_{m}\right)\one\left[j=k\right]-\check{p}_{jm}\check{s}_{km}\right]\tilde{g}_{km}\\
  & \phantom{=}\phantom{=}-2\check{\sigma}_{0}\check{\alpha}\check{\pi}\check{p}_{jm}\sum_{k\in\mathcal{J}_{m}}\check{s}_{km}\check{p}_{km}\left(\tilde{g}_{km}-\sum_{k'\in\mathcal{J}_{m}}\check{s}_{k'm}\tilde{g}_{k'm}\right)+O(\check{\sigma}^{2})\\
  & = 2\check{\sigma}_{0}\check{\pi}\left[\left(\check{p}_{jm}-\bar{p}_{m}\right)\tilde{g}_{jm}-\check{p}_{jm}\sum_{k\in\mathcal{J}_{m}}\check{s}_{km}\tilde{g}_{km}\right]\\
  & \phantom{=}\phantom{=}-2\check{\sigma}_{0}\check{\alpha}\check{\pi}\check{p}_{jm}\sum_{k\in\mathcal{J}_{m}}\check{s}_{km}\left(\check{p}_{km}-\bar{p}_{m}\right)\tilde{g}_{km}+O(\check{\sigma}^{2}),
\end{align*}
establishing the claims of the Proposition.

\subsection{Proof of Proposition \ref{prop:nonparametric}}

Write $\beta_{j}(\boldsymbol{x}_{m})=\expec{\xi_{m}\mid\boldsymbol{x}_{m}}$,
$\tilde{\xi}_{jm}=\xi_{jm}-\beta_{j}(\boldsymbol{x}_{m})$, and $
  \tilde{\mathcal{D}_{j}}(\boldsymbol{s}_{m},\boldsymbol{x}_{m})
  =
  \mathcal{D}_{j}(\boldsymbol{s}_{m},\boldsymbol{x}_{m})-\beta_{j}(\boldsymbol{x}_{m})
$.
Then we can rewrite equation (\ref{eq:np-model}) as
\[
  p_{jm}
  =
  \tilde{\mathcal{D}}_{j}(\boldsymbol{s}_{m},\boldsymbol{x}_{m})
  -
  \tilde{\xi}_{jm}
\]
with
\[
  \expec{\tilde{\xi}_{jm}\mid\boldsymbol{g}_{m},\boldsymbol{x}_{m}}
  =
  \expec{\xi_{jm}\mid\boldsymbol{g}_{m},\boldsymbol{x}_{m}}
  -
  \beta_{j}(\boldsymbol{x}_{m})
  =
  0,
\]
where the last equality uses Assumption \ref{assu:exogenous-shocks}.
This is a standard non-parametric IV problem of \citet{newey_instrumental_2003}.
The completeness assumption implies that $\tilde{\mathcal{D}}_{j}(\boldsymbol{s}_{m},\boldsymbol{x}_{m})$
and $\tilde{\xi}_{jm}$ are identified, meaning that $\mathcal{D}_{j}(\boldsymbol{s}_{m},\boldsymbol{x}_{m})$
and $\xi_{jm}$ are identified up to an additive term $\beta_{j}(\boldsymbol{x}_{m})$.

Inverting $
  \boldsymbol{\mathcal{D}}
  =
  \left(\tilde{\mathcal{D}_{j}}\right)_{j\in\mathcal{J}_{m}}
$ yields
\[
  \boldsymbol{s}_{m}
  =
  \tilde{\mathcal{\boldsymbol{\mathcal{D}}}}^{-1}
  \left(\boldsymbol{p}_{m}+\boldsymbol{\xi}_{m}-\boldsymbol{\beta}(\boldsymbol{x}_{m}),\boldsymbol{x}_{m}\right).
\]
Thus, cross-price elasticities are given by
\[
  \frac{d\boldsymbol{s}_{m}}{d\boldsymbol{p}_{m}'}
  =
  \frac{\partial\tilde{\mathcal{\boldsymbol{\mathcal{D}}}}^{-1}}{\partial\boldsymbol{p}_{m}'}
  \left(\tilde{\boldsymbol{\mathcal{D}}}(\boldsymbol{s}_{m},\boldsymbol{x}_{m}),\boldsymbol{x}_{m}\right),
\]
where all terms are point-identified.

\section{Monte Carlo Simulation Details}
\label{sec:monte_carlo_dgp_estimation}

\subsection{Baseline Data-Generating Process}

For each of 100 simulations, we generate a set of regions $r=1,\ldots,100$
in time periods $t=1,2$. Each market $m=(r,t)$ has $j=1,\ldots,15$
products and an outside good, $j=0$. Each product has $L_{1}=2$
time-invariant characteristics with random coefficients and an intercept
(with no random coefficient). The data-generating process is as follows:
\begin{itemize}
  \item Random coefficients:
    $\eta_{i\ell}\stackrel{iid}{\sim}N(0,\sigma_{\ell}^{2})$ for all $\ell=1,\ldots,L_{1}$.

  \item Observed characteristics:
    $x_{jr\ell}\stackrel{iid}{\sim}N(0,1)$ for all $\ell=1,\ldots,L_{1}$; $x_{jr0}=1$.

  \item Idiosyncratic preference shocks:
    $\varepsilon_{ijrt}\stackrel{iid}{\sim}T1EV(0,1)$.

  \item Unobserved taste shifters:
    $\xi_{jr1}\stackrel{iid}{\sim}N(0,1),\,\xi_{jr2}=0.9\xi_{jr1}+\sqrt{1-0.9^{2}}\cdot e_{jr2}$
    with $e_{jr2}\stackrel{iid}{\sim}N(0,1)$.

  \item Unobserved cost shifters:
    $\omega_{jr1}\stackrel{iid}{\sim}N(0,1),\,\omega_{jr2}=0.9\omega_{jr1}+\sqrt{1-0.9^{2}}\cdot w_{jr2}$
    with $w_{jr2}\stackrel{iid}{\sim}N(0,1)$.

  \item Observed cost shocks:
    $g_{jr1}=0,\,g_{jr2}\stackrel{iid}{\sim}N(0,0.2^{2})$.

  \item Marginal costs:
    $c_{jm}=\gamma^{\prime}x_{jm}+\omega_{jm}+g_{jm}$.

  \item Prices:
    $p_{jm}$ solve equation (\ref{eq:price-foc}). To characterize this
    solution, first note that shares do not directly depend on prices here:
    \begin{align*}
      \boldsymbol{S}(\boldsymbol{\delta}_{m};\sigma,\boldsymbol{x}_{m}^{(1)})
      &
      =
      \int s_{jmi}d\mathcal{\mathcal{P}}(\eta_{i};\sigma)
    \end{align*}
    for
    \[
      s_{jmi}
      =
      \frac{
        \exp\left(\delta_{jm}+\sum_{\ell=1}^{L_{1}}\eta_{i\ell}x_{jm\ell}\right)
      }{
        1+\sum_{k\in\mathcal{J}_{m}}\exp\left(\delta_{km}+\sum_{\ell=1}^{L_{1}}\eta_{i\ell}x_{km\ell}\right)
      }.
    \]
    Now write the derivative of $\boldsymbol{S}(\boldsymbol{\delta}_{m};\sigma,\boldsymbol{x}_{m}^{(1)})$
    in terms of own and cross price effects. Noting $\dfrac{d}{dp_{jm}}\delta_{jm}=\alpha$, we have
    \begin{align*}
      \dfrac{d\boldsymbol{S}(\boldsymbol{\delta}_{m};\sigma,\boldsymbol{x}_{m}^{(1)})}{d\boldsymbol{p}_{m}'}
      &
      =
      \dfrac{d\boldsymbol{S}(\boldsymbol{\delta}_{m};\sigma,\boldsymbol{x}_{m}^{(1)})}{d\boldsymbol{\delta}_{m}}
      \dfrac{d\boldsymbol{\delta}_{m}}{d\boldsymbol{p}_{m}}
      =
      \Lambda(\boldsymbol{\delta}_{m};\sigma,\boldsymbol{x}_{m}^{(1)})
      -
      \Gamma(\boldsymbol{\delta}_{m};\sigma,\boldsymbol{x}_{m}^{(1)})
    \end{align*}
    where $\Lambda_{jk}=0$ for $j\ne k$ and
    \begin{align*}
      \Lambda_{jj}(\boldsymbol{\delta}_{m};\sigma,\boldsymbol{x}_{m}^{(1)})
      &
      =
      \int\alpha s_{jmi}d\mathcal{\mathcal{P}}(\eta_{i};\sigma),
      \\
      \Gamma_{jk}(\boldsymbol{\delta}_{m};\sigma,\boldsymbol{x}_{m}^{(1)})
      &
      =
      \int\alpha s_{jmi}s_{kmi}d\mathcal{\mathcal{P}}(\eta_{i};\sigma).
    \end{align*}
    \citealp{Conlon2020} note the fixed point of the following mapping
    gives the same solution:
    \[
      \boldsymbol{p}_{m}
      \mapsfrom
      \boldsymbol{c}_{m}
      +
      \Lambda(\boldsymbol{\delta}_{m};\sigma,\boldsymbol{x}_{m}^{(1)})
      \Gamma(\boldsymbol{\delta}_{m};\sigma,\boldsymbol{x}_{m}^{(1)})
      (\boldsymbol{p}_{m}-\boldsymbol{c}_{m})
      -
      \Lambda(\boldsymbol{\delta}_{m};\sigma,\boldsymbol{x}_{m}^{(1)})^{-1}
      \boldsymbol{S}(\boldsymbol{\delta}_{m};\sigma,\boldsymbol{x}_{m}^{(1)}).
    \]
    We find the fixed point of this mapping for each market $m$ and check
    that it satisfies Equation (\ref{eq:price-foc}).\footnote{
      If it does not then we solve for a fixed point of equation (\ref{eq:price-foc})
      directly. In a small number of cases (less than 0.5 markets per simulation),
      we do not find a solution and drop the market from the sample.
    }

    \item Parameters:
      $\sigma_{\ell}=4$ for all $\ell=1,\ldots,L_{1}$, $\alpha=-0.2-4\exp(0.5)$,
      $\beta_{0}=35,\beta_{1}=\beta_{2}=2,\gamma_{0}=5,\gamma_{1}=\gamma_{2}=1$.
\end{itemize}

We compute market shares of each product using equation (\ref{eq:share-1000draws})
based on 1,000 draws of the random coefficient vectors that are the
same across markets.

\subsection{Computing and Inverting Market Shares}
\label{alg:share-inversion}

We approximate $\mathcal{S}_{j}(\boldsymbol{\delta}_{rt};\sigma,\boldsymbol{x}_{r}^{(1)})$
using 250 elements of a 2-dimensional Halton sequence $\widetilde{h}_{i}=(\widetilde{h}_{i}^{1},\widetilde{h}_{i}^{2})$.
Let $(h_{i}^{1},h_{i}^{2})=(\Phi^{-1}(\widetilde{h}_{i}^{1}),\Phi^{-1}(\widetilde{h}_{i}^{2}))$,
with $\Phi^{-1}$ denoting the inverse of the standard normal CDF, and
\[
  \mathcal{S}_{j}\left(\boldsymbol{\delta}_{rt};\sigma,\boldsymbol{x}_{r}^{(1)}\right)
  =
  \dfrac{1}{250}\sum_{i=1}^{250}s_{rtji},\quad s_{rtji}
  =
  \frac{
    \exp\left(\delta_{jr2}
    +
    \sum_{\ell=1}^{2}\sigma_{\ell}h_{i}^{\ell}x_{jr}\right)
  }{
    1+\sum_{k\in\mathcal{J}_{r}}\exp\left(\delta_{kr2}+\sum_{\ell=1}^{2}\sigma_{\ell}h_{i}^{\ell}x_{kr}^{\ell}\right)
  }.
\]
To generate $\widetilde{h}_{i}^{\ell}$, we use the reverse-radix
scrambling algorithm in \citet{kociswhiten1997halton} and skip the
first 1,000 draws. Various derivatives of $\mathcal{S}_{j}$ that
we will use below (e.g., $\partial\mathcal{S}_{j}/\partial\sigma_{\ell}$)
are approximated in the same way, with $s_{rtji}$ replaced by its
derivatives.

We compute $\delta_{jrt}=\mathcal{D}_{j}(\boldsymbol{s}_{rt};\sigma,\boldsymbol{x}_{r}^{(1)})$
as the fixed point of the contraction mapping
\[
  \delta_{jrt}^{(\iota+1)}
  \mapsfrom
  \delta_{jrt}^{(\iota)}
  +
  \log s_{jrt}
  -
  \log\mathcal{S}_{j}(\boldsymbol{\delta}_{rt}^{(\iota)};\sigma,\boldsymbol{x}_{r}^{(1)}),
\]
with the starting point $\delta_{jrt}^{(0)}=\log(s_{jrt}/s_{0rt})$.
To compute the fixed point, we use the “SQUAREM” method of
\citet{Varadhan2008Simple} with a maximum of 10,000 iterations and
a tolerance of $\epsilon^{5/6}$, where $\epsilon$ is the double-precision
machine epsilon.

\subsection{Computing Instruments}
\label{sub:computing-instruments}

We focus here on the recentered IVs since the characteristic-based
IVs are given in Section \ref{sec:Monte-Carlo-Simulations}. To implement
the recentered shift-share instrument from equation (\ref{eq:nabla_firstorder}),
we first note that $\mathcal{S}_{j}(\cdot)$ and $\mathcal{D}_{j}(\cdot)$
do not explicitly depend on price in this simulation. We further assume
the researcher knows $\expec{g_{jm}\mid\boldsymbol{x}_{m},\boldsymbol{q}_{m}}=0$,
such that $\tilde{g}_{jm}=g_{jm}$. We obtain the pass-through coefficient
$\check{\pi}$ from a simple linear regression of $\Delta p_{jr}$
on $\widetilde{g}_{jr2}$, and leave the discussion of other preliminary
parameter values $(\check{\alpha},\check{\sigma})$ to below. Thus
we have for $\ell=1,\dots,L_{1}$:
\begin{equation}
  Z_{jr2\ell}^{\text{\text{SSIV}}}
  =
  \check{\alpha}\check{\pi}\sum_{k,k^{\prime}\in\mathcal{J}_{r}}
  \frac{\partial^{2}}{\partial s_{k^{\prime}r}\partial\sigma_{\ell}}
  \mathcal{D}_{j}\left(\boldsymbol{s}_{r1};\check{\sigma},\boldsymbol{x}_{r}^{(1)}\right)
  \cdot
  \frac{\partial}{\partial\delta_{kr}}
  \mathcal{S}_{k^{\prime}}(\boldsymbol{\delta}_{r1};\check{\sigma},\boldsymbol{x}_{r}^{(1)})
  \cdot
  g_{kr2}.
  \label{eq:mcappendix-ssiv}
\end{equation}

To compute the derivatives, we first apply the implicit function theorem
to equation (\ref{eq:shares}) and find
\begin{align*}
  \partial_{s_{r1}^{\prime}}
  \mathcal{D}\left(\boldsymbol{s}_{r1};\check{\sigma},\boldsymbol{x}_{r}^{(1)}\right)
  & = \left[\partial_{\delta^{\prime}}\mathcal{S}(\boldsymbol{\delta}_{r1};\check{\sigma},\boldsymbol{x}_{r}^{(1)})\right]^{-1},
  \\
  \partial_{\sigma_{\ell}}\mathcal{D}\left(\boldsymbol{s}_{r1};\check{\sigma},\boldsymbol{x}_{r}^{(1)}\right)
  & = - \left[\partial_{\delta^{\prime}}\mathcal{S}(\boldsymbol{\delta}_{r1};\check{\sigma},\boldsymbol{x}_{r}^{(1)})\right]^{-1}
  \partial_{\sigma_{\ell}}\mathcal{S}(\boldsymbol{\delta}_{r1};\check{\sigma},\boldsymbol{x}_{r}^{(1)})
\end{align*}
for $
  \boldsymbol{\delta}_{r1}
  =
  \mathcal{D}\left(\boldsymbol{s}_{r1};\check{\sigma},\boldsymbol{x}_{r}^{(1)}\right)
$.
Next, we differentiate $
  \partial_{s_{r1}^{\prime}}
  \mathcal{D}\left(\boldsymbol{s}_{r1};\check{\sigma},\boldsymbol{x}_{r}^{(1)}\right)
$
with respect to $\sigma_{\ell}$ to obtain
\begin{align*}
  \partial_{\sigma_{\ell}}
  \partial_{s_{r1}^{\prime}}
  \mathcal{D}\left(\boldsymbol{s}_{r1};\check{\sigma},\boldsymbol{x}_{r}^{(1)}\right)
  & = -
  \left[\partial_{\delta^{\prime}}\mathcal{S}(\boldsymbol{\delta}_{r1};\check{\sigma},\boldsymbol{x}_{r}^{(1)})\right]^{-1}
  \left[
    \dfrac{d}{d\sigma_{\ell}}\partial_{\delta^{\prime}}
    \mathcal{S}\left(\mathcal{D}(\boldsymbol{s}_{r1};\check{\sigma},\boldsymbol{x}_{r}^{(1)});\check{\sigma},\boldsymbol{x}_{r}^{(1)}\right)
  \right]
  \left[\partial_{\delta^{\prime}}\mathcal{S}(\boldsymbol{\delta}_{r1};\check{\sigma},\boldsymbol{x}_{r}^{(1)})\right]^{-1}
\end{align*}
where the total derivative with respect to $\sigma_{\ell}$ is:
\[
  \dfrac{d}{d\sigma_{\ell}}\partial_{\delta^{\prime}}\mathcal{S}\left(\mathcal{D}(\boldsymbol{s}_{r1};\check{\sigma},\boldsymbol{x}_{r}^{(1)});\check{\sigma},\boldsymbol{x}_{r}^{(1)}\right)
  =
  \sum_{k\in\mathcal{J}_{r}}
    \partial_{\delta_{k}}\partial_{\delta'}
    \mathcal{S}(\boldsymbol{\delta}_{r1};\check{\sigma},\boldsymbol{x}_{r}^{(1)})
    \cdot
    \partial_{\sigma_{\ell}}
    \mathcal{D}_{k}\left(\boldsymbol{s}_{r1};\check{\sigma},\boldsymbol{x}_{r}^{(1)}\right)
  +
  \partial_{\sigma_{\ell}} \partial_{\delta^{\prime}}
  \mathcal{S}(\boldsymbol{\delta}_{r1};\check{\sigma},\boldsymbol{x}_{r}^{(1)}).
\]

The recentered exact formula instrument is given by
\begin{equation}
  Z_{jr2\ell}^{FIV}
  =
  \frac{\partial}{\partial\sigma_{\ell}}
  \mathcal{D}_{j}\left(\mathcal{S}(\boldsymbol{\delta}_{r1}+\check{\alpha}\check{\pi}\boldsymbol{g}_{r2};\check{\sigma},\boldsymbol{x}_{r}^{(1)});\check{\sigma},\boldsymbol{x}_{r}^{(1)}\right)-\dfrac{1}{20}\sum_{c=1}^{20}\frac{\partial}{\partial\sigma_{\ell}}\mathcal{D}_{j}\left(\mathcal{S}(\boldsymbol{\delta}_{r1}+\check{\alpha}\check{\pi}\boldsymbol{g}_{r2}^{(c)};\check{\sigma},\boldsymbol{x}_{r}^{(1)});\check{\sigma},\boldsymbol{x}_{r}^{(1)}\right),\label{eq:mcappendix-fiv}
\end{equation}
where $\boldsymbol{g}^{(c)}$ is obtained by randomly permuting the
shocks $\boldsymbol{g}$ across products and markets.

\subsection{Estimation Procedure}
\label{sub:estimation-procedure}

Given any moment condition $\expec{h(\varphi)}=0$ with the sample
analog $h_{N}(\varphi)=\dfrac{1}{N}\sum_{j,r}h_{jr}(\varphi)$ where
$N$ is the number of product-market pairs, we estimate $\varphi$
via non-linear GMM as
\begin{equation}
  \widehat{\varphi}
  =
  \arg\min_{\varphi}Q_{N}(\varphi)
  =
  \dfrac{1}{2}h_{N}(\varphi)^{\prime}W_{N}h_{N}(\varphi),
  \label{eq:GMM}
\end{equation}
where $N$ is the number of product-market pairs and $W_{N}$ is a
positive-definite weighting matrix. Appendix \ref{appx:chariv-estimation}
details the estimation using characteristic-based IVs; Appendices
\ref{appx:reciv-estimation} and \ref{appx:iterative-reciv-estimation}
detail the estimation using recentered IVs.

\subsubsection{Estimation with Characteristic-Based IVs}
\label{appx:chariv-estimation}

Let $Z_{jr2}$ be a vector collecting $g_{jr2}$, $x_{jr}$, and two
characteristic-based IVs for $\sigma=(\sigma_{1},\sigma_{2})$: either
BLP (sum of competitor characteristics) IVs or the \citet{Gandhi2015}
local or quadratic differentiation IVs. We use the second time period
to estimate the parameters $\varphi=(\sigma,\alpha,\beta)$, with
\begin{align*}
  h_{jr}(\varphi) & = \xi_{jr2}\cdot Z_{jr2}
\end{align*}
for $
  \xi_{jr2}
  =
  \mathcal{D}_{j}(\boldsymbol{s}_{r2};\sigma,\boldsymbol{x}_{r}^{(1)})
  -
  \alpha p_{jr2}
  -
  \beta'x_{jr}
$
and $\mathcal{D}_{j}(\boldsymbol{s}_{rt};\sigma,\boldsymbol{x}_{r}^{(1)})$
from Appendix \ref{alg:share-inversion}.

We rewrite the problem (\ref{eq:GMM}) as a numerical optimization
over $\sigma$, concentrating out $\alpha,\beta$. To do so, note
that the minimization over $\alpha,\beta$ is a linear IV-GMM problem
with the solution
\begin{equation}
  (\alpha(\sigma),\beta(\sigma)^{\prime})^{\prime}
  =
  \left(X^{\prime}ZW_{N}Z^{\prime}X\right)^{-1}
  X^{\prime}ZW_{N}Z^{\prime}\mathcal{D}\left(\sigma\right),
  \label{eq:mcappendix-concentrated-out}
\end{equation}
where $X$ collects $(p_{jr2},x_{jr}^{\prime})$, $Z$ collects $\left(Z_{jr2}^{\prime}\right)$,
and $\mathcal{D}(\sigma)$ collects $\mathcal{D}_{j}(\boldsymbol{s}_{r2};\sigma,\boldsymbol{x}_{r}^{(1)})$
across $jr$ pairs.

To solve $\min_{\sigma}Q_{N}(\sigma)\equiv Q_{N}\left(\sigma,\alpha(\sigma),\beta(\sigma)\right)$,
we use the following procedure:
\begin{enumerate}
  \item Take 50 points for $\sigma$ from $[0,10]^{2}$; we use the
    deterministic “$R_{d}$” sequence in \citet{Halchenko2020grid}. Choose
    $\sigma^{(0)}$ as the point that minimizes $Q_{N}(\sigma)$.

  \item Optimize over $\sigma$ via the Gauss-Newton regression algorithm,
    following the suggestion in \citet{Gandhi2015}:\footnote{
      To be more precise, we force our $\sigma$ to be strictly positive
      by using the softplus transformation $\sigma\equiv\log(1+\exp(\tilde{\sigma}))$
      and conducting the search over $\tilde{\sigma}$ (adjusting the derivatives by
      $\exp(\tilde{\sigma})/(1+\exp(\tilde{\sigma}))$ for the change
      of variables). Also, note that the Gauss-Newton regression can be
      derived using the first-order approximation $
        h_{N}(\sigma)
        \approx
        h_{N}(\sigma^{(\iota)})+H_{N}(\sigma^{(\iota)})(\sigma-\sigma^{(\iota)})
      $;
      the algorithm solves for the step size $b^{(\iota)}\equiv\sigma-\sigma^{(\iota)}$
      via linear GMM. Isolating $b$ from the first-order condition $
        0
        =
        \partial_{b}(h_{N}(\sigma^{(\iota)})+H_{N}(\sigma^{(\iota)})b)^{\prime}
        W_{N}
        (h_{N}(\sigma^{(\iota)})+H_{N}(\sigma^{(\iota)})b)
      $
      gives the expression.
    }
    \begin{itemize}
      \item At each iteration, set $\sigma^{(\iota+1)}=\sigma^{(\iota)}+b^{(\iota)}$, with
        \begin{align*}
          b^{(\iota)}
          &
          =
          -\left(H_{N}(\sigma^{(\iota)})^{\prime}W_{N}H_{N}(\sigma^{(\iota)})\right)^{-1}
          H_{N}(\sigma^{(\iota)})^{\prime}W_{N}h_{N}(\sigma^{(\iota)})
        \end{align*}
        where
        \[
          h_{N}(\sigma)
          \equiv
          h_{N}(\sigma,\alpha(\sigma),\beta(\sigma))
          =
          \frac{1}{N}
          \mathcal{D}\left(\sigma\right)^{\prime}
          \left(Z-ZW_{N}Z^{\prime}X(X^{\prime}ZW_{N}Z^{\prime}X)^{-1}X^{\prime}Z\right)
        \]
        and
        \[
          H_{N}(\sigma)^{\prime}\equiv\partial_{\sigma}h_{N}(\sigma)^{\prime}
          =
          \frac{1}{N}
          \left(\partial_{\sigma}\mathcal{D}(\sigma)^{\prime}\right)
          \left(Z-ZW_{N}Z^{\prime}X(X^{\prime}ZW_{N}Z^{\prime}X)^{-1}X^{\prime}Z\right).
        \]

      \item Continue updating $\sigma^{(\iota)}$ while $\Vert b^{(\iota)}\Vert>\epsilon^{1/2}$ and $\iota<100$.
    \end{itemize}

  \item If $\iota\ge100$ or $\left\Vert\dfrac{d}{d\sigma}Q_{N}(\widehat{\sigma})\right\Vert > \epsilon^{1/3}$,
    discard step 2 and minimize $Q_{N}(\sigma)$ via a BFGS-based algorithm
    (as suggested in \citealp{Conlon2020}) with the same starting parameter
    $\sigma^{(0)}.$ We use MATLAB's implementation via \texttt{fmincon} with a
    lower bound of $0$ for $\sigma$.
\end{enumerate}
We note that while the system of moment conditions is just-identified
and thus the choice of the positive definite matrix $W_{N}$ should
be irrelevant, in practice we set $W_{N}=\left(\frac{1}{N}\sum_{j,r}Z_{jr2}Z_{jr2}^{\prime}\right)^{-1}$.

\subsubsection{Continuously Updating Estimation with Recentered IVs}
\label{appx:reciv-estimation}

Let $Z_{jr2}$ be a vector collecting $g_{jr2}$ and either the shift-share
or exact prediction IVs in Section \ref{sec:General-Approach}. We
use the sample analog of the moment condition in first-differences, with
\begin{align*}
  h_{jr}(\theta) = \Delta\xi_{jr}\cdot Z_{jr2},
\end{align*}
where $
  \Delta\xi_{jr}
  =
  \Delta\mathcal{D}_{j}(\boldsymbol{s}_{rt};\sigma,\boldsymbol{x}_{r}^{(1)})
  -
  \alpha\Delta p_{jr}
$,
and $\mathcal{D}_{j}(\boldsymbol{s}_{rt};\sigma,\boldsymbol{x}_{r}^{(1)})$
is computed as in Appendix \ref{alg:share-inversion}. Note that we
do not estimate $\beta$ in this analysis. To estimate $\theta$ we
follow steps similar to Appendix \ref{appx:chariv-estimation}:
\begin{enumerate}
  \item Take 50 points from $[0,10]^{2}$ in the same way.
    Define $\alpha(\sigma)$ as the regression slope of
    $\Delta\mathcal{D}_{j}(\boldsymbol{s}_{rt};\sigma,\boldsymbol{x}_{r}^{(1)})$
    on $\Delta p_{jr}$ (and an intercept) instrumented with $g_{jr2}$.  Choose
    $\sigma^{(0)}$ that minimizes $Q_{N}(\sigma,\alpha(\sigma))$ among the 50
    points while recomputing $Z_{jr2}$ at each $(\sigma,\alpha(\sigma))$ via
    equation (\ref{eq:mcappendix-ssiv}) or equation (\ref{eq:mcappendix-fiv}). Set
    $\alpha^{(0)}=\alpha(\sigma^{(0)})$.\footnote{
      Note that, unlike Appendix \ref{appx:chariv-estimation},
      $\alpha(\sigma)\ne\arg\min_{\alpha}Q_{N}(\sigma,\alpha)$.
      Here the instrument depends on $\alpha$, such that it is not possible
      to concentrate out $\alpha$. Rather, $\alpha(\sigma)$ is a reasonable
      starting point that sets one of the moments to zero:
      $\expec{\Delta\xi_{jr}\cdot g_{jr2}}=0$. We do not use $\alpha(\sigma)$ in Step 2.
  }

  \item Estimate $\theta$ using Gauss-Newton regression, searching
    over both $\sigma$ and $\alpha$:\footnote{
      Like in Appendix \ref{appx:chariv-estimation}, we use the softplus
      transformation for $\sigma$. In addition, we force $\alpha$ to be negative
      by setting $\alpha\equiv-\log(1+\exp(\widetilde{\alpha}))$ and searching
      over $\tilde{\alpha}$ (adjusting the derivatives accordingly).
    }
    \begin{itemize}
      \item At each step, set $\theta^{(\iota+1)}=\theta^{(\iota)}+b^{(\iota)}$, with
        \begin{align*}
          b^{(\iota)} &
          =
          -
          \left(H_{N}(\theta^{(\iota)})^{\prime}W_{N}H(\theta^{(\iota)})\right)^{-1}
          H_{N}(\theta^{(\iota)})^{\prime}W_{N}h_{N}(\theta^{(\iota)})
        \end{align*}
        and $H_{N}(\theta)$ computed numerically using finite-differences (to
        account for the fact the instrument also changes with the parameters).

      \item Continue updating $\theta^{(\iota)}$ while $\Vert b^{(\iota)}\Vert>\epsilon^{1/2}$ and $\iota<100$.
    \end{itemize}

  \item If $\iota\ge100$ or $\Vert\partial_{\theta}Q_{N}(\widehat{\theta})\Vert>\epsilon^{1/3}$,
    discard step 2 and minimize $Q_{N}(\theta)$ using a BFGS-based algorithm
    with the same starting parameter $\theta^{(0)}.$
\end{enumerate}
Like in Appendix \ref{appx:chariv-estimation}, the system of moment
conditions here is just-identified; we set $W_{N}=I$ so that estimation
will not require recomputing the weighting matrix at each iteration.

\subsubsection{Iterative Estimation with Recentered IVs}
\label{appx:iterative-reciv-estimation}

Iterative estimation is similar to Appendix \ref{appx:reciv-estimation},
but the instrument at each step depends on the parameter estimates
from the previous step. This allows us to concentrate out $\alpha$
and use analytical derivatives. We use the following procedure:
\begin{enumerate}
  \item Choose $\sigma^{(0)}$ and $\alpha^{(0)}$ as in Step 1 in Appendix \ref{appx:reciv-estimation}.

  \item Compute $Z_{jr2}^{(\iota)}$ via equation (\ref{eq:mcappendix-ssiv}) or
    equation (\ref{eq:mcappendix-fiv}) using $
      \widehat{\theta}^{(\iota)}
      =
      (\widehat{\sigma}^{(\iota)},\widehat{\alpha}^{(\iota)})
    $.

  \item Obtain estimates $\widehat{\theta}^{(\iota+1)}$ as follows:
    \begin{itemize}
      \item Concentrate out $\alpha$ as
        \[
          \alpha^{(\iota)}(\sigma)
          =
          \left(X^{\prime}Z^{(\iota)}W_{N}Z^{(\iota)}{}^{\prime}X\right)^{-1}
          X^{\prime}Z^{(\iota)}W_{N}Z^{(\iota)}{}^{\prime}\Delta\mathcal{D}\left(\sigma\right),
        \]
        where $X$ collects $\Delta p_{jr}-\overline{\Delta p}$, $Z^{(\iota)}$
        collects $Z_{jr2}^{(\iota)}-\overline{Z^{(\iota)}}$, and
        $\Delta\mathcal{D}(\sigma)$ collects
        $\Delta\mathcal{D}_{j}(\boldsymbol{s}_{rt};\sigma,\boldsymbol{x}_{r}^{(1)})-\overline{\Delta\mathcal{D}(\sigma)}$
        across $jr$ pairs (with bars denoting sample averages). Note that
        concentrating $\alpha$ out is possible because the instruments $Z_{jr2}^{(\iota)}$
        are fixed based on $\widehat{\theta}^{(\iota)}$.

      \item Starting from $\hat{\sigma}^{(\iota)}$, search over $\sigma$ to
        minimize $Q_{N}^{(\iota)}(\sigma,\alpha^{(\iota)}(\sigma))$, which
        depends on $\iota$ via the instruments $Z_{jr2}^{(\iota)}$. Use
        Gauss-Newton regression iterations as in Step 2 of Appendix
        \ref{appx:chariv-estimation}.  If failed, discard the
        estimates and use a BFGS-based algorithm as in Step 3 of of
        Appendix \ref{appx:chariv-estimation}. Save the result as
        $\hat{\sigma}^{(\iota+1)}$.

      \item Set $\hat{\alpha}^{(\iota+1)}=\alpha^{(\iota)}\left(\hat{\sigma}^{(\iota+1)}\right)$.
    \end{itemize}

  \item Continue updating $\widehat{\theta}^{(\iota)}$ while $
    \left\Vert \widehat{\sigma}^{(\iota+1)}-\widehat{\sigma}^{(\iota)}\right\Vert
    >
    \epsilon^{1/2}
  $ and $\iota<100$.
\end{enumerate}

As in Appendix \ref{appx:reciv-estimation}, we set $W_{N}=I$.

\end{document}